\documentclass[a4paper,11pt, onecolumn]{article}

\usepackage{amsmath} 		
\usepackage{amssymb} 		
\usepackage{amsthm} 		
\usepackage{bm} 			
\usepackage{eucal} 			
\usepackage{latexsym} 		
\usepackage{mathrsfs} 		
\usepackage{textcomp} 		
\usepackage{theorem} 		
\usepackage{siunitx} 		
\usepackage{slashed} 		

\usepackage{array} 			
\usepackage{booktabs} 		
\usepackage{caption} 		
\usepackage{float} 			
\usepackage{graphics}		
\usepackage{graphicx} 		
\usepackage{longtable} 		
\usepackage{lscape} 		
\usepackage{subfigure} 		
\usepackage{wrapfig} 		

\usepackage{ae} 			
\usepackage[english]{babel} 	
\usepackage[T1]{fontenc} 	
\usepackage[active]{srcltx} 		

\usepackage{color} 			
\usepackage{fancybox} 		
\usepackage{fancyhdr} 		
\usepackage[top=2.5cm,left=2.5cm,right=2.5cm,bottom=2.5cm]{geometry} 		
\usepackage{lettrine} 		
\usepackage[parfill]{parskip}    
\usepackage{rotating} 		
\usepackage{setspace} 		

\usepackage{cite} 			

\usepackage{acronym} 		
\usepackage[toc,page,title,header]{appendix} 	
\usepackage{hyperref} 		
\usepackage{makeidx} 						
\usepackage{epstopdf} 		

\title{\bf An analysis of the influence of background subtraction and quenching on jet observables in heavy-ion collisions}
\author{{\normalsize Liliana Apolin\'ario$^{a,b}$, N\'estor Armesto$^a$ and Leticia Cunqueiro$^{c,d}$} \\ \\ \\
{\normalsize \it $^a$ Departamento de F\'{\i}sica de Part\'{\i}culas and  IGFAE,}\\
{\normalsize \it Universidade de Santiago de Compostela, 15706 Santiago de Compostela, Galicia-Spain}\\ \\
{\normalsize \it $^b$ CENTRA, Instituto Superior T\'ecnico, Universidade T\'ecnica de Lisboa,} \\
{\normalsize \it Av. Rovisco Pais, P-1049-001
Lisboa, Portugal}\\ \\
{\normalsize \it $^c$ Laboratori Nazionali di Frascati, INFN, Via E. Fermi, 40, 
00044 Frascati (Roma), Italy}\\ \\
{\normalsize \it $^d$ PH Department, CERN, CH-1211 Gen\`eve 23,
Switzerland}\\ \\ \\
{\normalsize E-mails: {\tt lilianamarisa.cunha@usc.es, nestor.armesto@usc.es,}}\\ {\normalsize {\tt leticia.cunqueiro.mendez@cern.ch}.}
}
                            
\begin{document}

\maketitle

\begin{abstract} 

Subtraction of the large background in reconstruction is a key ingredient in jet studies in high-energy heavy-ion collisions at RHIC and the LHC. Here we address the question to which extent the most commonly used subtraction techniques are able to eliminate the effects of the background on the most commonly discussed observables at present: single inclusive jet distributions, dijet asymmetry and azimuthal distributions. We consider two different background subtraction methods, an area-based one implemented through the FastJet package and a pedestal subtraction method, that resemble the ones used by the experimental collaborations at the LHC. We also analyze different ways of defining the optimal parameters in the second method. We use a toy model that easily allows variations of the background characteristics: average background level and fluctuations and azimuthal structure, but cross-checks are also done with a Monte Carlo simulator. Furthermore, we consider the influence of quenching using Q-PYTHIA on the dijet observables with the different background subtraction methods and, additionally, we examine the missing momentum of particles. The average background level and fluctuations affect both single inclusive spectra and dijet asymmetries, although differently for different subtraction setups. A large azimuthal modulation of the background has a visible effect on the azimuthal dijet distributions. Quenching, as implemented in Q-PYTHIA, substantially affects the dijet asymmetry but little the azimuthal dijet distributions. Besides, the missing momentum characteristics observed in the experiment are qualitatively reproduced by Q-PYTHIA.

\end{abstract}

\section{Introduction}
\label{intro}

\par The characterization of the medium produced in ultrarelativistic heavy-ion collisions through observables whose production involves a large perturbative scale - hard probes - is one of the main subjects in high-energy nuclear physics at present. Among the different hard probes, the suppression of energetic particles and of jet-like correlations generically called jet quenching, see  \cite{dEnterria:2009am,Wiedemann:2009sh,Majumder:2010qh,Armesto:2011ht} for recent reviews, lies among the most prominent ones. Indeed, at the Relativistic Heavy Ion Collider (RHIC) the suppression of high transverse momentum particles and of non-photonic electrons from heavy flavor decays, and the suppression of back-to-back correlations, has been observed \ \cite{Adler:2006bv,Adare:2008ae,Abelev:2009gu,Adams:2006yt,Adare:2006hc,Abelev:2006db}. The same observables are currently under study at the Large Hadron Collider (LHC) \cite{Aamodt:2010jd,:2012eq,CMS:2012aa,ALICE:2012ab,Aamodt:2011vg} where the suppression of high transverse momentum charged particles, charmed mesons, and of back-to-back correlations, have been measured.

\par While studies at hadron level have been crucial in order to establish the existence and to gain understanding on the jet quenching phenomenon, the study of jets was proposed  long ago as a complementary possibility\cite{Baier:1999ds,Salgado:2003rv}. Specifically, single hadron spectra are supposed to be mostly sensitive to the medium-induced energy loss of the leading parton coming from a hard scattering, while jet-related observables should offer information about the medium modifications on the QCD branching process. The latter are expected to be affected by  potential biases in a different manner than the former and, thus, they offer a possibility to additionally constrain the mechanism of energy loss and  characterize the medium produced in the collisions.

\par Jet studies at RHIC \cite{Putschke:2008wn,Salur:2008hs,Bruna:2009em,Ploskon:2009zd,Lai:2011zzb} face several difficulties both at the detector level and due to kinematical limitations. On the other hand, in the last two years several jet-related analysis have been performed at the LHC \cite{Aad:2010bu,:2012is,Chatrchyan:2011sx,Chatrchyan:2012nia,Chatrchyan:2012gt,Chatrchyan:2012gw} that have triggered great interest and a large experimental, phenomenological and theoretical activity. Summarizing, the results show: (i) a larger imbalance of the transverse energy of leading and subleading jets in PbPb collisions than in pp and increasing with centrality, which indicates the existence of medium-induced energy loss; (ii) a similar  azimuthal distribution between leading and subleading jets in central PbPb collisions to that in pp, apparently pointing to the absence of sizable medium-induced broadening in transverse momentum; (iii) an excess of soft particles at large angles with respect to the subleading jet in PbPb collisions and increasing with increasing dijet momentum imbalance, compared to Monte Carlo expectations which reproduce pp data; (iv) a lack of sizable modifications of the hard jet fragmentation  (i.e. the fragmentation into particles with energies close to the jet energy) between pp and PbPb collisions. These observations look, at first sight, challenging for the standard explanation of jet quenching in terms of medium-induced gluon radiation in which energy loss and broadening are linked and the induced radiation is semi-hard. Several phenomenological works have appeared \cite{CasalderreySolana:2010eh,Qin:2010mn,He:2011pd,Young:2011qx,Lokhtin:2011qq,Renk:2012cx,Renk:2012cb} that claim to predict totally or partially the observed experimental results in terms of different mechanisms and discuss the compatibility of the observed results with those at hadron level \cite{Adler:2006bv,Adare:2008ae,Abelev:2009gu,Adams:2006yt,Adare:2006hc,Abelev:2006db,Aamodt:2010jd,:2012eq,CMS:2012aa,ALICE:2012ab,Aamodt:2011vg} and their standard explanations.

\par On the other hand, jet studies in hadronic collisions  demand  a procedure in order to separate jet constituents from the background of soft particles not coming from the shower of a hard parton, see \cite{Salam:2009jx} and refs. therein. This is true both in pp collisions due to pileup but becomes mandatory in heavy-ion collisions where the energy per unit area in pseudorapidity $\times$ azimuth becomes of the order of the energy of the jet, ${\cal O}(100\  {\rm GeV})$. Strategies to deal with this situation have been devised e.g. in \cite{Kodolova:2007hd,Cacciari:2010te,Cacciari:2012mu}. On the phenomenological side, the ideal situation consists in considering that the background subtraction applied to experimental data has completely removed the effects of the background on the jet energy and substructure. Assuming that this is the case and that jet and medium are decoupled, the only thing to be done in order to compare with data is computing the jet spectrum in elementary (pp) collisions without and with medium effects (as usually done  in the study of hadron spectra).

\par In a realistic situation, however, it may turn out that this ideal aim cannot be achieved. On a theoretical level, the coupling between jet and  medium may turn out to be so important that it cannot be neglected and the simple embedding of an elementary collision in a heavy-ion background does not work. On the reconstruction level, it may happen that the jet-background back-reaction (i.e. how the hard content of the jet changes if its hard constituents are clustered together with a soft background \cite{Cacciari:2010te}) cannot be controlled and the jet energy and substructure become altered in a way that cannot be corrected for a generic background already for jets not affected by medium effects. In this case the use of  realistic heavy-ion events for background, and the same reconstruction techniques used in the experimental analysis, is compulsory and the background subtraction becomes part of the jet definition. Indeed, it has been noted in \cite{Cacciari:2011tm} that the fluctuations in a single-event background may affect the dijet energy imbalance and thus the extraction of medium properties and the characterization of the energy loss mechanism. This point has triggered experimental studies \cite{Abelev:2012ej} on the effect of background fluctuations on the jet enegy resolution. Thus, a full phenomenological analysis of jet production in heavy-ion collisions would require: (a) the generation of medium-modified jet events and a realistic fluctuating background, either coupled or decoupled; (b) the reconstruction of the jets and subtraction of the background in a way as close as possible of the experimental analysis, eventually including detector effects like calorimeter granularity, particle momentum cut-offs and particle species dependence, or efficiencies.

\par Nevertheless, the mentioned procedure is very involved and our understanding of how to model a realistic fluctuating background and how to treat the coupling  between medium and hard probe is still unsatisfactory. As a consequence, none of the phenomenological analysis of jet observables in PbPb at the LHC available until now  follow the full procedure outlined above. Therefore, it becomes extremely important to understand how the different ingredients used in the analysis of jets in heavy-ion collisions affect the different observables. This is compulsory for a more precise extraction of medium properties and characterization of the mechanism responsible for the medium-induced modification of jets. It will also allow the design of new observables that may complement the existing ones in order to achieve these aims. 

\par In this work, we focus on the effects of two ingredients: background fluctuations in a given event and quenching (implemented as radiative energy loss in a given Monte Carlo model, Q-PYTHIA \cite{Armesto:2009fj}), on several jet observables like the dijet energy imbalance, the azimuthal distributions and the single jet spectra. Some considerations on the momentum imbalance in jet events using particles will also be presented. We will assume a totally decoupled scheme for the jets and the medium. For the jet signal, we generate pp collisions with different quenching strengths. For the background, we use a simple, flexible toy model that allows us to control the mean values of multiplicities \cite{Aamodt:2010cz,ATLAS:2011ag,Chatrchyan:2011pb} while varying the fluctuations of the background and other characteristics like the azimuthal asymmetries \cite{ALICE:2011ab,ATLAS:2011ah,Chatrchyan:2012xq}.

\par We use two background subtraction methods: an area-based background subtraction \cite{Cacciari:2007fd} through FastJet \cite{Cacciari:2011ma,Cacciari:2005hq}; and a pedestal subtraction that attempts to mimic the method used by CMS  \cite{Kodolova:2007hd,Chatrchyan:2011sx,Chatrchyan:2012nia}. In this latter case, two procedures for fixing the parameters in the method will be examined. For both methods, jets will be defined using the anti-$k_t$ algorithm \cite{Cacciari:2008gp} as done by ATLAS \cite{Aad:2010bu,:2012is}\footnote{ATLAS uses a new method for background subtraction in their recent analysis of the nuclear modification factor of jets \cite{:2012is}.} and in the most recent work by CMS \cite{Chatrchyan:2012nia}.

\par The paper is structured as follows: in Section \ref{background} we describe the observables, the toy model for the background and the methods for background subtraction. In Section \ref{bkgParam} we show the results for the observables for different choices of background parameters and for both background subtraction techniques. In Section \ref{quench} we introduce the quenching and show the results for the mentioned observables, including the distribution of missing momentum for particles in jet events. Our conclusions will be presented in Section \ref{conclu}.

\par Finally, let us stress that the aim of this work is not to criticize the experimental analyses. Both ATLAS and CMS \cite{Aad:2010bu,Chatrchyan:2011sx,Chatrchyan:2012nia,Chatrchyan:2012gt}, using widely different experimental apparatuses and analysis techniques and even observables, reach similar conclusions. Our aim is to investigate to which extent the approximation of neglecting the background effects on jet reconstruction in heavy-ion collisions fails, and to clarify which medium characteristics have to be more carefully modeled in order to make a sensible job when analyzing phenomenologically such observables. We also attempt to scrutinize how quenching may affect these observables when a background is introduced. For that, we mimic - admittedly in a simplified way - the methods used in the experimental analyses as it may turn out that they show different sensitivities to medium characteristics and quenching.

\section{Jet observables, background model and jet reconstruction}
\label{background}

\subsection{Jet observables}

\par In order to study jets in a heavy-ion environment, we will assume that they are fully decoupled from the medium. We will generate the jet signal via pp collisions at $\sqrt{s}=2.76$ TeV in Q-PYTHIA \cite{Armesto:2009fj} that is based on PYTHIAv6.4.18 \cite{Sjostrand:2006za}. We use the DW tune \cite{Albrow:2006rt}, a minimum $p_T$ in the hard scattering of 70 GeV/c (except for the spectrum of jets, where several files of PYTHIA were simulated with different bins of $p_T$ in the hard scattering, from 5 GeV/c to 302 GeV/c) and only QCD physics processes (i.e. PYTHIA settings {\tt MSEL=1}, {\tt CKIN(3)=70.0}).  We have checked that this minimum $p_T$ offers a compromise between minimizing the CPU time required for the simulations  and minimizing the biases in the distributions (observed in \cite{Cacciari:2011tm}) for the minimum $E_T$ of the leading jet that we will use, see below. Samples of $10^5$ pp events are generated in this way for each set of parameters\footnote{Note that Q-PYTHIA with no medium effects ($\hat q=0$) is identical to standard PYTHIA.}. For details of the simulation with quenching, we refer to Section \ref{quench}. Background subtraction is not performed for pp events as it gives a negligible effect.

\par We will examine the following observables (details of the kinematical cuts will be provided in Subsection \ref{reco}):

\begin{itemize}
	\item The inclusive jet spectrum in $E_T$.
	\item For the hardest and next-to-hardest jets in the event, with transverse energies $E_{T1}$ and $E_{T2}$ respectively, the distribution in azimuthal angle between them,

\begin{equation}
	\Delta \phi = | \phi_1 - \phi_2|
\end{equation}

and the dijet energy imbalance or asymmetry, defined as

\begin{equation}
	A_J=\frac{E_{T1}-E_{T2}}{E_{T1}+E_{T2}}\,.
	\label{aj}
\end{equation}

	\item The average missing transverse momentum defined as

\begin{equation}
	\left\langle \slashed{p}_T^\parallel \right\rangle = \sum_{i} -p_T^i \cos (\phi_i -\phi_{\rm leading\ jet} ),
	\label{eq:misspt}
\end{equation}

where the sum runs over all charged particles in the event with transverse momenta $p_T^i$ and azimuthal angle $\phi_i$. Note that the expression above has a sign that sets the projection of particles on the hemisphere of the leading jet to give a negative contribution to the sum.

\end{itemize}

\subsection{Toy model for the background}
\label{toymodel}

\par Background subtraction is needed in order to attempt to define the jet characteristics when the jet is produced together with an underlying event. In order to study the influence of the background subtraction method on the different jet observables, we will use a toy model for generating particles uniformly in pseudorapidity $\eta$ and azimuthal angle $\phi$  along the full detector acceptance, with the following distribution in transverse momentum $p_T$ which smoothly matches a thermal-like spectrum to a power law:
\begin{equation}
	f(p_T) \propto  \left\{
	\begin{array}{lr}
		\text{e}^{-p_T/T}, & p_T \leq \alpha T, \\
		\text{e}^{- \alpha} \left( \frac{ \alpha T }{ p_T } \right)^\alpha, & p_T > \alpha T.
	\end{array} \right.
	\label{thermal}
\end{equation}
Here $\alpha = 6$ is a power suggested by perturbative calculations and $T$ is a 'temperature' which determines the exponential behavior of the soft part of the spectrum. We generate in this way $N$ particles with a mean value corresponding to a multiplicity  $dN/d\eta = 2100$ which is allowed to fluctuate from event to event following a Gaussian distribution with a dispersion of a 4 \% of the mean value. In this way we attempt to mimic the $0\div 10$ \% centrality class \cite{Aamodt:2010cz,ATLAS:2011ag,Chatrchyan:2011pb} in the experimental analyses \cite{Aad:2010bu,:2012is,Chatrchyan:2011sx,Chatrchyan:2012nia,Chatrchyan:2012gt}.

\begin{figure}[htbp]
	\begin{center}
		\subfigure[Average energy density per unit area, $\rho$.]{
			\label{fig:SpecFJ1}
			\includegraphics[width=0.47\textwidth]{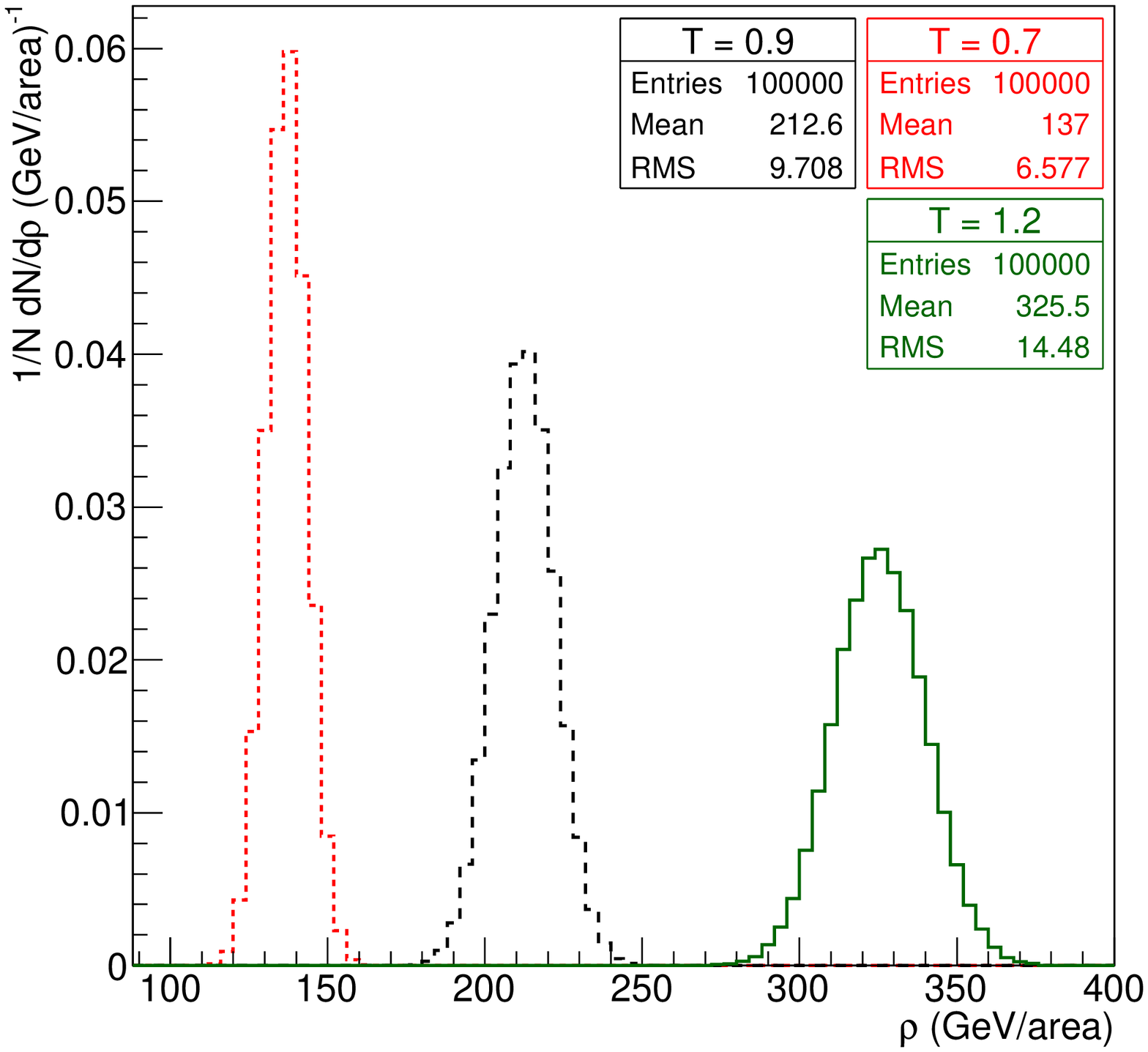}
		}
		\subfigure[Region-to-region fluctuation on a jet of R = 0.4, $\sigma_{jet}$.]{
			\label{fig:SpecCMS1}
			\includegraphics[width=0.47\textwidth]{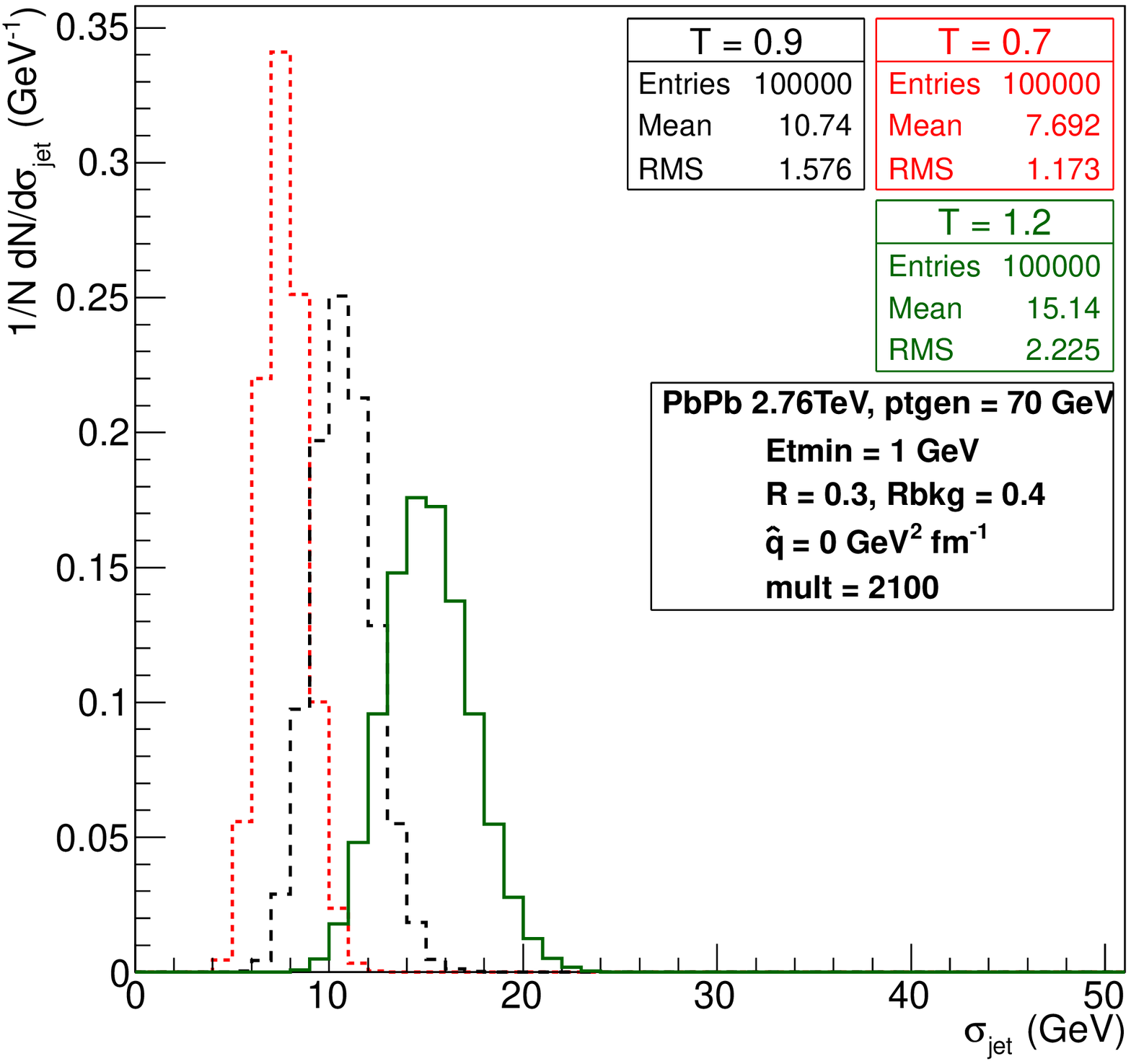}
		}
		\caption{Energy density per unit area, $\rho$ (left) and region-to-region fluctuations $\sigma_{jet}$ on a jet of R = 0.4  (right), for different values of temperature $T=0.7$ (red dotted), 0.9 (black dashed) and 1.2  GeV (green solid) in the background toy model.}
		\label{fig:RhoSigma}
	\end{center}
\end{figure}

\par In Eq. (\ref{thermal}), the temperature is used as a free parameter to control the main characteristics of the background, the average level of energy deposition per unit area ($\rho$) and fluctuations ($\sigma_{jet}$) in $\eta\times \phi$. These two quantities are computed through FastJet v.2.4.2 with active areas, see details in the following Subsection. The corresponding values\footnote{Our $\sigma_{jet}$ values are in rough agreement with the ones measured by ALICE \cite{Abelev:2012ej} and ATLAS \cite{:2012is}, although our average energy density per unit area, $\rho$ is larger than the one measured in the experiments. ALICE, using only charged tracks, a minimum cut of $p_{Tmin} = 1$ GeV and reconstructed jets using the $k_t$-algorithm with $R = 0.4$, finds $\sigma_{jet}$ to be between $8.5$ and $ 8.8$ GeV. Scaling by $\sqrt{1.5}$ to account for neutral particles (assumed uncorrelated with charged), this would correspond to $\sigma_{jet} \simeq 10.8$ GeV. As for ATLAS, the comparison is not so straightforward since the effective cut-off for charged particles is $p_{Tmin} \simeq 0.5$ GeV and  fluctuations from the calorimeter are included in the experimental value. Nevertheless, taking the definition that is closer to the area occupied by a jet with $R = 0.4$, ATLAS gives $\sigma_{jet} \simeq 12.5$ GeV.} can be seen in Fig. \ref{fig:RhoSigma}.

Additionally, this simple toy model allows the introduction of an event-by-event azimuthal modulation in the form of elliptic $v_2$ and triangular $v_3$ flow that we will do, in Subsection \ref{flow}, in order to check the sensitivity of the reconstruction to those additional fluctuations that exist in a real event \cite{ALICE:2011ab,ATLAS:2011ah,Chatrchyan:2012xq}.

\par It should be stressed that  this model is not realistic: the transverse momentum spectrum does not describe experimental data \cite{Aamodt:2010jd,:2012eq,CMS:2012aa} no particle species dependence is considered, the $\eta$-distribution is flat and, although the experimental value of $\sigma_{jet}$ lies within the range that we consider, our value of $\rho$ is larger than the experimental one \cite{Abelev:2012ej} for all considered temperatures. In any case, in this work we do not attempt to provide an extraction of the medium characteristics from jet observables at the LHC, but to identify some potential challenges for phenomenological analysis. Therefore the flexibility of the model and the fast generation of backgrounds possible through it gives a substantial advantage for this study, though it presents limitations for the missing transverse momentum that we will comment in Subsection \ref{misspt}. The use of some Monte Carlo simulator e.g. PSM \cite{Amelin:2001sk}, that we employed in a preliminary stage of this work, does not provide such flexibility and the results for jet observables were compatible when the values of $\rho$ and $\sigma_{jet}$ in the toy model and in the Monte Carlo  are similar {(see Appendix \ref{appA} for some discussion)}.

\subsection{Jet Reconstruction}
\label{reco}

\par Jets are defined within the Les Houches accords \cite{Buttar:2008jx} for particles species and lifetimes. Except for the analysis of the missing transverse momentum, jets are reconstructed using particles with a minimum cut for the transverse energy of $E_{Tmin} = 1$ GeV and contained in $|\eta| < 3$. The jet finding algorithm used is anti-$k_t$ with a resolution parameter set to $R = 0.3$. For dijet event reconstruction, we require the leading jet to have $E_{T1} \geq 120$ GeV and the sub-leading jet $E_{T2} \geq 30$ GeV. The dijet asymmetry has an additional cut, $\Delta \phi > 2 \pi /3$, to select only dijet pairs in opposite direction. As for the missing transverse momentum, the dijet pair follows previous constraints, but the projection is made using only charged particles with a minimum cut of $E_{Tmin} = 0.5$ GeV, and the acceptance is tightened to $|\eta| < 2.4$ for particles and $|\eta| < 1.6$ for the dijet pair. Also, the jet resolution parameter in this case is changed to $R = 0.4$.

\par For the background subtraction, two different techniques are applied: an area-based method using the FastJet package \cite{Cacciari:2010te,Cacciari:2007fd} and a pedestal subtraction method. In the case of the area-based method, jets are found with the $k_t$-algorithm ($R = 0.4$) over a full stripe in $|\eta| < 2$. The event-wise energy density per unit area, $\rho$ is estimated by taking the median of the ratio $p_{T}^{rec}/A^{rec}$ for all $k_{t}$ clusters, where $A^{rec}$ and $p_{T}^{rec}$ are the area and momentum of the reconstructed cluster. To reduce the influence of true jets on the background estimate, the two hardest clusters are removed from the median calculation. The fluctuations are computed through the 1-$\sigma$ dispersion (below the median) from the distribution of the ratio $p_{T}^{rec}/A^{rec}$ for all $k_{t}$ clusters, see \cite{Cacciari:2010te,Cacciari:2007fd} for details.

\subsubsection{Pedestal background subtraction method}
\label{cmsproc}

 \par {In order to apply this method, we define a grid resembling the calorimeter segmentation of CMS: for $-3 \leq \eta \leq -1.74$ and $1.74 \leq \eta \leq 3$, we divided the $\eta$ axis in 13 bins and the $\phi$ axis in 36 bins giving each cell an area of $\Delta \eta \times \Delta \phi \simeq 0.0969 \times  0.174$. For the most central part of the calorimeter, $ -1.74 < \eta < 1.74$, the $\eta$ axis was divided in 40 bins whereas the $\phi$ axis into 72 bins, resulting in an area of $\Delta \eta \times \Delta \phi \simeq 0.0870 \times 0.0873$ for each cell.
 
\par The subtraction procedure consists on the following steps:

\begin{enumerate}

\item Each cell is filled with the transverse energy of the input particles that have an $E_{Tmin} > $1 GeV. For each bin of $\eta$, all the cells in $\phi$ are summed and the average transverse energy and dispersion are computed as:
\begin{eqnarray}
	\left\langle E_{T}^{tower} (\eta) \right\rangle &=& \frac{ \sum_{i} E_{Ti}^{tower} }{ \# \phi \text{ bins } }\,, \\
	\left\langle \sigma_{T}^{tower} (\eta) \right\rangle &=& \sqrt{ \left\langle E_{T}^{tower} (\eta)^2 \right\rangle - \left\langle E_{T}^{tower} (\eta) \right\rangle^2 }\,.
\end{eqnarray}

\item For each cell, the average cell energy and dispersion in the corresponding $\eta$ stripe are subtracted and set to zero if the result is negative:
\begin{equation}
	E_{T}^{tower*} = {\rm max}\left[E_{T}^{tower} - \left\langle E_{T}^{tower} (\eta) \right\rangle - \kappa \left\langle \sigma_T^{tower} (\eta) \right\rangle,0\right],
	\label{eq:CMS_correction}
\end{equation}
where $E_{T}^{tower}$ is the original energy of the cell and $E_{T}^{tower*}$ the corrected one (note that this implies noise reduction).

\item  Using only particles that are inside cells with a non-zero $E_{T}^{tower*}$, jets with a transverse energy higher than a cut, $E_T > E_{T,jets}$, are reconstructed.

\item In order to remove true jets from the background estimate, the cells/particles contained in the reconstructed jets  with $E_T > E_{T,jets}$ are removed from the event and step 1 is repeated; a final estimate of background energy and dispersion in each $\eta$ stripe is thus obtained\footnote{One sample of the resulting values of $\left\langle E_{T}^{tower} (\eta) \right\rangle$ and $\left\langle \sigma_{T}^{tower} (\eta) \right\rangle$ for a toy background with moderate fluctuations ($T = 0.9$ GeV) can be seen in Figure \ref{fig:CMS}. Taking $\left\langle \sigma_{T}^{tower} (\eta) \right\rangle \sim 2$ GeV from the barrel of the calorimeter and scaling by  $\sqrt{number\ of\ cells}$ occupied by a jet with $R = 0.3$ that is $\sim \sqrt{37}$, this would correspond to $\sigma_{jet} \sim 12$ GeV. Comparing to the values presented by CMS\cite{CMS_RMS_RHO} for the most central collisions (that include fluctuations from the calorimeter and consider an effective $p_{Tmin} \simeq 0.5$ GeV), $RMS \sim 9$ GeV, one can observe that our value is larger but in rough agreement with the experimental one.}.

\item Finally, step 2 is repeated using the initial values of $E_{T}^{tower}$ and new corrected energies are found. Again, cells with negative transverse energy are set to zero and the final list of jets is found from the  cells with $E_T>0$.

\end{enumerate}

\par The two parameters in the method\footnote{Another possibility for tuning the method would be changing the $E_{Tmin}$ of the considered particles. We do not explore such possibility here as we want to compare the pedestal method with the FastJet area-based one with the same particles included in the jet reconstruction for both methods.} are $\kappa$ and $E_{T,jets}$. The former defines the amount of fluctuation that are removed and have a large effect due to the zeroing of the cells that becomes negative in step 2. The latter sets the limit above which reconstructed jets are considered true signal jets so that their constituent cells are removed from the background estimate. Their optimal values depend on the toy model parameters. In our case, we fix them in two different ways.

\begin{itemize}

\item We fix $E_{T,jets}=30$ GeV and vary $\kappa$ in order to get the better reconstruction of the single inclusive jet spectrum at high $E_T$ (the specific value of $E_T$ depends on the background parameter $T$), similarly in spirit as done in \cite{Kodolova:2007hd}. In this way, the optimal reconstruction of the jet energy is achieved, but for values of $\kappa$ that are larger than the value $\kappa=1$ used in \cite{Kodolova:2007hd,Chatrchyan:2011sx,Chatrchyan:2012nia}.

\item We fix $\kappa=1$ (as in \cite{Kodolova:2007hd,Chatrchyan:2011sx,Chatrchyan:2012nia}) and determine an optimal value of $E_{T,jets}$ by comparing the values of the background estimates in all  $\eta$ stripes ($\left\langle E_{T}^{tower} (\eta) \right\rangle$  and $\left\langle \sigma_{T}^{tower} (\eta) \right\rangle$) given by the subtraction method, to the corresponding values when the toy model is purely underlying event, without generation of  a hard component. Proceeding in this way, this method results quite independent of how the jets are generated for the comparison i.e. of whether we use quenched or unquenched jets and of the (presently uncertain) embedding of the jet in the medium, as only the background parameters and not the jet spectra are considered. But it results in quite a deficient reconstruction of the single inclusive jet spectrum. We present the results of this procedure in Appendix \ref{appb} but discuss its results compared to the previous procedure in the main part of the manuscript.

\end{itemize}

\par We refrain from varying simultaneously $\kappa$ and $E_{T,jets}$ in order to understand how the details of the reconstruction affect the different observables. It turns out that the key aspect is not this simultaneous variation but  the way of fixing them i.e. whether you choose a better reconstruction of the jet energy or of the input background.

\section{Effects of the different characteristics of the underlying event}
\label{bkgParam}

\subsection{Background energy and event-by-event fluctuations}
\label{fluct}

\par The effect of background fluctuations on jet observables has extensively been discussed in \cite{Cacciari:2011tm}.  In this section we explore the response of the different reconstruction/subtraction techniques to background fluctuations and background energy, and their impact on the inclusive jet spectrum, dijet asymmetry and azimuthal correlations.

\par The inclusive jet spectrum in heavy-ion collisions is affected over the entire $E_{T}$ range by background. E.g. the convolution of a steeply falling perturbative jet spectrum with the fluctuations measured by ALICE \cite{Abelev:2012ej} in central PbPb collisions using a cut of $p_{Tmin} = 0.15$ GeV, leads to an enhancement of the jet yield of a factor $\sim 10$ even for high $E_{T}> 60$ GeV. This enhancement is reduced to a factor of $\sim 1.3$ when $p_{Tmin} = 2$ GeV.

\begin{figure}[htbp]
	\begin{center}
		\subfigure[Reconstructed inclusive jet spectra  using the FastJet area method.]{
			\label{fig:SpecFJ}
			\includegraphics[width=0.48\textwidth]{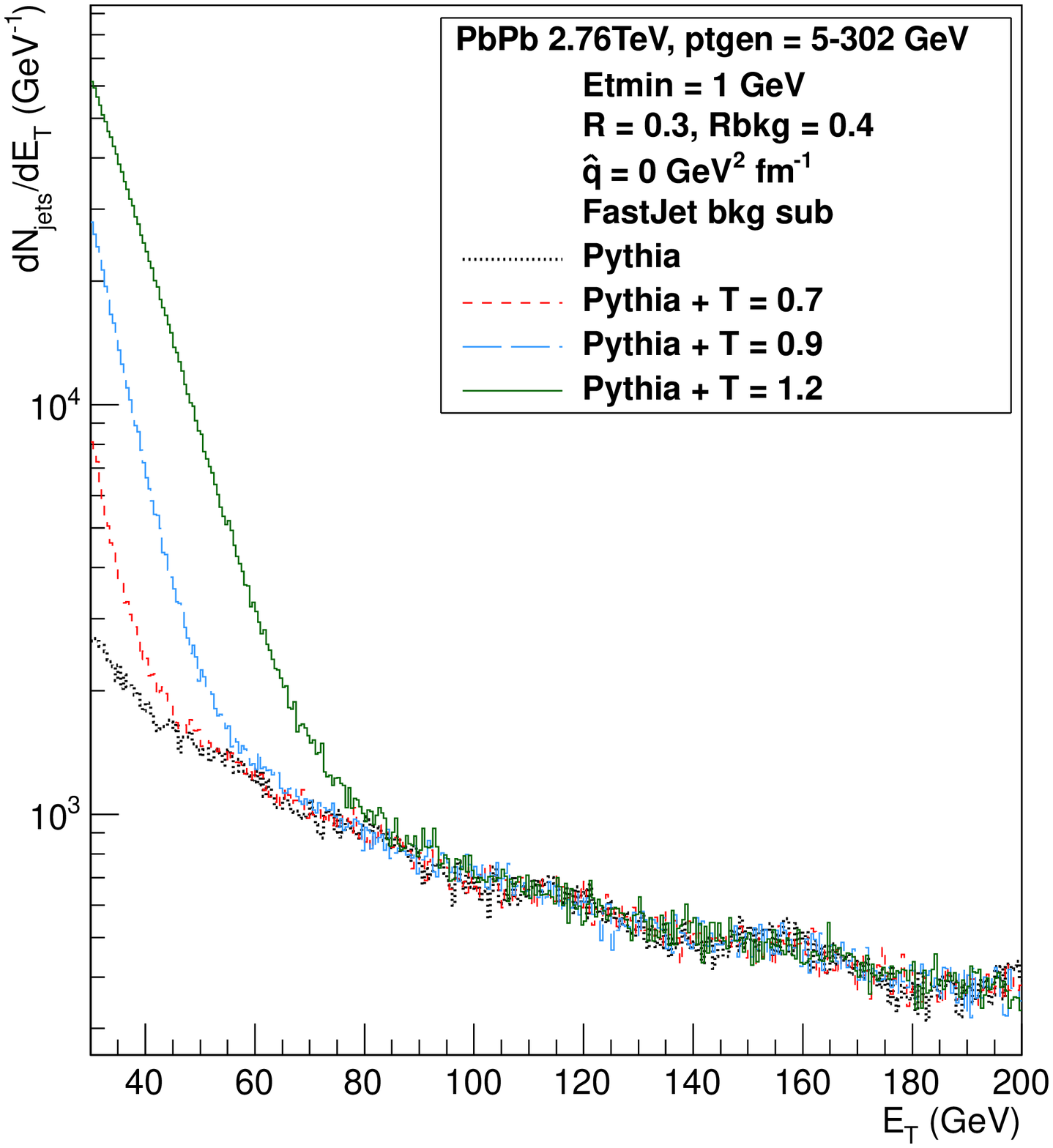}
		}
		\subfigure[Reconstructed inclusive jet spectra using a pedestal technique.]{
			\label{fig:SpecCMS}
			\includegraphics[width=0.48\textwidth]{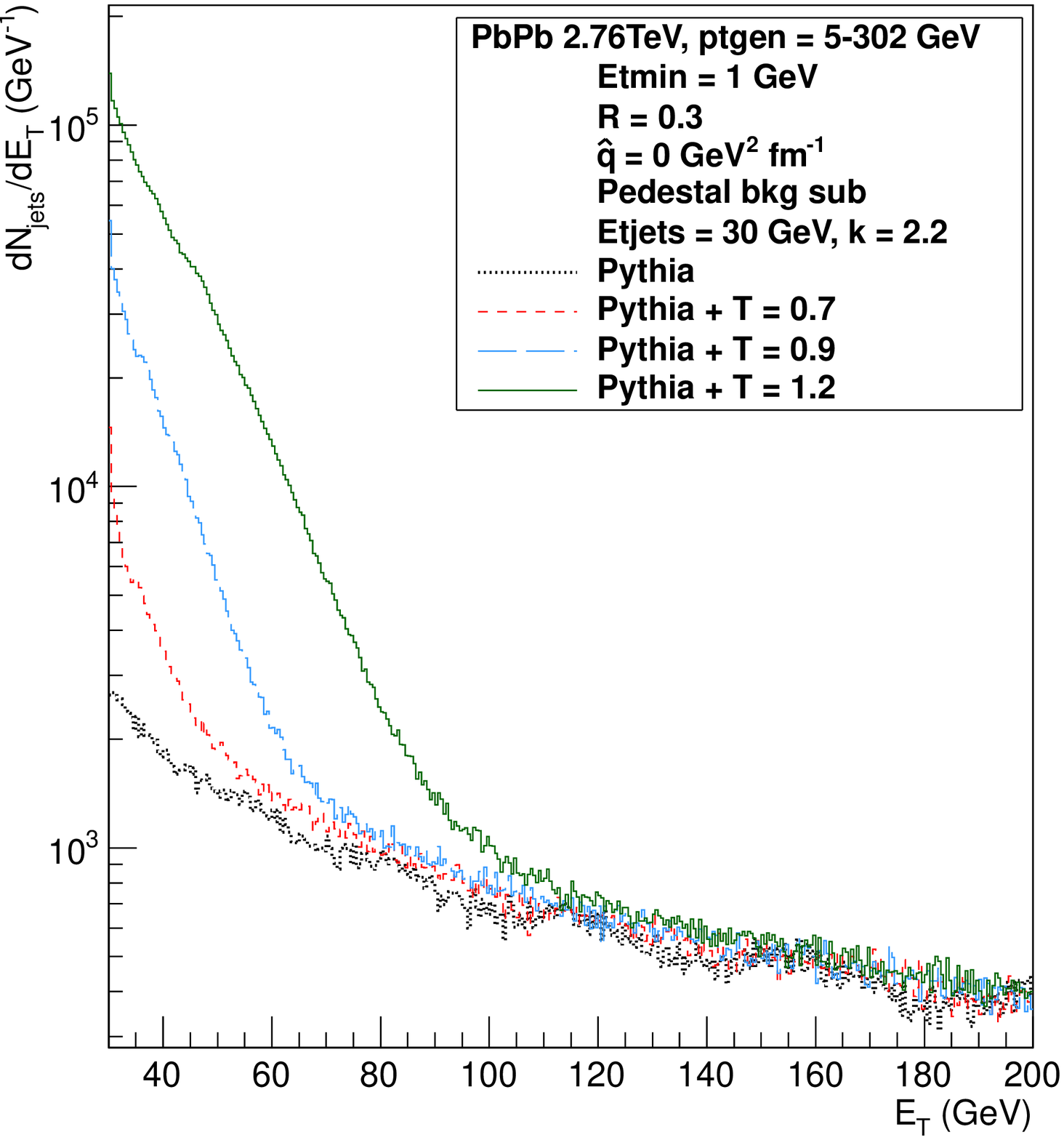}
		}
		\caption{Inclusive jet spectrum of pure (unquenched) PYTHIA events (black dotted line) and spectrum of subtracted jets for PYTHIA embedded in  the toy model background  with a $T = 0.7$ (red dashed line), $T = 0.9$ (blue long dashed line) and $T = 1.2$ GeV (green solid line).}
	\end{center}
\end{figure}

\par Figure \ref{fig:SpecFJ} shows a comparison between the reconstructed inclusive jet spectrum for PYTHIA events (black dotted line) and for PYTHIA events embedded in different configurations of background (red dashed, blue long dashed and green solid). In this figure the background subtraction was done using the area-based FastJet method. As one can observe the background causes an enhancement of the jet yield with respect to PYTHIA that persists up to rather large\footnote{One can observe that all spectra are in agreement for $E_{T} \geq 90$ GeV. For smaller transverse energies, however, there is an overestimation of the jet population, which increases with increasing background values. The energy at which the jet subtracted spectrum deviates from the PYTHIA result is around 45 GeV for $T = 0.7$, 60 GeV for $T = 0.9$ and 90 GeV for $T = 1.2$ GeV. These are approximately the average levels of energy deposition inside a cone of $R = 0.3$  for each value of $T$. Thus, the reconstruction method fails at this point since the amount of background jets at these energies increase and it is not able to distinguish between a jet coming from the hard event or a pure background jet.} values of $E_{T}$.

\par The inclusive spectra for the case of the pedestal subtraction method is shown in Figure \ref{fig:SpecCMS}, using $E_{T,jets}=30$ GeV and varying $\kappa$. The result of this procedure is similar to that with the area subtraction\footnote{In both cases, for $E_T>140$ GeV the input and reconstructed spectra differ less than 10 \%.}, although the value of $E_T$ below which the reconstructed spectrum exceeds the input one is larger for the pedestal method, particularly for the largest $T$. The optimal values of $\kappa$ are sizably larger than 1. In the following, we  use $\kappa=2.2$ for all values of $T$. On the other hand, for the second procedure in the pedestal method indicated in Subsection \ref{cmsproc}, the impact of background fluctuations is stronger and the jet yield is enhanced with respect to PYTHIA up to $E_{T} \sim 200$ GeV for all considered temperatures, see Appendix \ref{appb}.

\par Concerning the dijet asymmetry, background fluctuations can induce a momentum imbalance of the dijet pair by differently shifting the energy of the two jets. Furthermore, an incorrect estimation of $\rho$ will originate a possible shift in the jet energy that affects both jets equally in the same event. Since the observable $A_J$ is normalized to the dijet transverse energy, this can induce modifications in the distribution.

\begin{figure}[htbp]
	\begin{center}
		\subfigure[Dijet asymmetry using FastJet.]{
			\label{fig:AjFluctFJ}
			\includegraphics[width=0.48\textwidth]{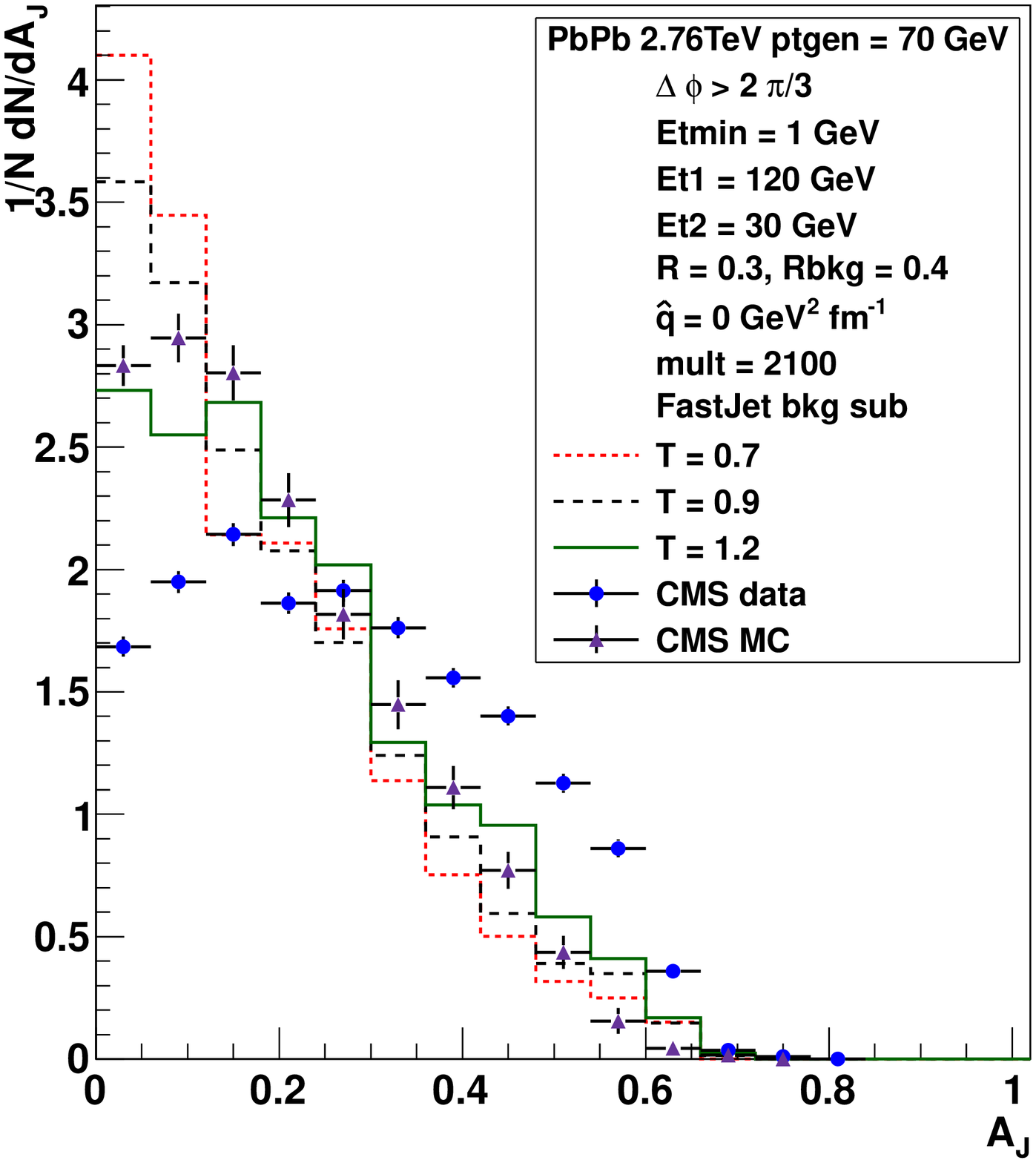}
		}
		\subfigure[Dijet asymmetry using a pedestal technique.]{
			\label{fig:AjFluctCMS}
			\includegraphics[width=0.48\textwidth]{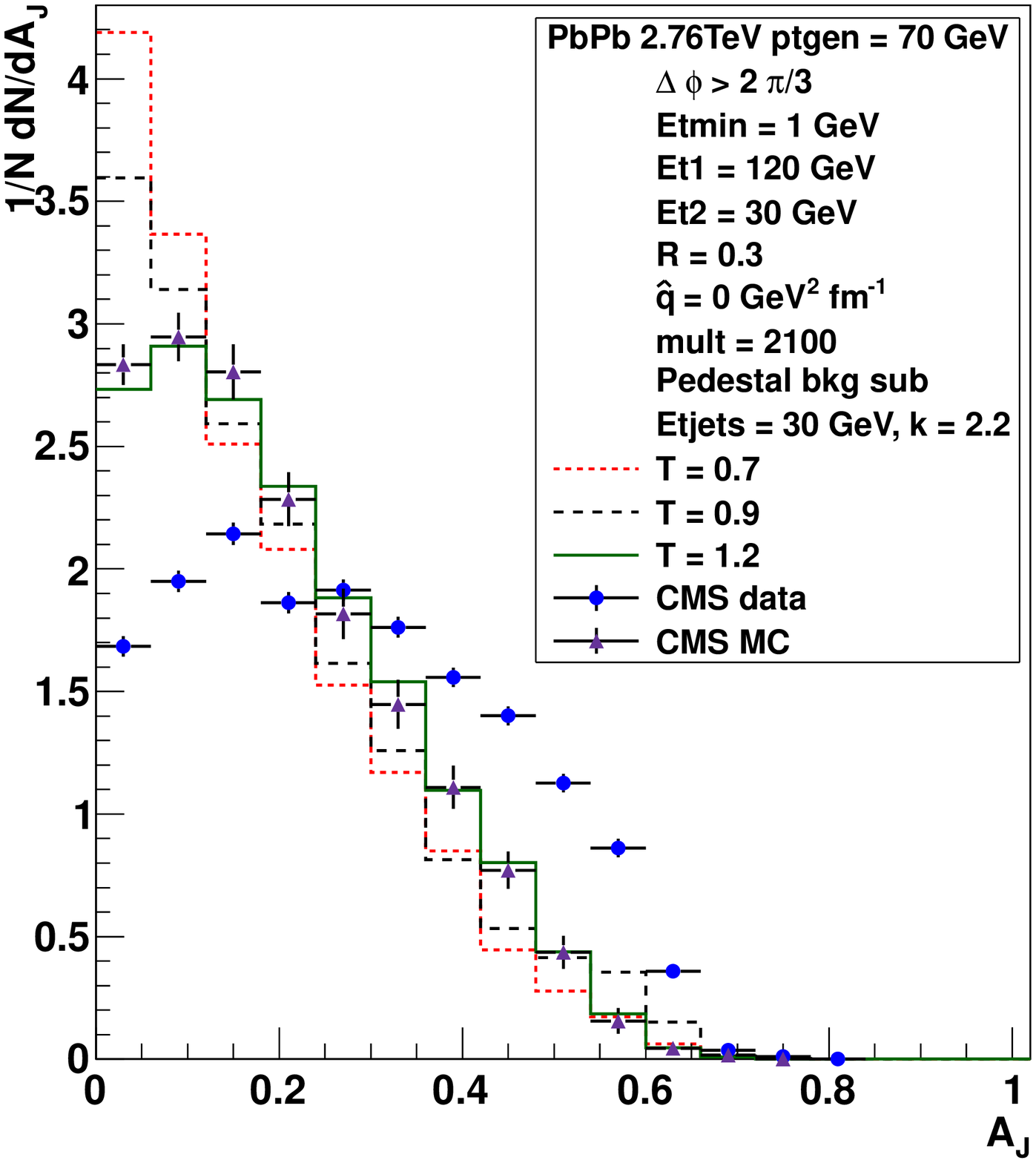}
		}
		\caption{Dijet asymmetry ($A_J$) for a simulation using Q-PYTHIA with $\hat{q} = 0$ embedded in a background with different $T$'s. The red dotted lines correspond to a background with $T = 0.7$, the black dashed ones to a $T = 0.9$ and the green solid ones to a $T = 1.2$ GeV. The blue points are the CMS data with the corresponding error bars and the purple triangles the CMS Monte Carlo \cite{Chatrchyan:2012nia}.}
		\label{fig:AjFluct}
	\end{center}
\end{figure}

\begin{figure}[!h]
	\begin{center}
		\subfigure[Dijet azimuthal correlation using the FastJet area-based method.]{
			\label{fig:DphiFluctFJ}
			\includegraphics[width=0.48\textwidth]{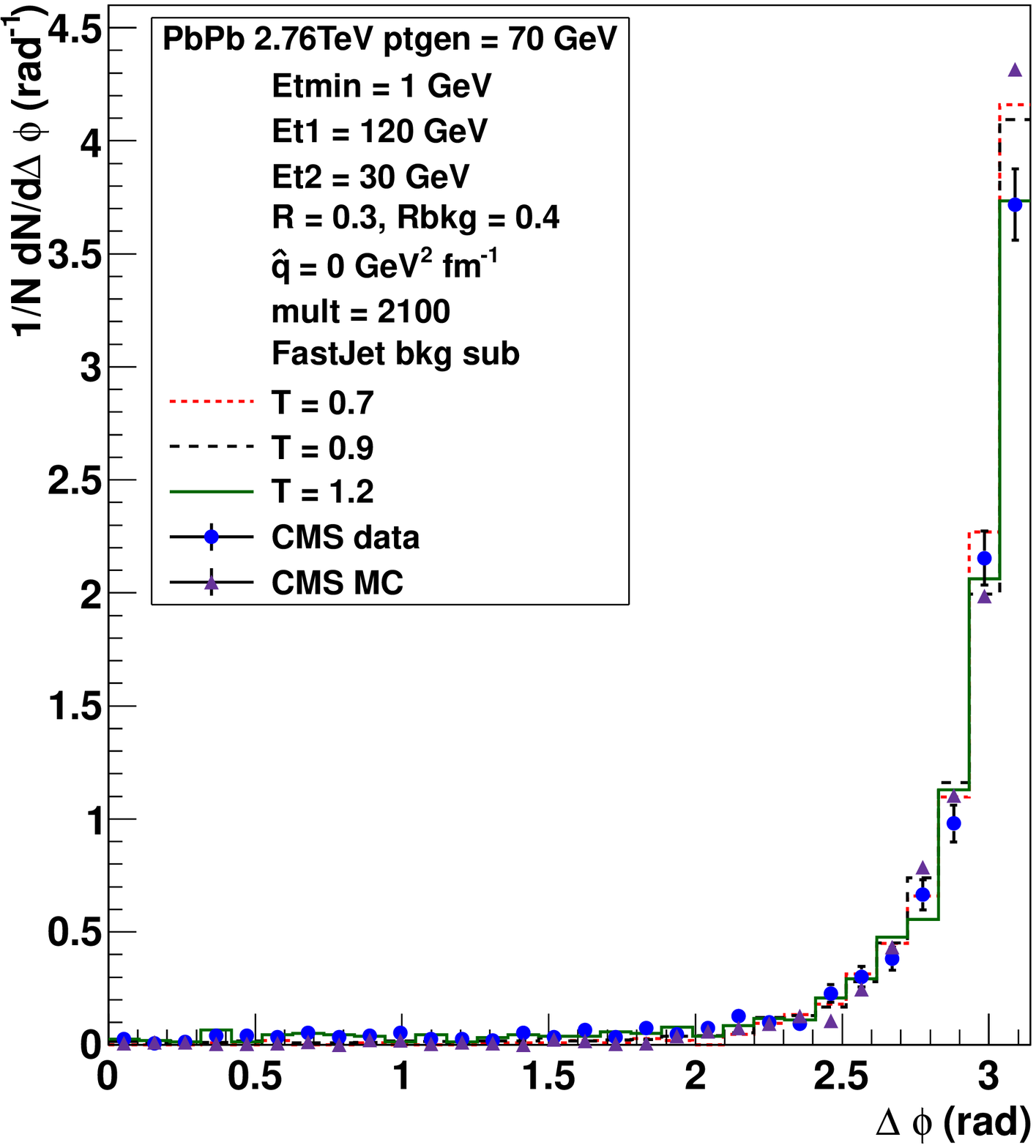}
		}
		\subfigure[Dijet azimuthal correlation using a pedestal technique.]{
			\label{fig:DphiFluctCMS}
			\includegraphics[width=0.48\textwidth]{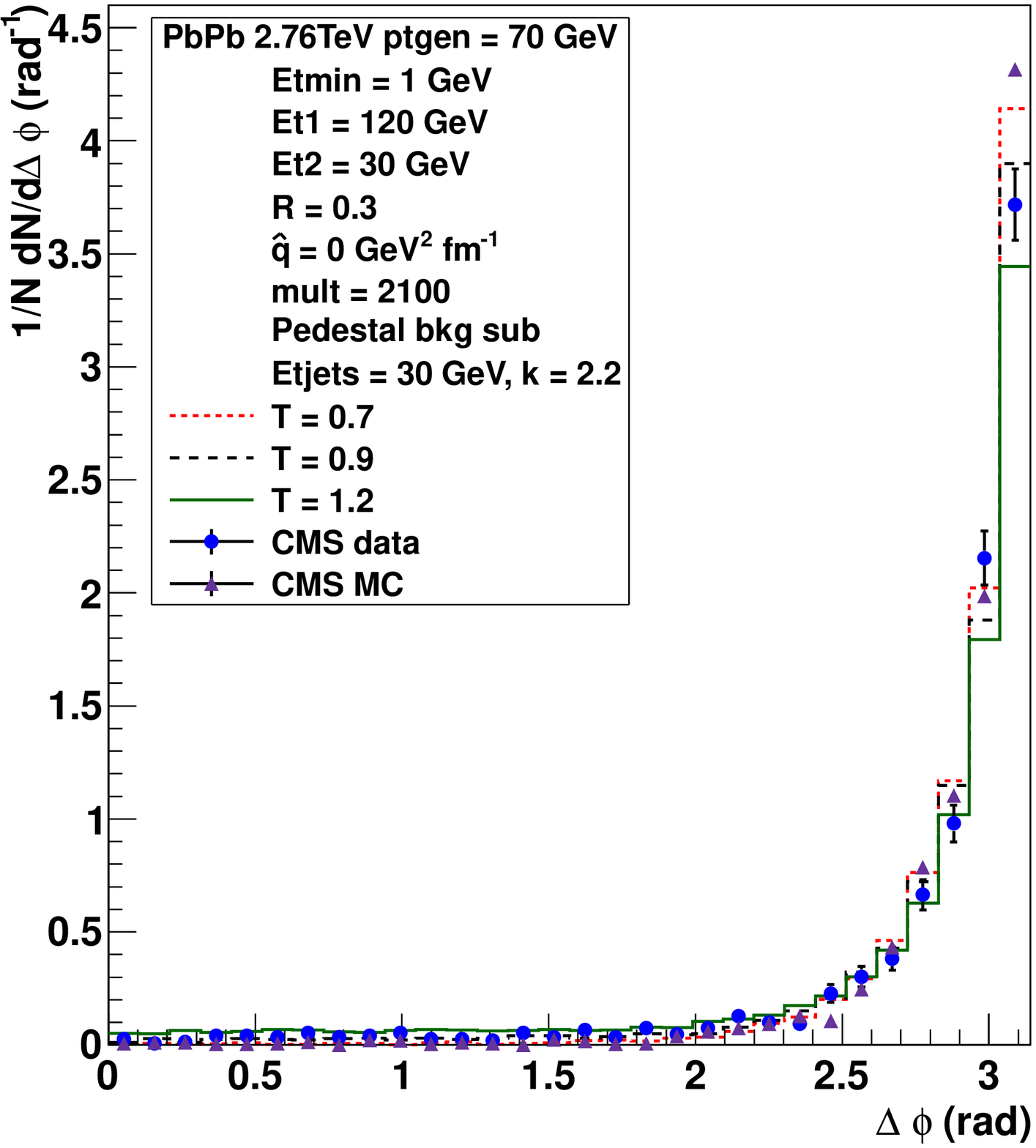}
		}
		\caption{Dijet azimuthal correlation ($\Delta \phi$) for a simulation using Q-PYTHIA with $\hat{q} = 0$ embedded in a background with different $T$'s. The red dotted lines correspond to a background with $T = 0.7$, the black dashed ones to a $T = 0.9$ and the green solid ones to a $T = 1.2$ GeV. The blue points are the CMS data with the corresponding error bars and the purple triangles the CMS Monte Carlo \cite{Chatrchyan:2012nia}.}
		\label{fig:DphiFluct}
	\end{center}
\end{figure}

\par In Figure \ref{fig:AjFluctFJ} the dijet asymmetry is shown for different values of $T$ using the FastJet area-based background subtraction method. The momentum imbalance of the dijet pair increases with fluctuations, but the effect is mild for small-to-moderate fluctuations. For $\sigma_{jet} \sim 15$ GeV, there are significant changes in the shape and mean value of the distribution with respect to the cases of lower average background fluctuations and a good agreement with the CMS Monte Carlo (PYTHIA pp events embedded in HYDJET \cite{Lokhtin:2005px}, see \cite{Chatrchyan:2012nia}) is found. When the pedestal subtraction method with $E_{T,jets}=30$ GeV and  $\kappa=2.2$ is used  (see Figure \ref{fig:AjFluctCMS}), a similar effect is observed, which suggests that the effect of the fluctuations on this observable can actually be understood in terms of the single inclusive spectrum. On the other hand,  for the second procedure in the pedestal method indicated in Subsection \ref{cmsproc}  ($\kappa=1$) we observe that the asymmetry is reduced with increasing $T$, see Appendix \ref{appb}. This may be linked to the increasing shift of the single inclusive distributions discussed there.

\par The effect of the background subtraction on the azimuthal correlation of the dijet pair is also explored. The results using FastJet area-based subtraction method are shown in Figure \ref{fig:DphiFluctFJ}. The $\Delta \phi$ distance between the dijet is stable under changes on the background main characteristics, $\rho$ and $\sigma_{jet}$. When using the pedestal method with $E_{T,jets}=30$ GeV and  $\kappa=2.2$ (Figure \ref{fig:DphiFluctCMS}), the azimuthal correlation shows similar features to those observed with the area-based subtraction, though a small pedestal in the whole $\Delta \phi$ range appears  for the highest fluctuations - an effect that becomes more pronounced for the second procedure in the pedestal method indicated in Subsection \ref{cmsproc},  see Appendix \ref{appb}.

\subsection{Flow}
\label{flow}

\par Now we turn to introducing flow in our simulation in order to understand its relevance for the observables that we want to describe. To consider the effects of flow in our toy model we modulate the distribution of particles in azimuth according to:
\begin{equation}
	\frac{dN}{d\phi} \propto 1 + \sum_{n} v_{n} (p_T) \cos (n \phi).
\end{equation}
We include up to the third component $v_3$ in the previous expression and take the $p_{T}$ integrated values\footnote{These values correspond to semi-peripheral collisions, see \cite{ALICE:2011ab}} to be $v_2(p_T) = <v_2> = 0.1$ and $v_3 (p_T) = <v_3> = 0.03$. While these values are larger than the experimentally measured ones for central collisions \cite{ALICE:2011ab}, we use them in order to explore the potential sensitivity of the jet observables. Additionally, a random reaction plane (RP) is defined for each event. Note that we assume that PYTHIA jets and their constituents do not flow and are uncorrelated with the reaction plane.

\par By introducing the flow components, the effective value of the fluctuations change. In Table \ref{table1} the effective values of $\sigma_{jet}$ for each background $T$ and each configuration of flow parameters are shown. The largest difference is observed for $T = 1.2$ GeV, where the fluctuations increase by $\sim 5$ GeV/area. In the following, we use this background configuration to examine jet observables.

\begin{table}[htdp]
	\begin{center}
	\begin{tabular}{|c|c|c|c|}
		\hline
		Temperature (GeV) & $v_2 = 0.0$, $v_3 = 0.0$ & $v_2 = 0.1$, $v_3 = 0.0$ &$v_2 = 0.1$, $v_3 = 0.03$ \\
		\hline
		$T = 0.7$ & 7.69 & 9.16 & 9.26 \\
		$T = 0.9$ & 10.74 & 13.31 & 13.47 \\
		$T = 1.2$ & 15.14 & 19.36 & 19.67 \\
		\hline
	\end{tabular}
	\end{center}
	\caption{Values of $\sigma_{jet}$ (in GeV) obtained using FastJet.}
	\label{table1}
\end{table}

\par Figure \ref{fig:AjFlow} shows the dijet asymmetry for three extreme cases ($v_{2}=0$,$v_{3}=0$;$v_{2}=0.1$,$v_{3}=0$;$v_{2}=0.1$,$v_{3}=0.03$) using the FastJet area-based (Figure \ref{fig:AjFlowFJ}) and the pedestal  (with $E_{T,jets}=30$ GeV and  $\kappa=2.2$, Figure \ref{fig:AjFlowCMS}) methods for background subtraction. The dijet asymmetry shows negligible dependence on flow, see also the results for  the second procedure in the pedestal method in Appendix \ref{appb}.

\begin{figure}[!h]
	\begin{center}
		\subfigure[Dijet asymmetry using the area-based FastJet method.]{
			\label{fig:AjFlowFJ}
			\includegraphics[width=0.48\textwidth]{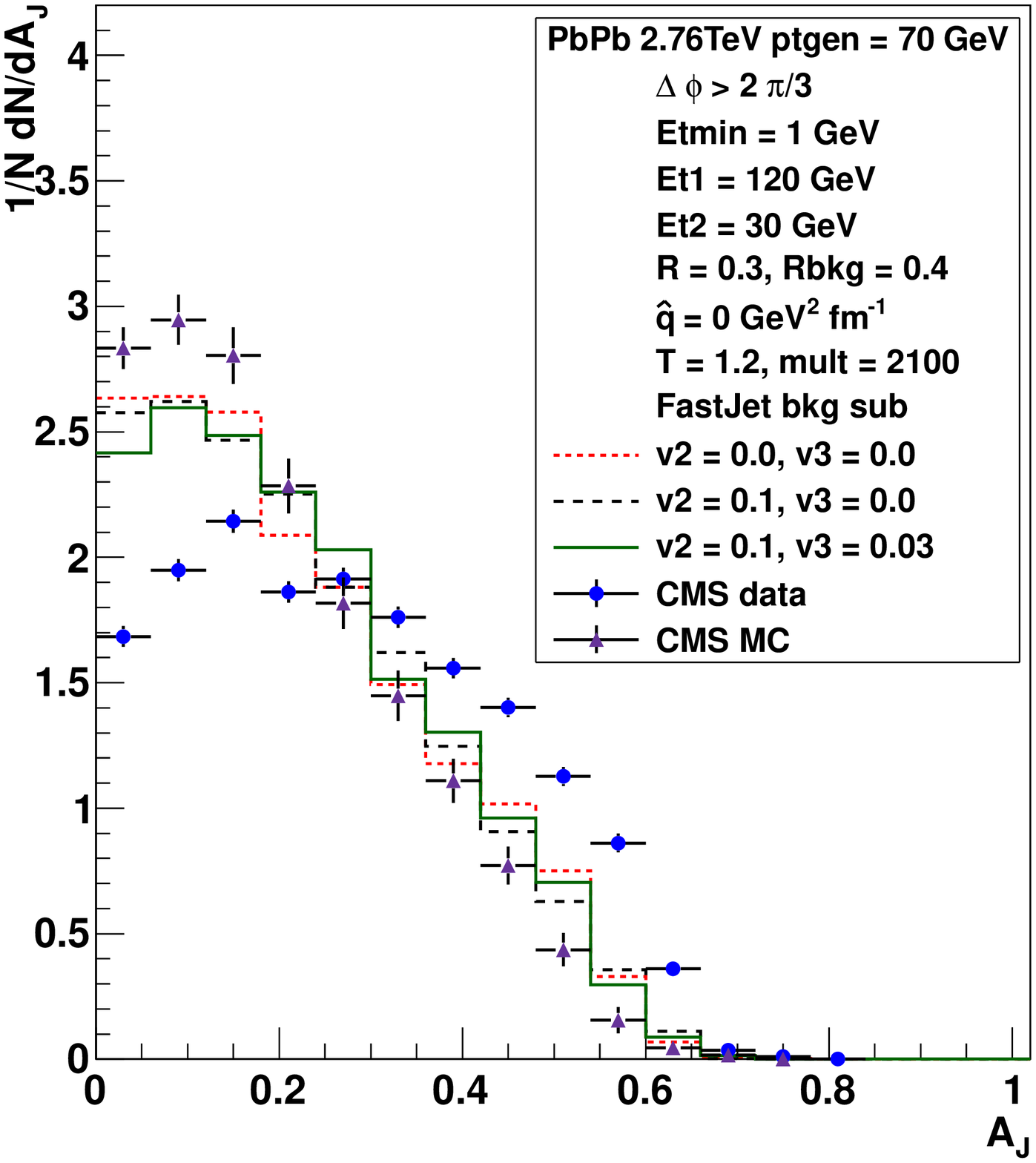}
		}
		\subfigure[Dijet asymmetry using a pedestal technique.]{
			\label{fig:AjFlowCMS}
			\includegraphics[width=0.48\textwidth]{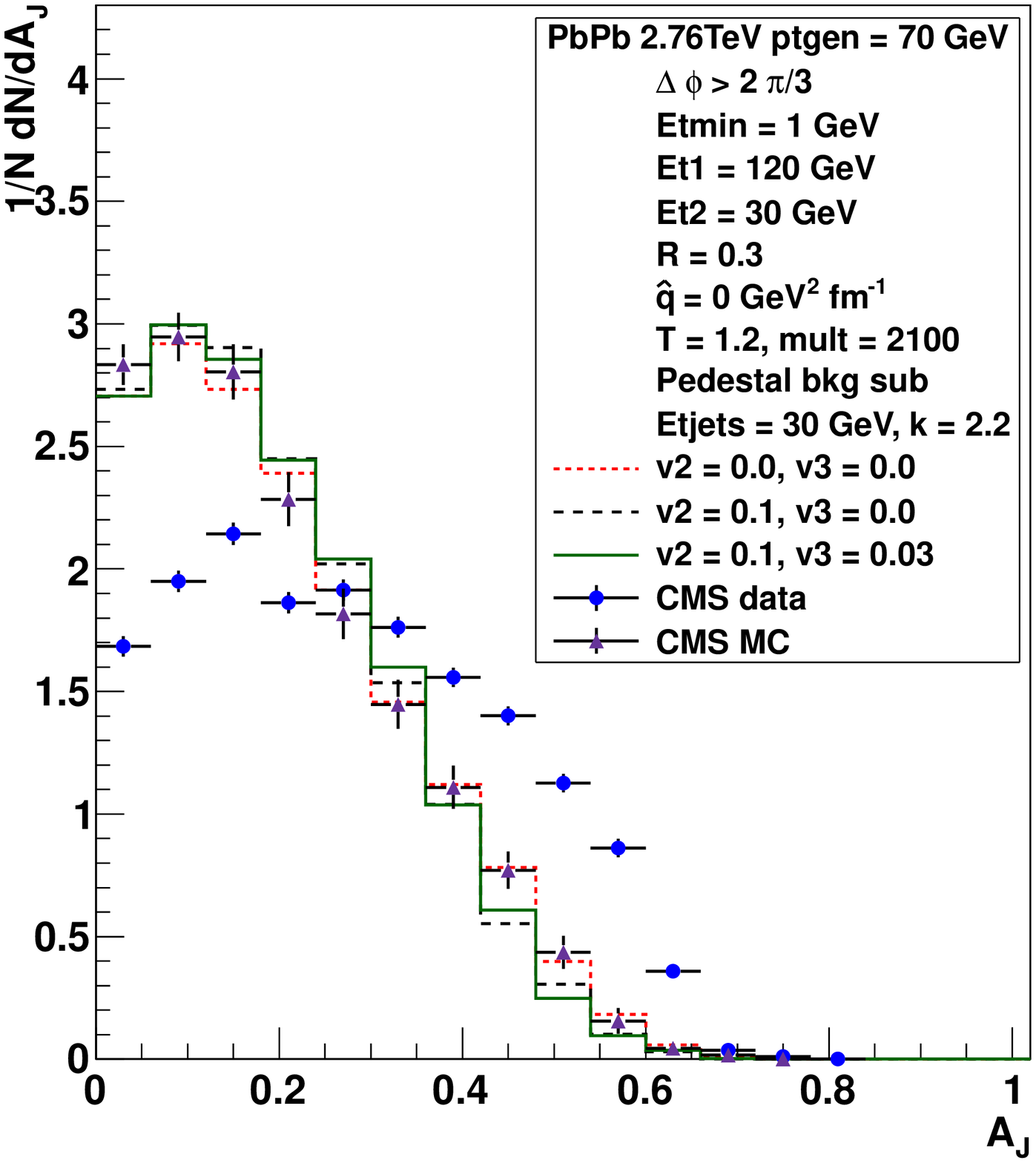}
		}
		\caption{Dijet asymmetry ($A_J$) for a simulation using Q-PYTHIA with $\hat{q} = 0$ embedded in a background with $T = 1.2$ GeV. The red dotted lines correspond to a simulation without flow, the black dashed ones with an elliptic flow component ($v_2 = 0.1$) and the green solid ones with an additional triangular flow component ($v_2 = 0.1, v_3  = 0.03$). The blue dots are the CMS data with the corresponding error bars and the purple triangles the CMS Monte Carlo \cite{Chatrchyan:2012nia}.}
		\label{fig:AjFlow}
	\end{center}
\end{figure}

\par The azimuthal correlation, as shown in Figures \ref{fig:DphiFlowFJ} and \ref{fig:DphiFlowCMS} for FastJet area-based and pedestal subtraction (with $E_{T,jets}=30$ GeV and  $\kappa=2.2$) methods respectively, shows a bump near $\Delta \phi \sim 0$. This bump is more evident for  the second procedure in the pedestal method, see Appendix \ref{appb}. Note that since the azimuthal flow is a modulation in $\phi$, the corresponding fluctuations are not local or random. They have a symmetry that follows the Fourier components. Due to a higher concentration of particles in certain areas of the phase space, the jet finding algorithm will reconstruct mostly the jets on that regions. Thus, the transverse energy is recovered, not affecting significantly  the dijet momentum imbalance. But the number of events in which the two leading jets are close in azimuth (but not too close to be merged by the clustering algorithm) increases, inducing a bump at $\Delta \phi \sim 0$. This effect can be seen in Figure \ref{fig:Corr}, that shows the correlation of the azimuthal angle of the leading jet, $\phi_1$, and the results for the dijet asymmetry using the pedestal subtraction method\footnote{Results are similar to those obtained with the Fastjet area-based subtraction method.} Without flow (Figure \ref{fig:Corr1}), all events have a leading jet distribution approximately uniform over all phase space. When a $v_2$ component is introduced, the leading jet is more likely to be found at $\phi = \phi_{RP} = 0$ and $\phi = \pi$ for the events of small asymmetry, as shown in Figure \ref{fig:Corr2}. The same will happen with the subleading jet, since here is where the concentration of particles is higher. In Figure \ref{fig:Corr3}, a $v_3$ component is also introduced. Its main effect is to reduce or destroy the leading jet-subleading jet symmetry along the reaction plane. If other Fourier components are taken into account, the correlation between $A_J$ and $\phi_{1(2)}$ might become reduced.

\begin{figure}[htbp]
	\begin{center}
		\subfigure[Dijet azimuthal correlation using FastJet.]{
			\label{fig:DphiFlowFJ}
			\includegraphics[width=0.48\textwidth]{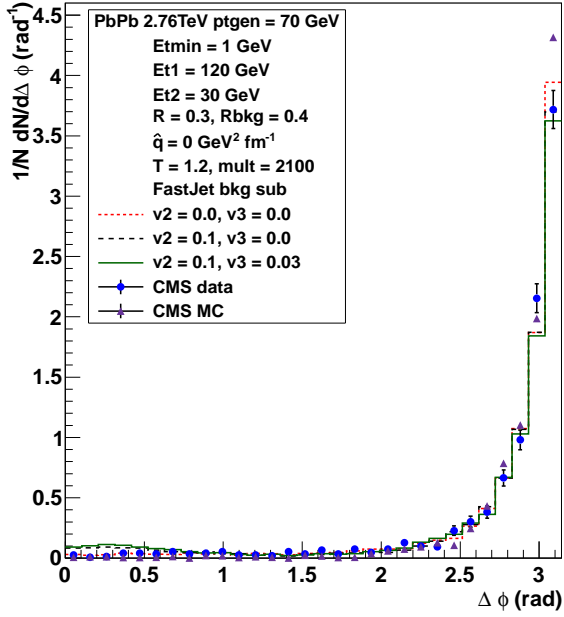}
		}
		\subfigure[Dijet azimuthal correlation using a pedestal technique.]{
			\label{fig:DphiFlowCMS}
			\includegraphics[width=0.48\textwidth]{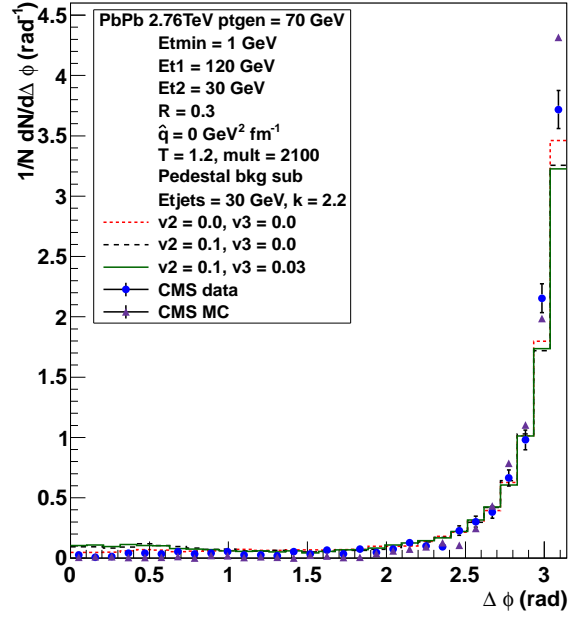}
		}
		\caption{Dijet azimuthal correlation ($\Delta \phi$) for a simulation using Q-PYTHIA with $\hat{q} = 0$ embedded in a background with $T = 1.2$ GeV. The red dotted lines correspond to a simulation without flow, the black dashed ones with an elliptic flow component ($v_2 = 0.1$) and the green solid ones with an additional triangular flow component ($v_2 = 0.1, v_3  = 0.03$). The blue dots are the CMS data with the corresponding error bars and the purple triangles the CMS Monte Carlo \cite{Chatrchyan:2012nia}.}
		\label{fig:DphiFlow}
	\end{center}
\end{figure}

\begin{figure}[htbp]
	\begin{center}
		\subfigure[Results for $v_2 = 0, v_3 = 0$.]{
			\label{fig:Corr1}
			\includegraphics[width=0.31\textwidth]{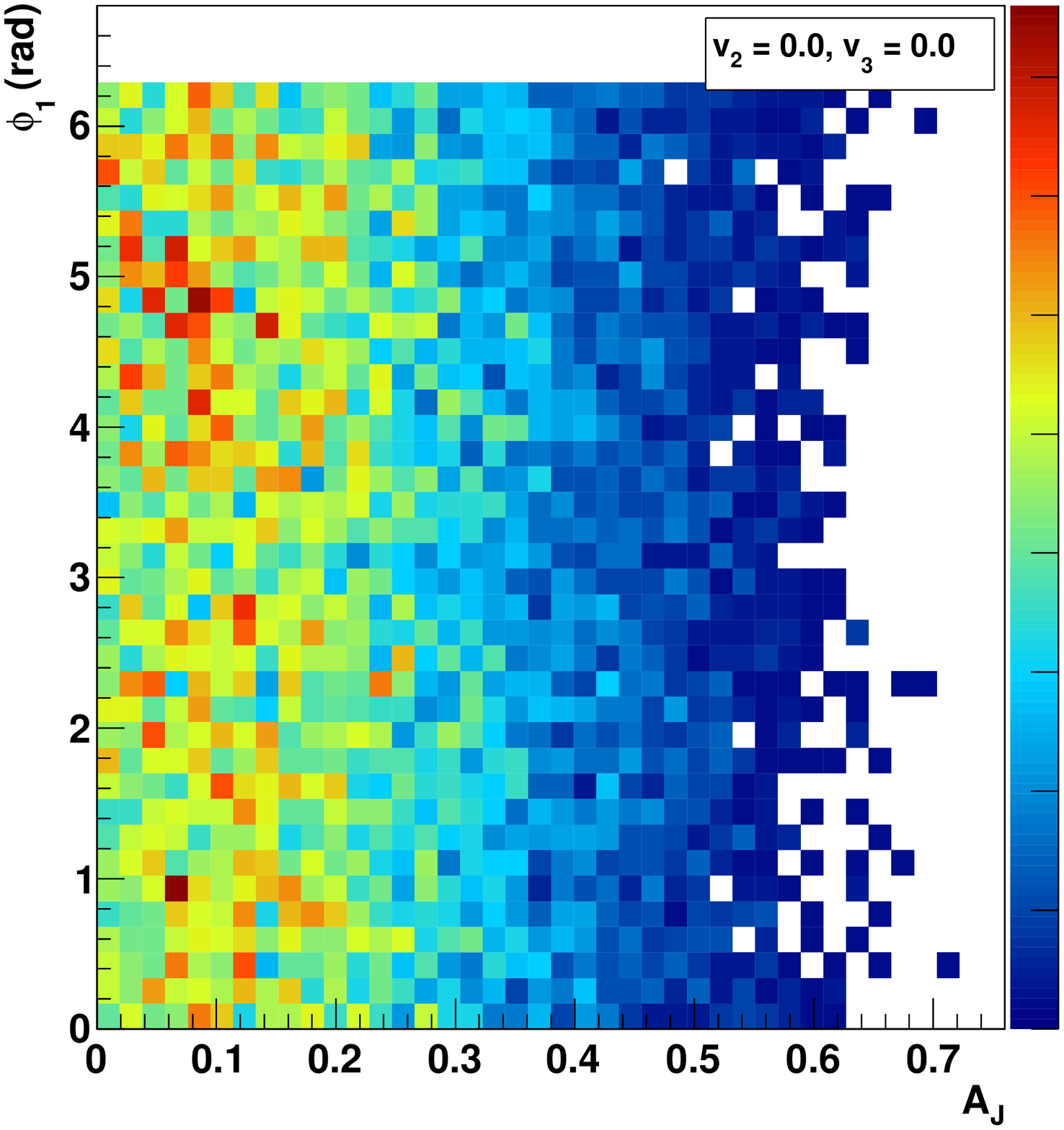}
		}
		\subfigure[Results for $v_2 = 0.1, v_3 = 0$.]{
			\label{fig:Corr2}
			\includegraphics[width=0.31\textwidth]{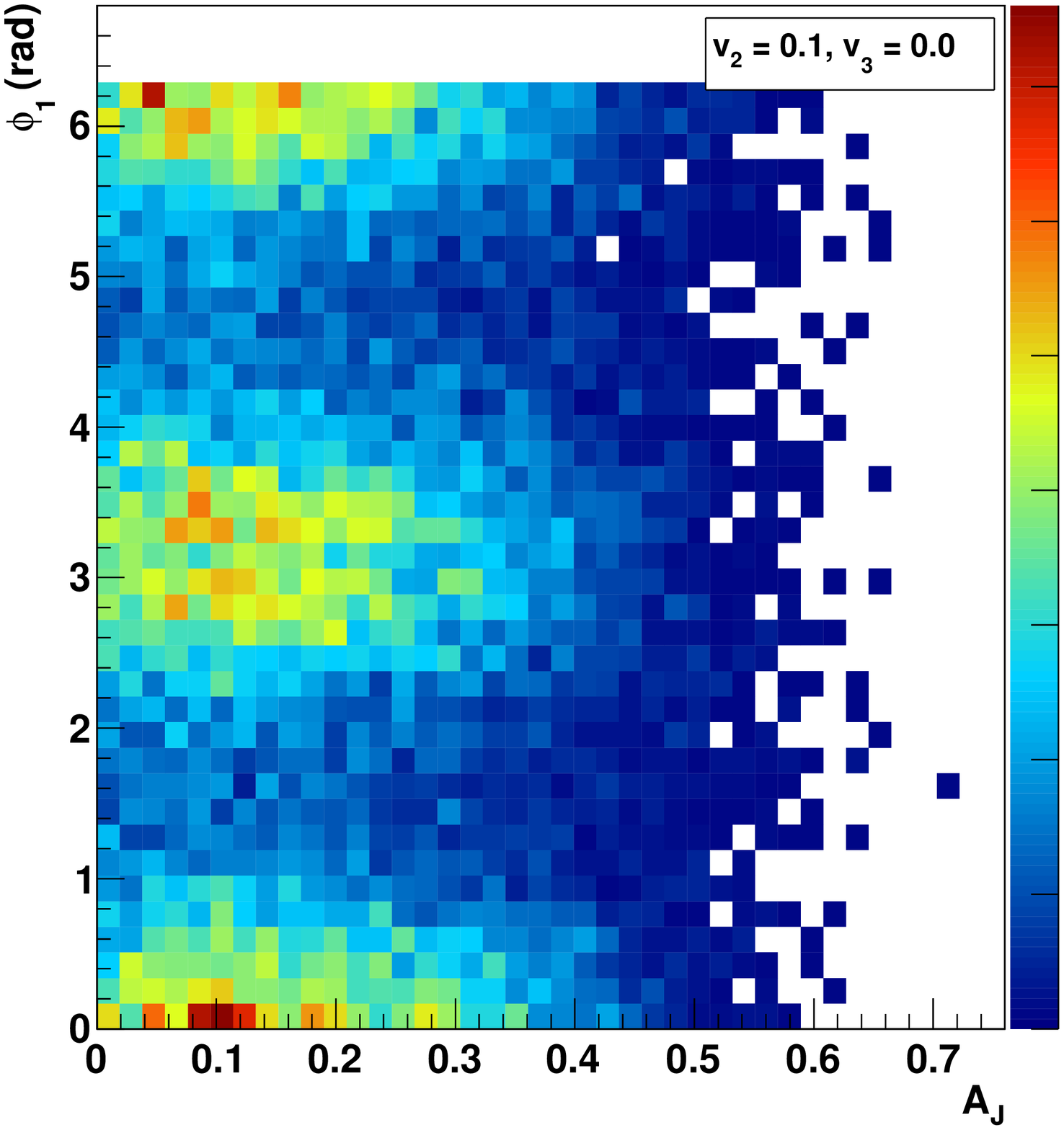}
		}
		\subfigure[Results for $v_2 = 0.1, v_3 = 0.03$.]{
			\label{fig:Corr3}
			\includegraphics[width=0.31\textwidth]{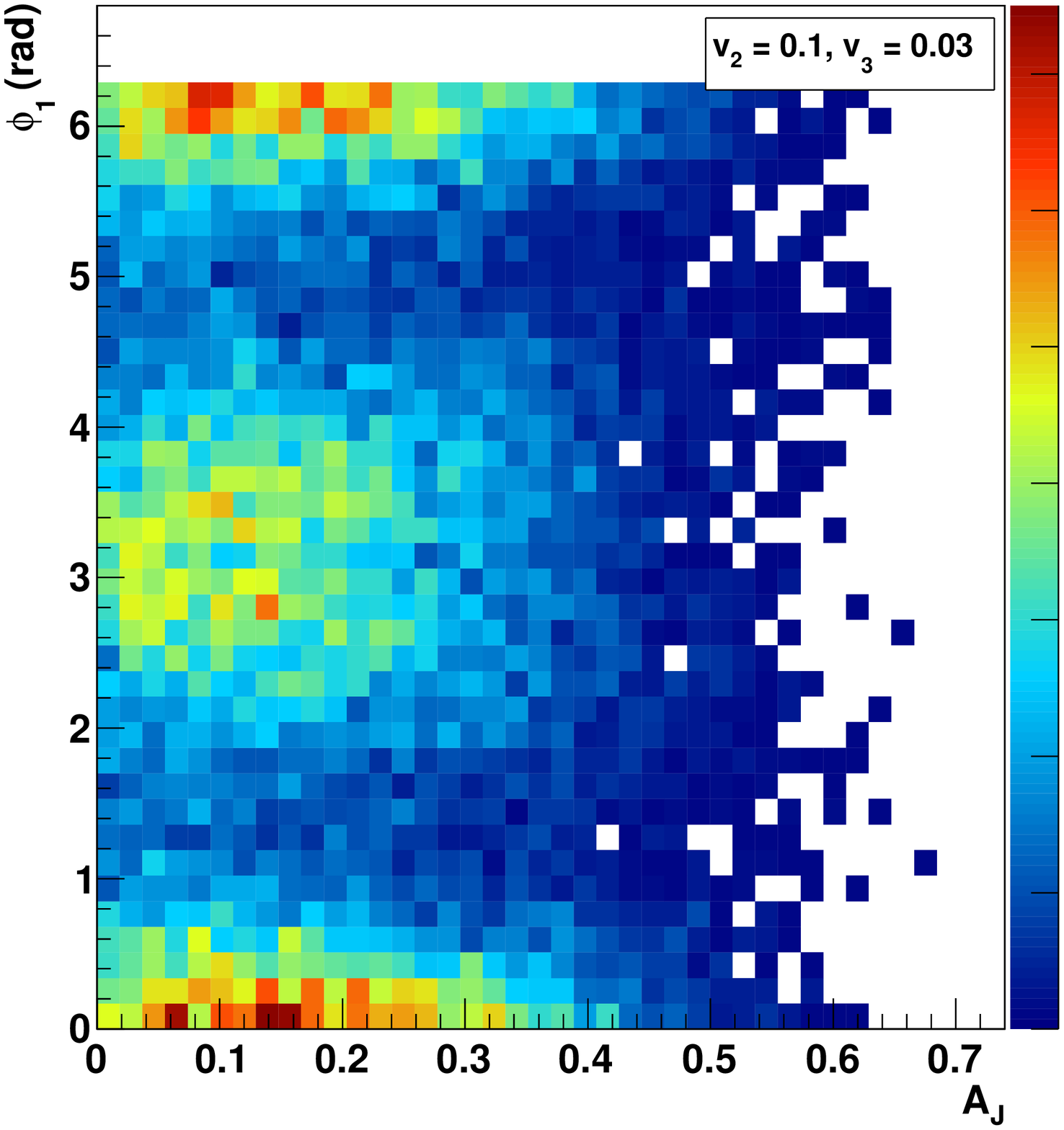}
		}
		\caption{Correlation between $A_J$ and $\phi_1$ for a simulation using Q-PYTHIA with $\hat{q} = 0$ embedded in a background with $T = 1.2$ GeV with a reaction plane fixed to $\phi_{RP} = 0$. The background subtraction was made using a pedestal method. The colour gradation from deep blue to deep red denotes increasing correlations.}
		\label{fig:Corr}
	\end{center}
\end{figure}

\par Two comments are in order: First, the pedestal subtraction method seems to be very sensitive to fluctuations at the level of the inclusive spectrum. The combinatorial excess of jets up to high jet $E_T$ for the highest $T$ is probably behind the pedestal observed in the previous Subsection  as well as the flow dependence in Fig. \ref{fig:Corr} (though some effect is also visible for the FastJet area-based subtraction method).  Second, the effects that were observed in this Subsection are for unrealistically large values of flow and  background but, in any case, they point to the importance of not only the average  fluctuations but also of their azimuthal distribution. A careful consideration of this feature when making azimuthally differential studies seems unavoidable\footnote{The possibility of a $v_2$-modulated background is considered in the experimental analysis in \cite{:2012is}. Besides, the use of more sophisticated, local definitions of the regions where the background is extracted \cite{Cacciari:2010te}, should result in a better consideration of flow in background subtraction.}.

\section{Quenching}
\label{quench}

\par In the previous Section the effects of  background characteristics and subtraction methods on jet observables were identified. Here we explore the influence of quenching by varying the transport coefficient $\hat{q}$ in Q-PYTHIA (we refer the reader to \cite{Armesto:2009fj} for a detailed description of the model and the physics contained in it). By increasing $\hat{q}$, the jet shower is accelerated with respect to that in vacuum due to the increasing number of medium-induced splittings, and the shape of the splittings is modified resulting in an increase of semi-hard, large angle (non-collinear) emissions tamed by energy-momentum conservation. In our case, the medium density has been considered proportional to the nuclear overlap as in the PQM model \cite{Dainese:2004te}, see the details in that reference (thus the average value of $\hat q$ is obtained as in there)\footnote{The value of impact parameter was fixed to 3.3 fm, and the nuclear density distributions are Wood-Saxon with a radius of 6.34 fm and a nuclear thickness 0.545 fm.}. While this is not a realistic model for medium densities and evolution, it contains some of the potential biases, like surface bias, that must be included in any realistic medium modeling.

\par In what follows, the results will be presented for $\hat{q} = 4$ GeV$^2$ fm$^{-1}$ (black dashed lines) and $\hat{q} = 8$ GeV$^2$ fm$^{-1}$ (green solid lines), while keeping the ones for $\hat{q} = 0$  (red dotted lines) as a reference. Furthermore, flow is not considered due to its negligible impact on the asymmetry and the $T$ parameter is fixed to $T = 0.9$ GeV, which corresponds to fluctuations  $\sigma_{jet} \sim 11$ GeV.

\par Let us stress that we make no attempt to confront the results of Q-PYTHIA with data. This would demand considering other observables like nuclear modification factors of particles and jets and it is not the focus of this paper. As an example of one specific point that deserves further investigation within the model, the number of events that pass the cuts that are imposed, is a small fraction of the initial sample and decreases with increasing $\hat{q}$. This, as usual, points to the existence of some bias. For example, for a simulation with $\hat{q} = 8$ GeV$^2$ fm$^{-1}$, less than $20\, \%$ of the events that we get with $\hat{q} = 0$ fulfill the cuts. This aspect deserves further investigation that we leave for the future.

\subsection{Asymmetry}

\par Figure \ref{fig:Quench_FJ} shows the dijet asymmetry (Figure \ref{fig:Aj_Quench_FJ}) and  the dijet azimuthal correlation (Figure \ref{fig:Dphi_Quench_FJ}) using the area-based FastJet subtraction method for different strengths of quenching. The increase of $\hat{q}$ induces  momentum imbalance: for $\hat{q}=8$ GeV$^{2}$/fm, the $A_{J}$ distribution is in qualtitative agreement with experimental data (the agreement seems better when considering the difference between the CMS Monte Carlo and CMS data \cite{Chatrchyan:2012nia}).On the other hand, in spite of the shower being degraded in energy so that a significant $A_{J}$ is induced by quenching, the azimuthal dijet correlation is not very strongly modified, in apparent conflict with the generic link between energy loss and broadening \cite{dEnterria:2009am,Wiedemann:2009sh,Majumder:2010qh,Armesto:2011ht}.

\begin{figure}[htbp]
	\begin{center}
		\subfigure[{Dijet asymmetry $A_J$.}]{
			\label{fig:Aj_Quench_FJ}
			\includegraphics[width=0.48\textwidth]{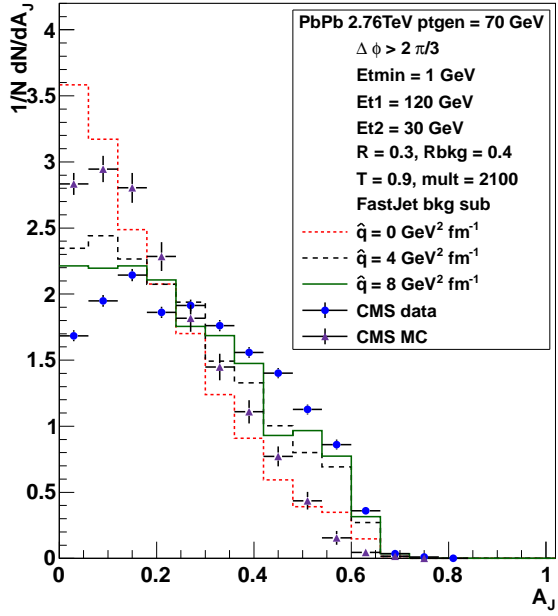}
		}
		\subfigure[{Dijet azimuthal correlation.}]{
			\label{fig:Dphi_Quench_FJ}
			\includegraphics[width=0.48\textwidth]{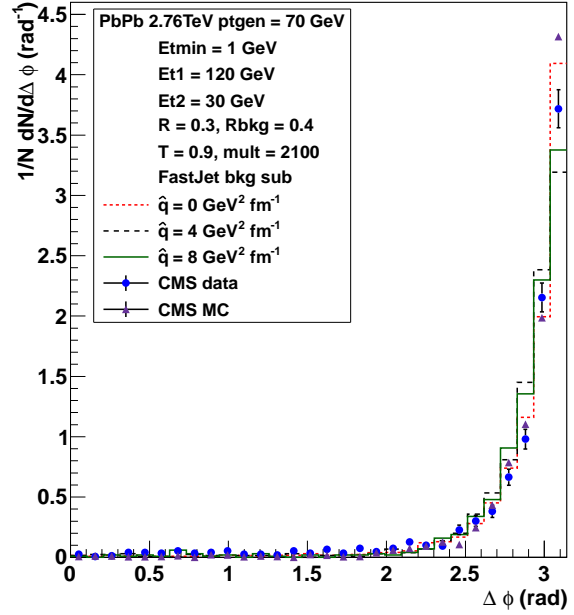}
		}
		\caption{{Dijet observables for a simulation using Q-PYTHIA with different $\hat{q}$   embedded in a background with $T = 0.9$ GeV ($\sigma_{jet} \simeq 11$ GeV). The red dotted lines corresponds to  $\hat{q} = 0$, the black dashed ones to $\hat{q} = 4$ GeV$^{2}$ fm$^{-1}$ and the green solid ones to $\hat{q} = 8$ GeV$^{2}$ fm$^{-1}$. The blue dots are the CMS data with the corresponding error bars, and the purple triangles the CMS Monte Carlo \cite{Chatrchyan:2012nia}. The background subtraction method is the area-based FastJet one.}}
		\label{fig:Quench_FJ}
	\end{center}
\end{figure}

\par Similar results are obtained for the pedestal method with $E_{T,jets}=30$ GeV and  $\kappa=2.2$, Figure \ref{fig:Quench_CMS}. When the second procedure in the pedestal subtraction method is used instead, see Appendix \ref{appb}, the momentum imbalance induced by quenching persists, but it is less significant than in the previous cases and the agreement with data is consequently poorer even for $\hat{q}=8$ GeV$^2$ fm$^{-1}$. The azimuthal correlations are slightly broader.

\begin{figure}[htbp]
	\begin{center}
		\subfigure[{Dijet asymmetry $A_J$.}]{
			\label{fig:Aj_Quench_CMS}
			\includegraphics[width=0.48\textwidth]{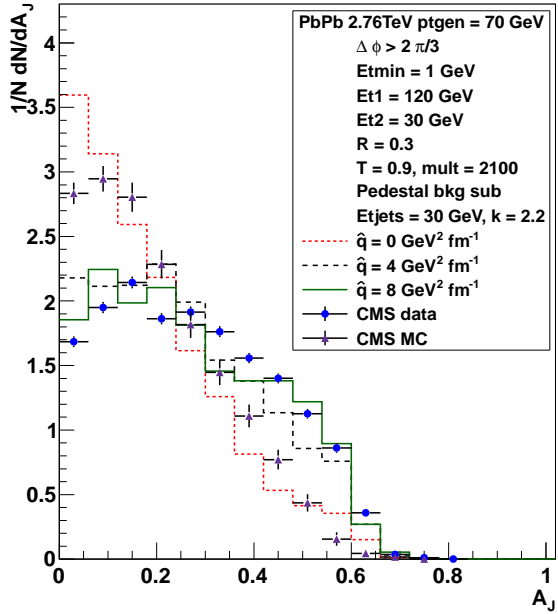}
		}
		\subfigure[{Dijet azimuthal correlation.}]{
			\label{fig:Dphi_Quench_CMS}
			\includegraphics[width=0.48\textwidth]{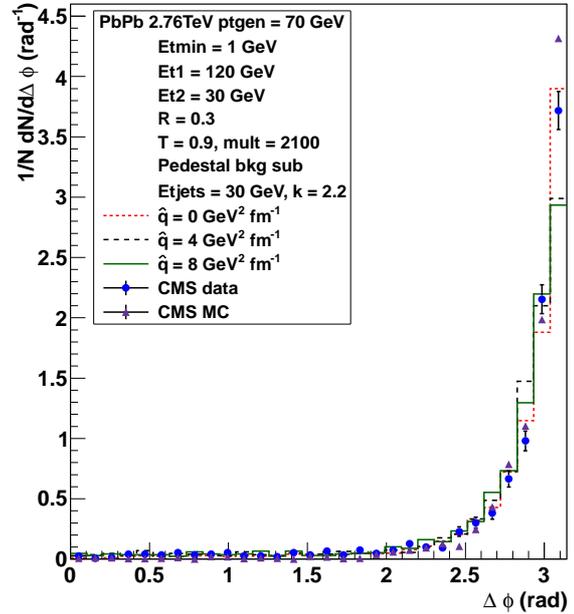}
		}
		\caption{{Dijet observables for a simulation using Q-PYTHIA with different $\hat{q}$   embedded in a background with $T = 0.9$ GeV ($\sigma_{jet} \simeq 11$ GeV). The red dotted lines corresponds to  $\hat{q} = 0$, the black dashed ones to $\hat{q} = 4$ GeV$^{2}$ fm$^{-1}$ and the green solid ones to $\hat{q} = 8$ GeV$^{2}$ fm$^{-1}$. The blue dots are the CMS data with the corresponding error bars, and the purple triangles the CMS Monte Carlo. The background subtraction was made using a pedestal method \cite{Chatrchyan:2012nia}.}}
		\label{fig:Quench_CMS}
	\end{center}
\end{figure}

\subsection{Missing transverse momentum}
\label{misspt}

\par Another observable of interest is the missing transverse momentum measured by CMS \cite{Chatrchyan:2011sx}. Given a dijet pair, the projection of the charged tracks onto the dijet axis is computed as indicated in Equation (\ref{eq:misspt}). All charged particles with $p_T > 0.5$ GeV/c and within $|\eta| < 2.4$ enter the sum. CMS measures (\ref{eq:misspt}) to be zero for all dijet asymmetries and different centralities, meaning that the momentum balance of the dijet pairs is recovered. CMS further explores the jet structure by computing the missing $p_{T}$  with only the tracks that are contained inside or outside a cone of radius $R=0.8$ around the leading and subleading jet axis. Several relevant observations in PbPb collisions, which becomes more and more significant with increasing momentum imbalance,  are extracted from the study:

\begin{itemize}
	\item For $R<0.8$ there is a net negative $\left\langle \slashed{p}_T^\parallel \right\rangle$ dominated by tracks with $p_{T}>8$ GeV.  For $R>0.8$ there is a net positive contribution dominated by tracks with low $0.5<p_{T}<1$ GeV.
	\item Both contributions sum up to give a net zero $\left\langle \slashed{p}_T^\parallel \right\rangle$.
	\item The subleading jet structure is significantly softer in data than in the CMS Monte Carlo.
\end{itemize}

\par In summary, CMS measures dijet events with large momentum imbalance. In those events, the core of the subleading jets is degraded in energy and this energy is recovered at large angles in the form of  soft particles. This suggests a mechanism for energy loss that transports soft particles up to very large angles. Such a mechanism is not implemented as such in Q-PYTHIA, as commented above. Nevertheless, here we will show the results that this quenching model provides. {The reason, apart from its intrinsic interest, is that it provides an example of an observable that considers all particles without background subtraction (except for the definition of the leading jet axis that hopefully is little affected as shown by the experimental azimuthal correlations). Note that our toy model simulates a system with global variables similar to those measured in the experiment: multiplicity and average background fluctuations. However, the distribution of particles in momentum and the  range of their correlations is unconstrained. The observable defined in equtation (\ref{eq:misspt}) studies the track structure with respect to the dijet axis  and is sensitive to these  details not considered in our toy model. In the absence of a truly realistic background model we rather present our results for Q-PYTHIA alone,  thus aiming for - at most - a qualitative study of the observable.

\par In Figure \ref{fig:MissPt_qhat0}, we show the results for the average missing transverse momentum using Q-PYTHIA with $\hat{q} = 0$. Each $p_T$ bin contribution is associated to a different color. In Figure \ref{fig:MissPt_qhat0_all}, where the full phase space for the projection is considered, there is a higher amount of hard particles ($p_T > 8$ GeV) in the direction of the leading jet. Those are essentially balanced by particles with a transverse momentum $p_T > 2$ GeV and only a small fraction of the available energy is carried by the softest particles.  When only particles inside a cone of $R = 0.8$ around the leading and subleading jets are considered (Figure \ref{fig:MissPt_qhat0_in}), the momentum imbalance is due to an excess of hard particles in the direction of the leading jet. Outside this cone (Figure \ref{fig:MissPt_qhat0_out}) the composition of the event is also essentially hard ($p_T>4$ GeV/c).  These features are in qualitative agreement with the CMS Monte Carlo \cite{Chatrchyan:2011sx}.

\begin{figure}[htbp]
	\begin{center}
		\subfigure[Projection onto the leading and subleading jet axis for the whole phase space.]{
			\label{fig:MissPt_qhat0_all}
			\includegraphics[width=0.31\textwidth]{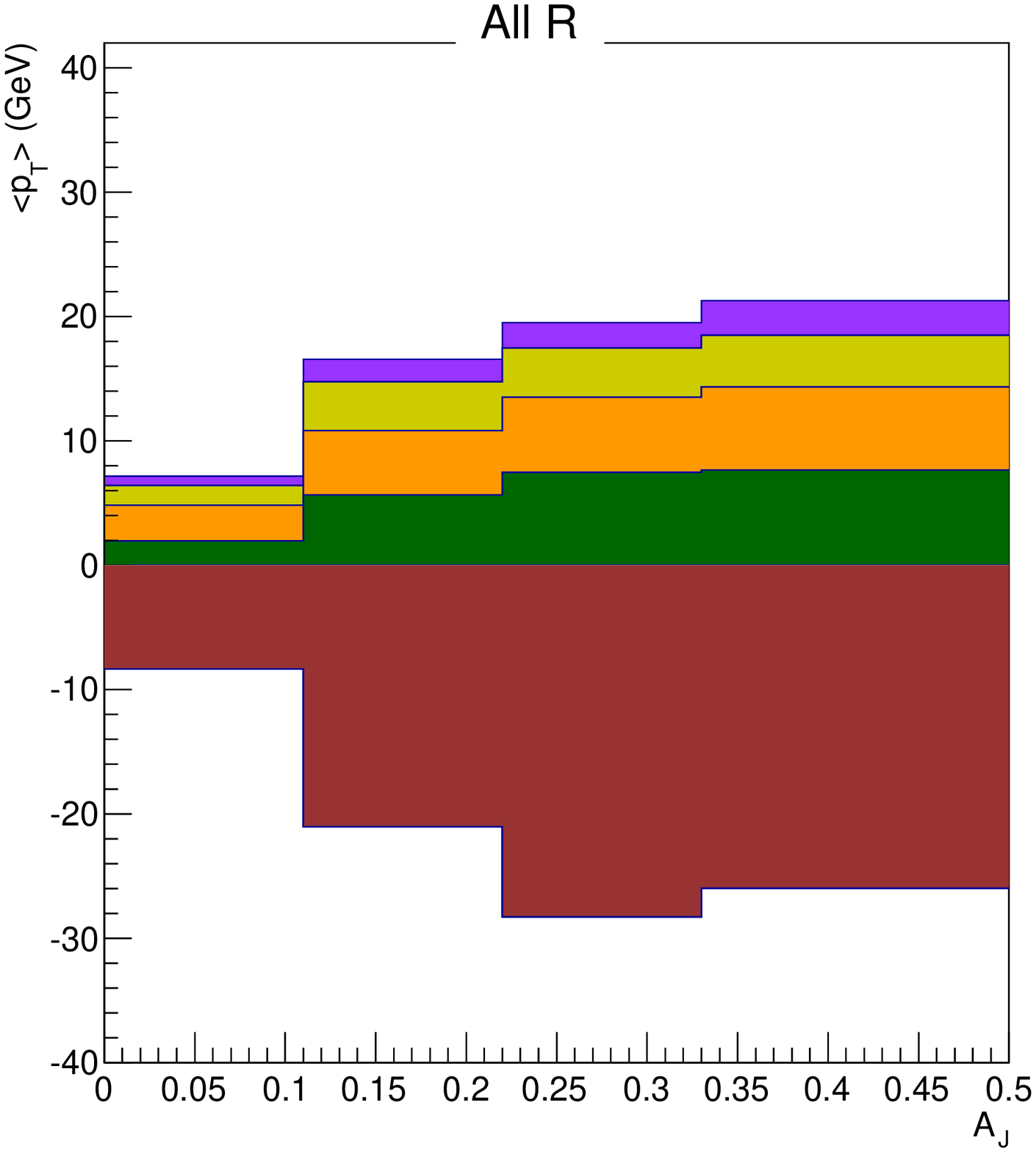}
		}
		\subfigure[Projection onto the leading and subleading jet axis for particles inside a cone with $R = 0.8$.]{
			\label{fig:MissPt_qhat0_in}
			\includegraphics[width=0.31\textwidth]{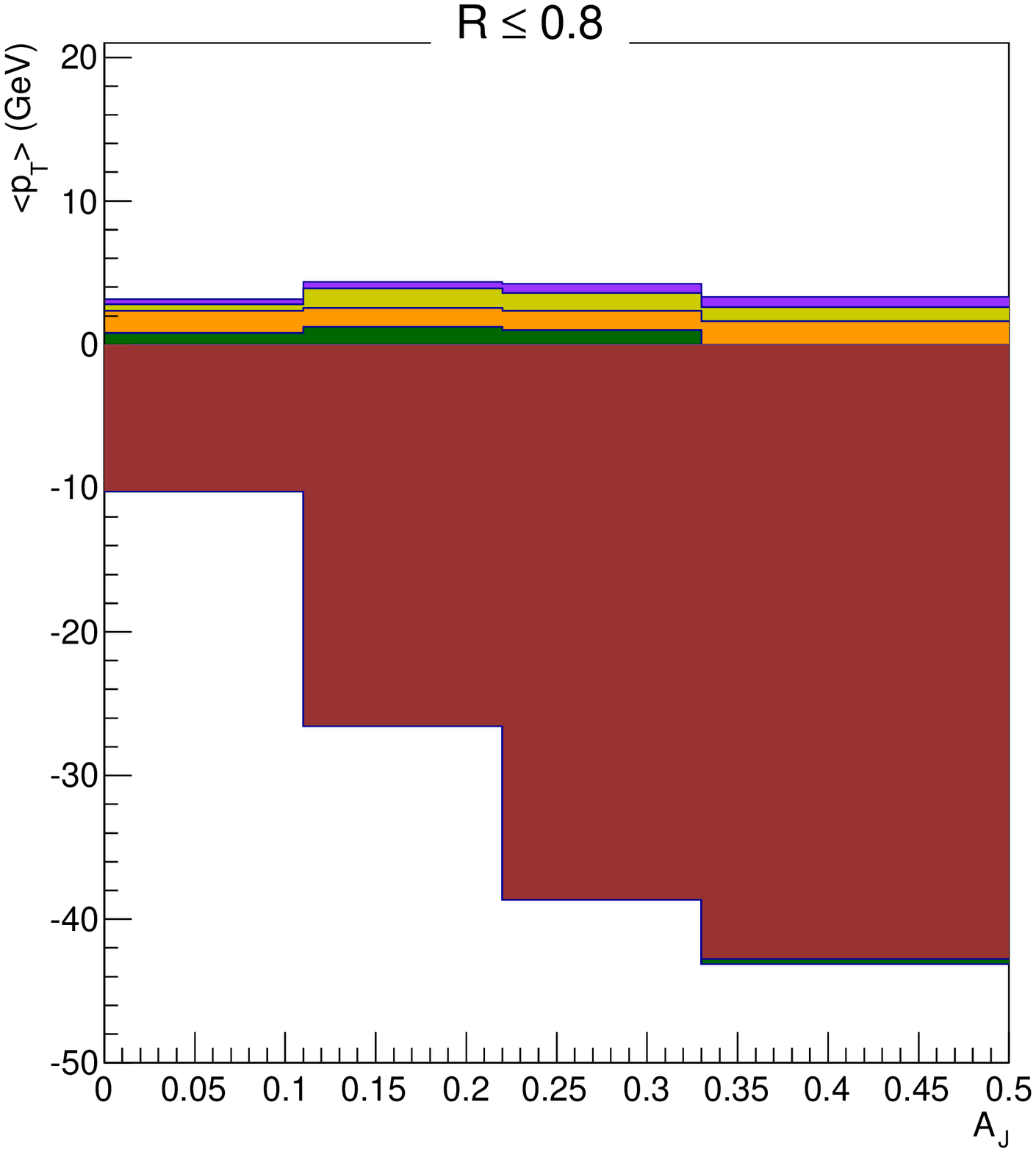}
		}
		\subfigure[Projection onto the leading and subleading jet axis for particles outside a cone with $R = 0.8$.]{
			\label{fig:MissPt_qhat0_out}
			\includegraphics[width=0.31\textwidth]{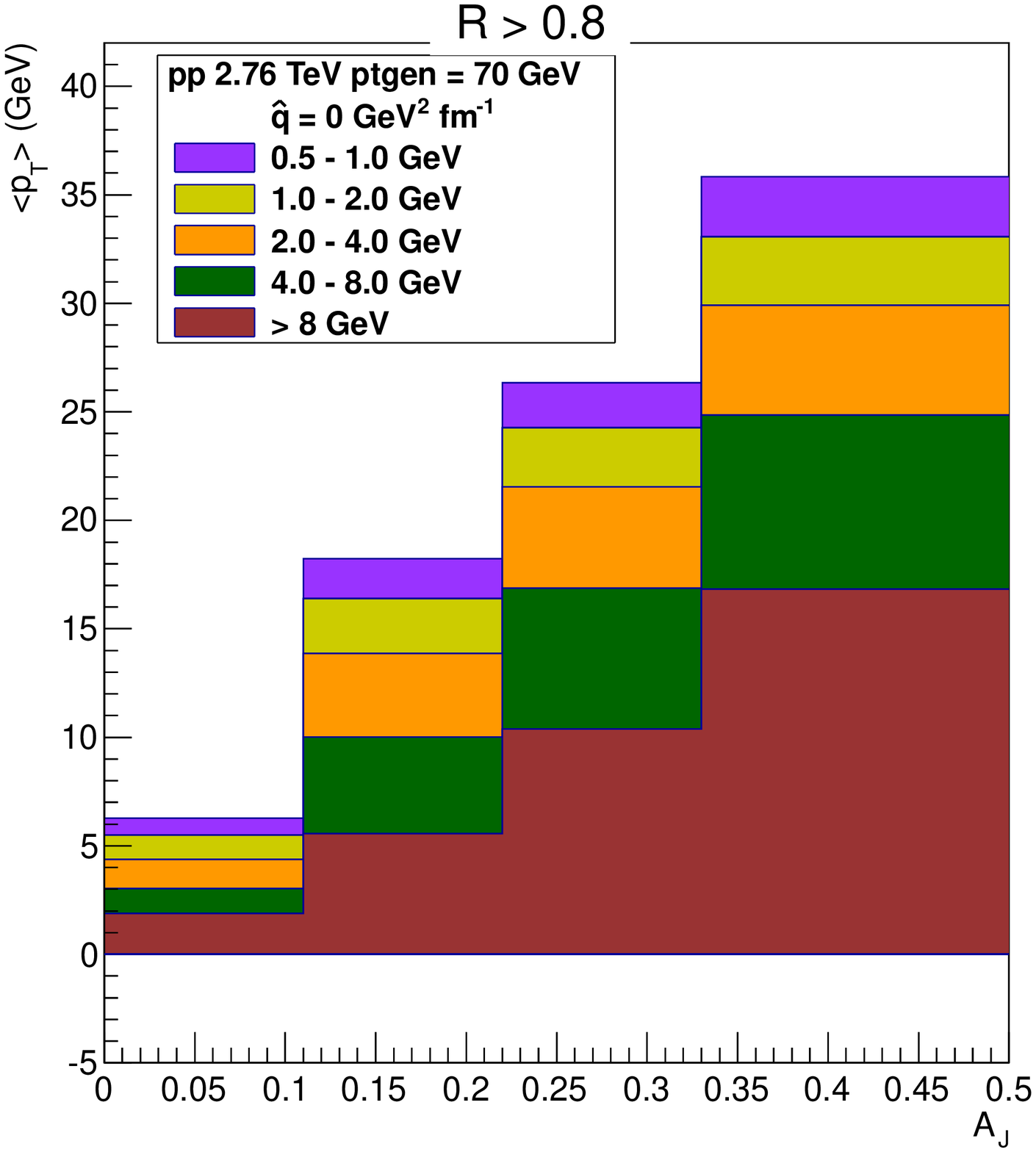}
		}
		\caption{Average missing transverse momentum $\left\langle \slashed{p}_T^\parallel \right\rangle$ for Q-PYTHIA simulated events with $\hat{q} = 0$.}
		\label{fig:MissPt_qhat0}
	\end{center}
\end{figure}

\par Taking our results for $\hat{q} = 0$ as reference, we now explore the effects of quenching: we consider Q-PYTHIA with a $\hat{q} = 8$ GeV$^2$ fm$^{-1}$ in Figure \ref{fig:MissPt_qhat8}. The global structure changes and the projections onto the leading and subleading axis are softened. Note that for $\hat{q}=0$  there was an excess of hard particles at $R>0.8$ with respect to the subleading jet. For $\hat{q} = 8$ GeV$^2$ fm$^{-1}$, this excess of momentum is driven by particles with $p_{T} \in [0.5, 4]$ GeV/c. This softening of the particle composition, mostly noticeable  at large angles with respect to the subleading jet, is in qualitative agreement with data.

\begin{figure}[htbp]
	\begin{center}
		\subfigure[Projection onto the leading and subleading jet axis for the whole phase space.]{
			\label{fig:MissPt_qhat8_all}
			\includegraphics[width=0.31\textwidth]{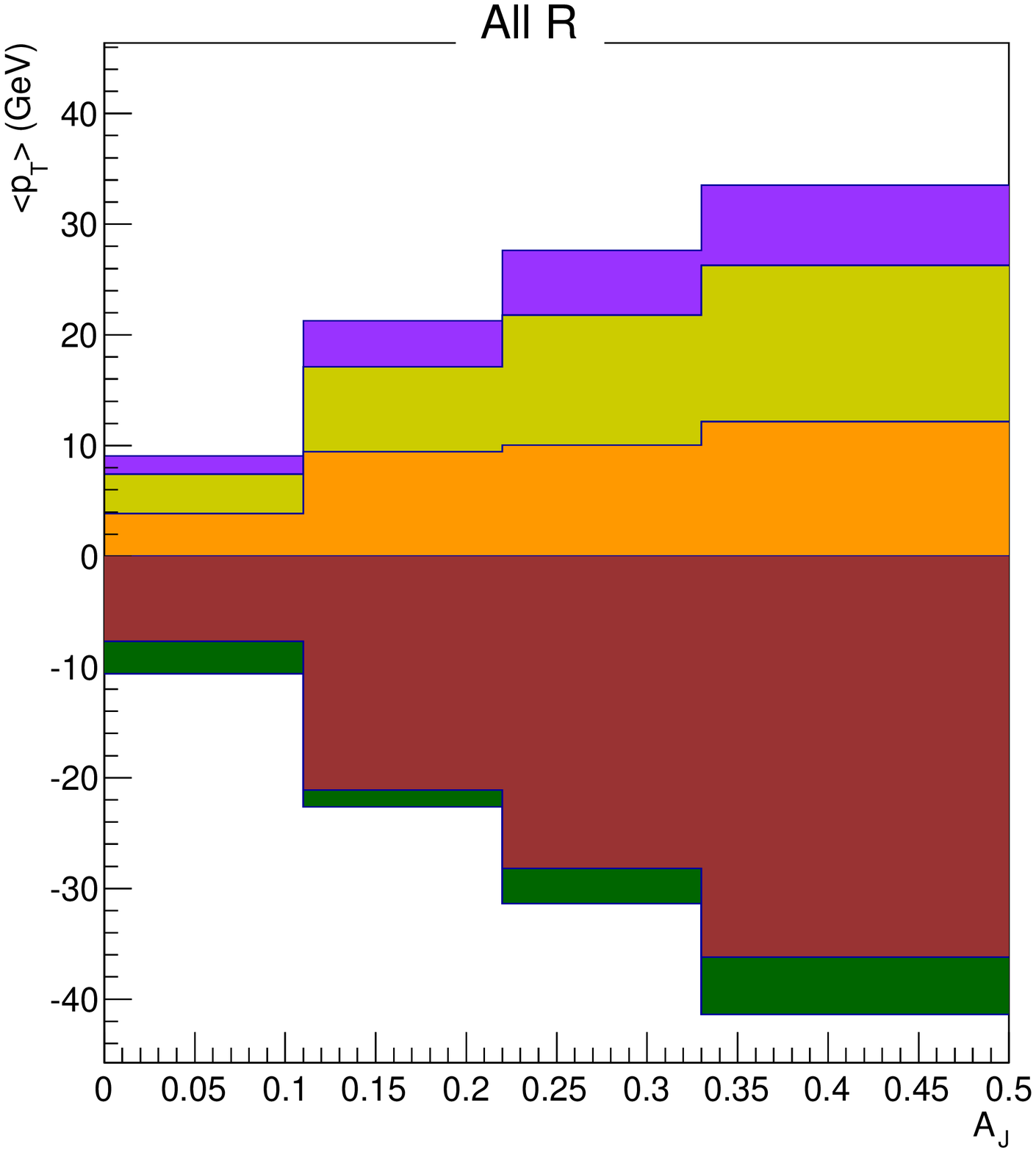}
		}
		\subfigure[Projection onto the leading and subleading jet axis for particles inside a cone with $R = 0.8$.]{
			\label{fig:MissPt_qhat8_in}
			\includegraphics[width=0.31\textwidth]{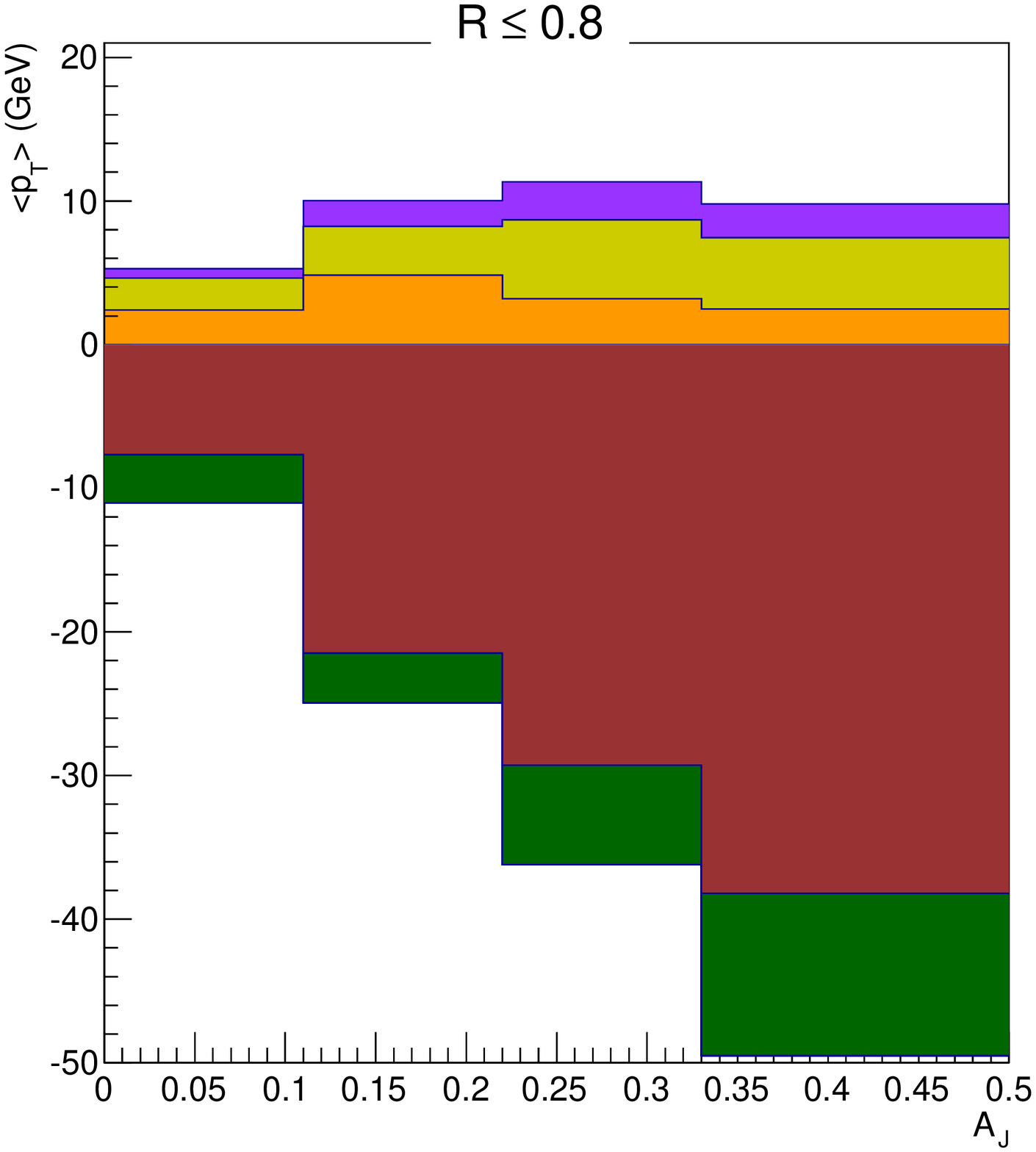}
		}
		\subfigure[Projection onto the leading and subleading jet axis for particles outside a cone with $R = 0.8$.]{
			\label{fig:MissPt_qhat8_out}
			\includegraphics[width=0.31\textwidth]{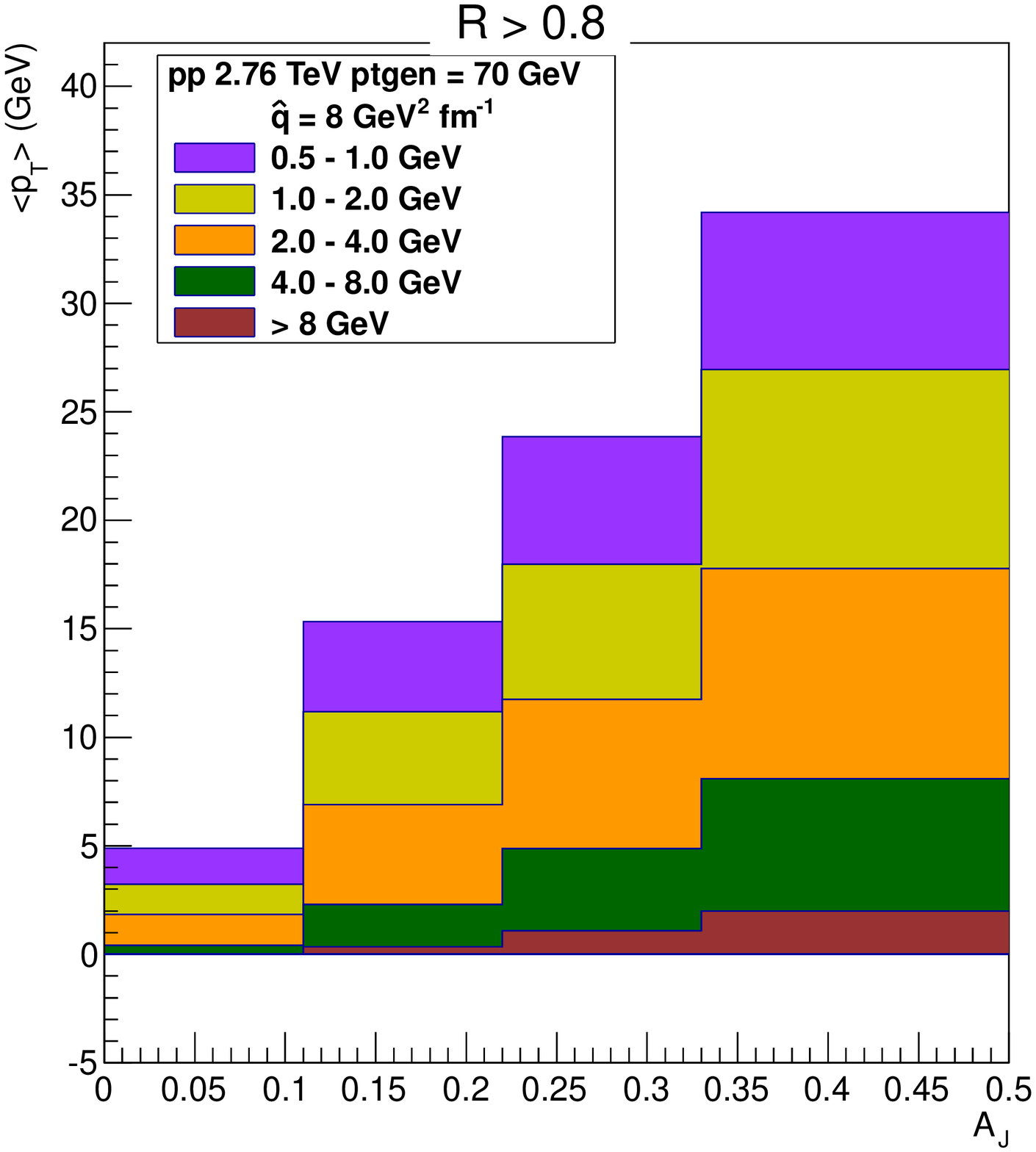}
		}
		\caption{Average missing transverse momentum $\left\langle \slashed{p}_T^\parallel \right\rangle$ for Q-PYTHIA simulated events with $\hat{q} = 8$ GeV$^{2}$ fm$^{-1}$.}
		\label{fig:MissPt_qhat8}
	\end{center}
\end{figure}

\par In order to discuss this finding, let us consider  events with a large asymmetry $A_J > 0.3$. Both the Q-PYTHIA simulation with $\hat{q} = 0$ (proton-proton) and the CMS simulation contain a hard contribution outside the subleading jet cone. This can only come from either hard emissions at large angles from the subleading jet that are reconstructed as a third jet at $\Delta R > 0.8$ or from a hard scattering with 3 hard particles (real "3 jet-like" structures). The latter  are suppressed by the cut in $\Delta \phi$ and, in any case, they are not considered in PYTHIA that only contains lowest-order matrix elements. Now we turn on quenching: by definition, the subleading jet suffers a larger energy degradation that implies a larger amount of radiation in  radiative energy loss scenarios, and the jet-finding algorithm will most probably reconstruct two or more smaller jets instead of a single broad jet. This combination of semi-hard jet multiplication at large angles (but still inside the dijet cone as hard particles lead the jet reconstruction in the anti-$k_t$ algorithm) with the fact that these semi-hard jets are further forced to radiate in-medium,   may lead to an overall softer composition even outside the cone formed by the dijet pair.

\begin{figure}[htbp]
	\begin{center}
		\subfigure[Average number of jets with a minimum $p_T> 8$ GeV for the whole phase space.]{
			\label{fig:Njets_all}
			\includegraphics[width=0.31\textwidth]{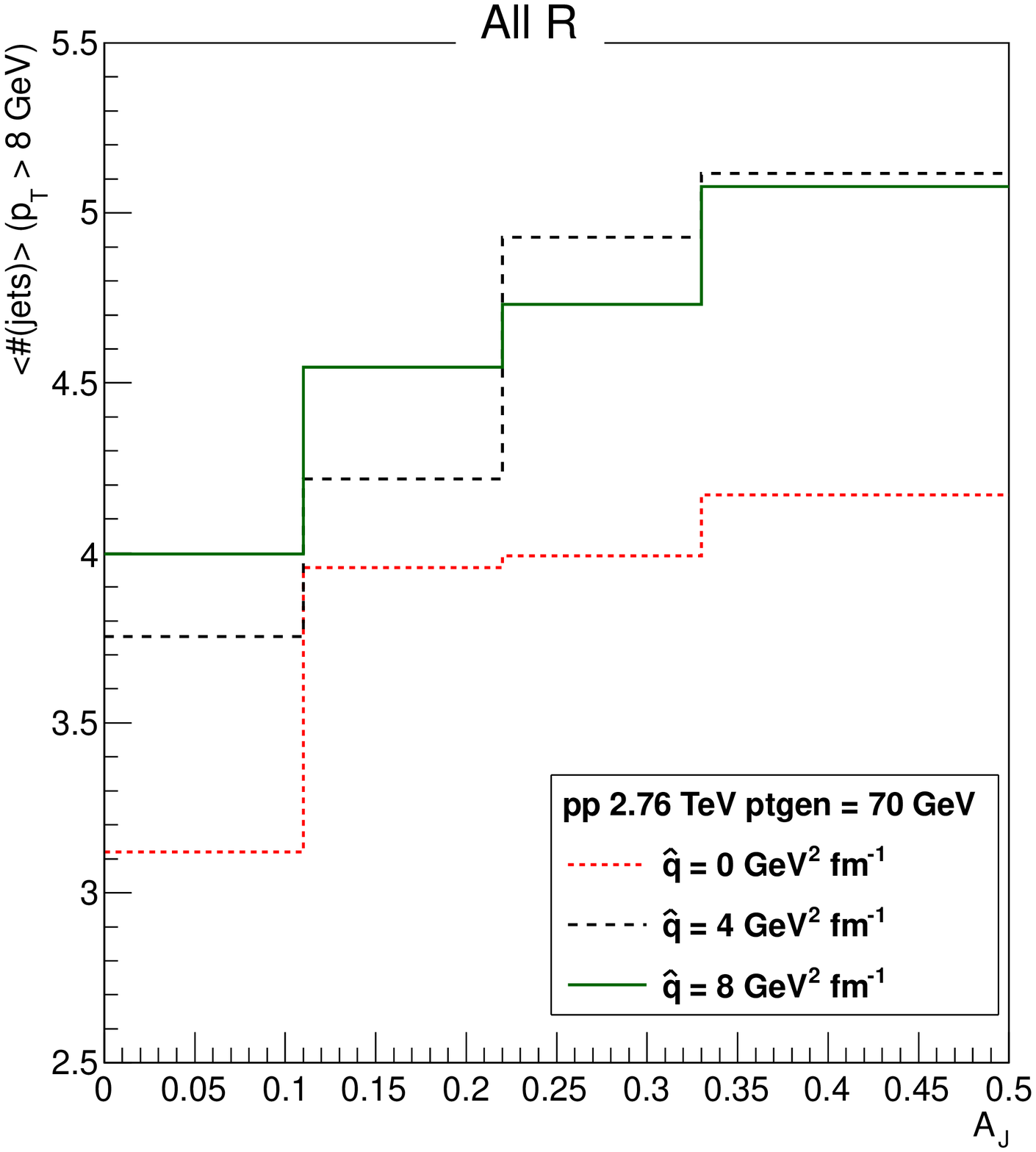}
		}
		\subfigure[Average number of jets with a minimum $p_T > 8$ GeV inside a cone with $R = 0.8$.]{
			\label{fig:Njets_in}
			\includegraphics[width=0.31\textwidth]{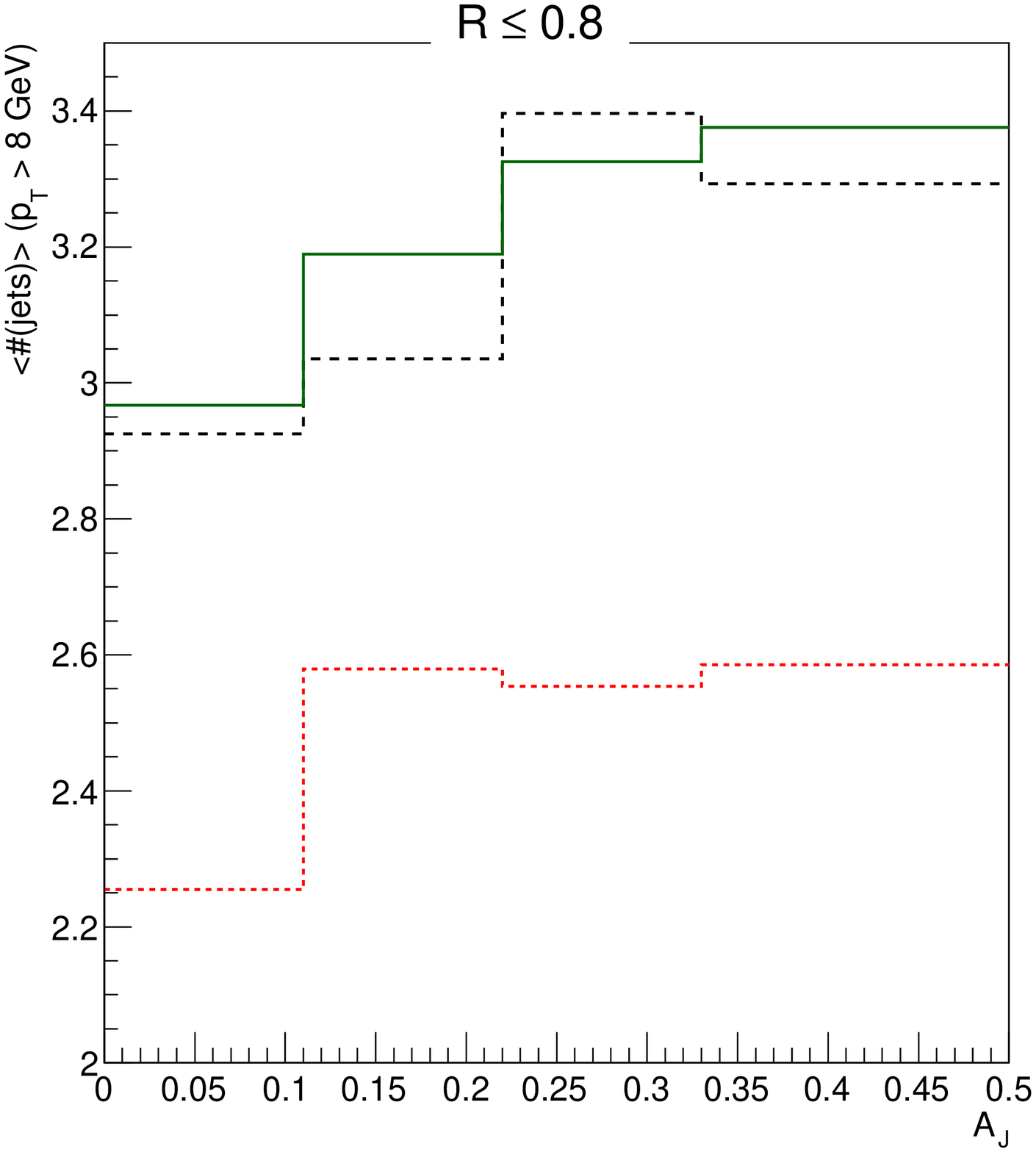}
		}
		\subfigure[Average number of jets that have at least one particle with a minimum $p_{T,track} > 8$ GeV outside a cone with $R = 0.8$.]{
			\label{fig:Njets_out}
			\includegraphics[width=0.31\textwidth]{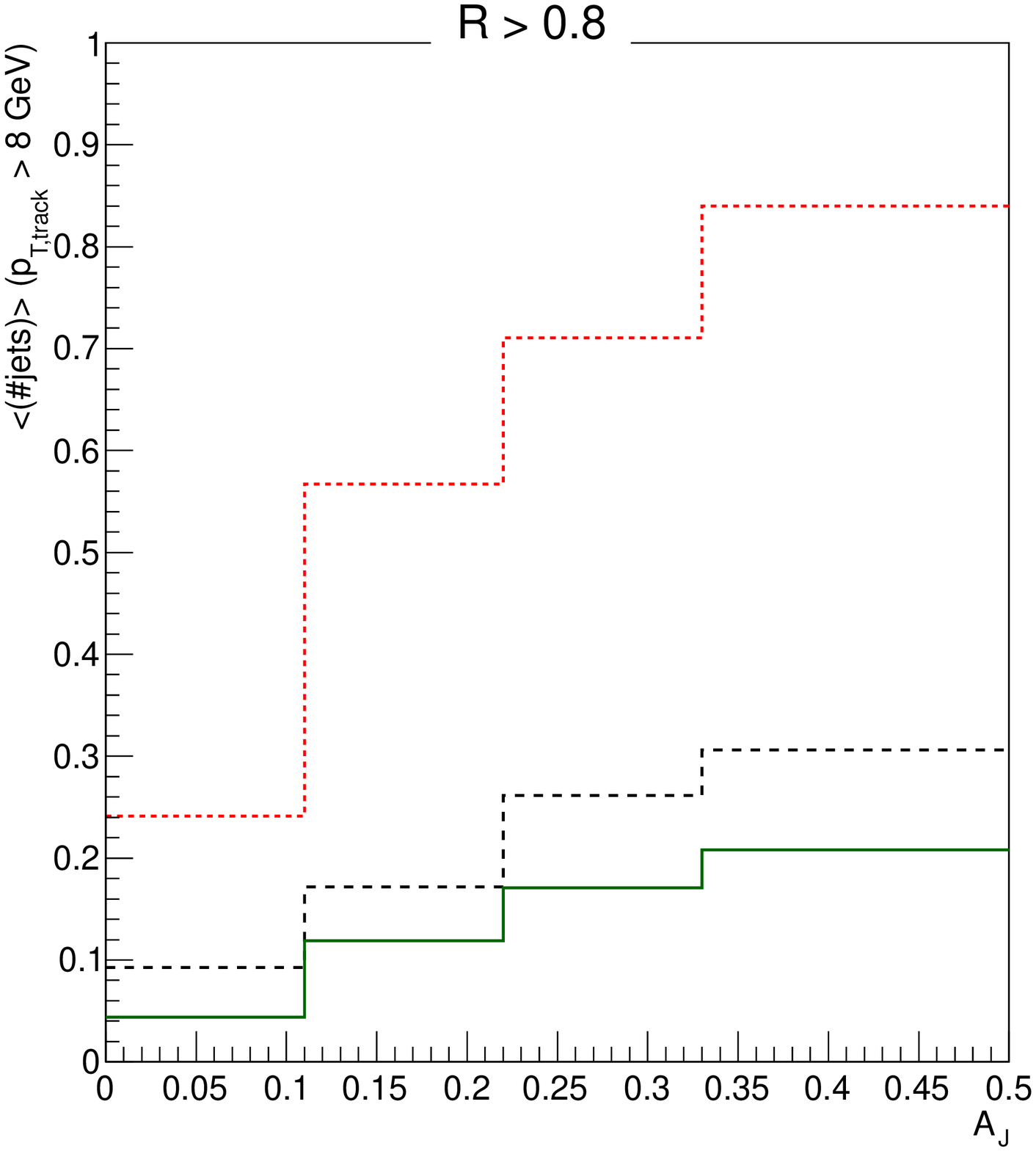}
		}
		\caption{Average number of jets, $<\#(jets)>$, for Q-PYTHIA simulated events with different $\hat{q}$. The red dotted lines corresponds to $\hat{q} = 0$, the black dashed ones to $\hat{q} = 4$ GeV$^{2}$ fm$^{-1}$ and the green solid ones to $\hat{q} = 8$ GeV$^{2}$ fm$^{-1}$. }
		\label{fig:Njets}
	\end{center}
\end{figure}

\par Such considerations are supported by Figure \ref{fig:Njets_all}, where the average number of jets with a transverse momentum $p_T > 8$ GeV is shown for all phase space. The red dotted lines corresponds to a simulation of Q-PYTHIA using $\hat{q} = 0$, the black dashed ones to $\hat{q} = 4$ GeV$^2$ fm$^{-1}$ and the green solid ones to $\hat{q} = 8$ GeV$^2$ fm$^{-1}$. The number of jets is found to increase with the asymmetry and with  $\hat{q}$. With quenching, the jet finding algorithm reconstructs more jets than in vacuum. Inside a cone around the leading and subleading jet (Figure \ref{fig:Njets_in}) the same description is found: in proton-proton, on average only the dijet pair is reconstructed but, when medium effects are present, an additional jet with $p_T > 8$ GeV is identified. Outside the cone, however, if we compute the average number of jets that have at least one particle with a transverse momentum $p_{T,track} > 8$ GeV, we see that this number decreases with increasing $\hat{q}$ (Figure \ref{fig:Njets_out}). So, in general, medium effects produce more jets but with a softer composition.

\par One may argue that such considerations may become washed out by the presence of a background in PbPb collisions that may lead to a larger number of jets picking background, relatively soft constituents. Even in this case, the increase of the soft contribution at large angles with respect to the dijet should remain.

\section{Conclusions}
\label{conclu}

\par In this work we address the question of the effects of jet reconstruction and background subtraction in high-energy heavy-ion collisions on different jet observables. Our aim is to gain insight on how these issues affect the understanding and detailed characterization of the produced  medium through present jet observables (using the experimental data on the dijet asymmetry and azimuthal correlation in \cite{Chatrchyan:2012nia} and on the missing transverse momentum in \cite{Chatrchyan:2011sx} as references).

\par For this purpose, see Section \ref{background}, we use a highly flexible toy model for the background - where particles are simulated according to a thermal spectrum matched to a power law at larger transverse momentum - that allows fluctuations both among different events and, more importantly, event-by-event. By changing the slope of the exponential function, $T$, we can set different values for the background fluctuations, $\sigma_{jet}$, and for the average level of energy deposition, $\rho$. The results of the toy model have also been checked and found in agreement with those using a detailed Monte Carlo simulator for the background, the PSM model. Jets are generated through pp events in PYTHIA for vacuum jets. In order to address possible interplays between a different  structure of  in-medium quenched jets, we also generated samples of jets with different degrees of quenching through pp collisions in Q-PYTHIA. We have studied two background subtraction techniques: the FastJet area-based method, where the estimation of the background parameters is made at jet level; and a pedestal method, where the background estimation is made at a calorimetric level and uses a pedestal subtraction. For the latter, two procedures have been used to fix the parameters, either enhancing the subtraction of background (i.e. using $\kappa>1$ for fixed $E_{T,jets}$)  or, see Appendix B, by considering that the background may contain harder and harder jets (i.e. varying $E_{T,jets}$ for fixed $\kappa=1$). Their influence on several jet observables: jet spectra, dijet azimuthal correlation and dijet asymmetry, was investigated in Section \ref{bkgParam}. The conclusions that we get are the following:

\begin{itemize}
	\item  Concerning the inclusive jet spectra: for the Fastjet procedure and for the pedestal method with $\kappa>1$, it mainly depends on  the average deposition of energy that sets a lower bound in $p_T$ for the reconstruction ability of the method; for the pedestal one with $\kappa=1$ it depends on the cut used to separate the particles coming from the hard event and the background particles, $E_{T,jets}$, that we have tuned in order to reconstruct the same level of background with and without embedded jets.
	\item The dijet asymmetry or momentum imbalance is affected by the background fluctuations in the Fastjet method and for the pedestal method with $\kappa>1$, with increasing fluctuations going in the same direction that pp to increasingly central PbPb collisions observed in data (as already observed in \cite{Cacciari:2011tm}). As for the pedestal technique with $\kappa=1$, its sensitivity to background parameters is smaller for small fluctuations but may even work in opposite direction than the FastJet method for larger background parameters.
	\item The azimuthal correlations between hardest and next-to-hardest jets is little affected by different background parameters for the FastJet method, while for the pedestal technique some effect in the form of a pedestal at all azimuths is seen for the largest $\rho$ and $\sigma_{jet}$, a feature that we attribute to the appearance of fake jets due to the presence of the background.
	\item When including azimuthal structures $v_2$ and $v_3$ in the background, we find that 'realistic' values do not result in a significant change in the dijet asymmetry, but can induce modifications in the dijet azimuthal correlation. This feature is due to the azimuthal distribution of the dijet pair following the particle distribution modulation. As a consequence, a strong correlation between the dijet asymmetry and the dijet azimuthal coordinate may appear for large background parameters.
\end{itemize}

\par From this study, when using the toy model to simulate the background, or a more realistic Monte Carlo simulator, we conclude that for the FastJet background subtraction, an average $\rho$ and $\sigma_{jet}$ are sufficient to characterize a background, since no apparent dependency was found. As for the pedestal method, we find a higher sensitivity to the background intrinsic structure which requires a  tuning of the parameters in the method, specifically the value $E_{T,jets}$ that separates those jets whose constituents are included in the background estimation from those whose constituents are not included, and the value $\kappa$ that sets the level of background subtraction above the average and leads to larger empty cells for reconstruction. Note that in this respect  our results differ from the qualitative estimations in \cite{Cacciari:2011tm} due to the use of different parameters $\kappa$ and $E_{T,jets}$ and a different background model.

\par In Section \ref{quench}, we studied the effect of a different jet substructure using quenched jets through Q-PYTHIA with different transport coefficient parameters, $\hat{q}$, to simulate the medium effects for pp events that are then embedded in the background. We find significant changes depending on the background subtraction technique that was used. Specifically:

\begin{itemize}
	\item At the level of the dijet asymmetry, quenching increases it but this effect is sizably larger for the FastJet method and for the pedestal method with $\kappa>1$ for background subtraction, than for the pedestal one with $\kappa=1$.
	\item At the level of the dijet azimuthal correlation, we find no strong change with quenching for the FastJet method and for the pedestal method with $\kappa>1$, while the pedestal one with $\kappa=1$ results in a sizable modification due to quenching.
\end{itemize}

\par From all this, we conclude that a key feature in background subtraction is the criteria that it is considered to fix the parameters in the method. When the optimization of the reconstructed energy is considered, both the area-based method and the pedestal one show similar features.

\par We also investigated, in Subsection \ref{misspt}, the average missing transverse momentum observable by comparing Q-PYTHIA without background with the CMS results. We checked first a Q-PYTHIA simulation without medium effects ($\hat{q} = 0$) which results in qualitative agreement with the CMS simulation (PYTHIA events embedded in a HYDJET background \cite{Chatrchyan:2011sx}). Then, switching quenching on, we found that Q-PYTHIA has the same trend than CMS data for this observable: a softer composition in the subleading jet direction that persists even at large angles from the dijet direction. Considering that both this fact and the interpretation of the dijet asymmetry and azimuthal correlations as energy loss without broadening defy the 'standard' understanding of radiative medium-induced energy loss (in which energy loss and broadening are linked and radiation is semi-hard and takes place at large angles), we find this qualitative agreement between Q-PYTHIA and data noteworthy.

\par From our study, it seems unavoidable to conclude that the naive expectation that background subtraction methods are enough for phenomenological jet studies to extract medium characteristics without considering the background, becomes strongly weakened. Indeed, it seems that realistic - even real - background events and the use and detailed understanding of the background subtraction method used in each experiment are required in order to achieve the medium characterization through jet observables. We hope that it helps to trigger joint experimental-theoretical efforts in order to set standards for jet definition and reconstruction in heavy-ion collisions, paralleling the ones done in pp \cite{Buttar:2008jx}. As a bonus, we got some input on the limitations and unexpected features of quenching models compared to several observables, and on their interplay with background subtraction. Work along these directions will be subject of our future studies.

\section*{Acknowledgments}

\par We thank Patricia Conde Mui\~no, David d'Enterria, Yen-Jie Lee, Marco van Leeuwen, Guilherme Milhano, Gunther Roland, Gavin Salam, Carlos Salgado and Gregory Soyez for useful discussions and information during the elaboration of this paper, and Matteo Cacciari and Christof Roland  for discussions, information and a critical reading of the manuscript. We also thank Konrad Tywoniuk who participated in an early stage of this work. LA warmly thanks CERN PH Department for hospitality during stays during which part of this work was done. We gratefully acknowledge financial suppport by the European Research Council grant HotLHC ERC-2001-StG-279579 (LA and NA); by Ministerio de Ciencia e Innovaci\'on of Spain grants FPA2008-01177, FPA2009-06867-E and Consolider-Ingenio 2010 CPAN CSD2007-00042 (LA and NA); by Xunta de Galicia grant PGIDIT10PXIB \-206017PR (LA and NA);  by Funda\c c\~ao para a Ci\^encia e a Tecnologia
of Portugal under projects SFRH/BD/64543/2009 and CERN/FP/116379/2010 (LA); and by FEDER.

\appendix
\section{Comparison of the background models}
\label{appA}

\par The first studies that were done to understand the influence of the background main characteristics on the observables used a Monte Carlo (PSM \cite{Amelin:2001sk}) to simulate the background. In the end, this was substituted by a toy model, described in Subsection \ref{toymodel}, since it was easier to provide different values of $\rho$ and $\sigma_{jet}$. In order to test if the toy model was an valid tool for jet studies, we compare the inclusive jet subtracted spectrum (Figure \ref{fig:2Spec_PSMvsT07}) using both background models, with the same level of fluctuations ($\sigma_{jet,PSM} = 7.8$ GeV and $\sigma_{jet,toy} = 7.7$ GeV) and similar average level of contamination ($\rho_{PSM} = 115.6$ GeV/area and $\rho_{toy} = 137$ GeV/area). For the toy model, these correspond to $T=0.7$ GeV. In  Figure \ref{fig:2Spec_PSMvsT07_FJ}, the background subtraction was performed using FastJet with its standard parameters, indicated in Subsection \ref{reco}. We get a very good agreement between both  subtracted spectra (red dashed lines for the toy model and blue solid lines for PSM) and the original PYTHIA spectrum (black dotted lines). In fact, the subtraction seems to be effective up to smaller $p_T$ when using the toy model, which may come from the detailed shape $\rho$ distribution for each background. We also compare the effect of the background structure in the dijet asymmetry and azimuthal correlation, but no sizable change was found.

\begin{figure}[htbp]
	\begin{center}
		\subfigure[Jet subtracted spectrum using the FastJet area-based technique.]{
			\label{fig:2Spec_PSMvsT07_FJ}
			\includegraphics[width=0.30\textwidth]{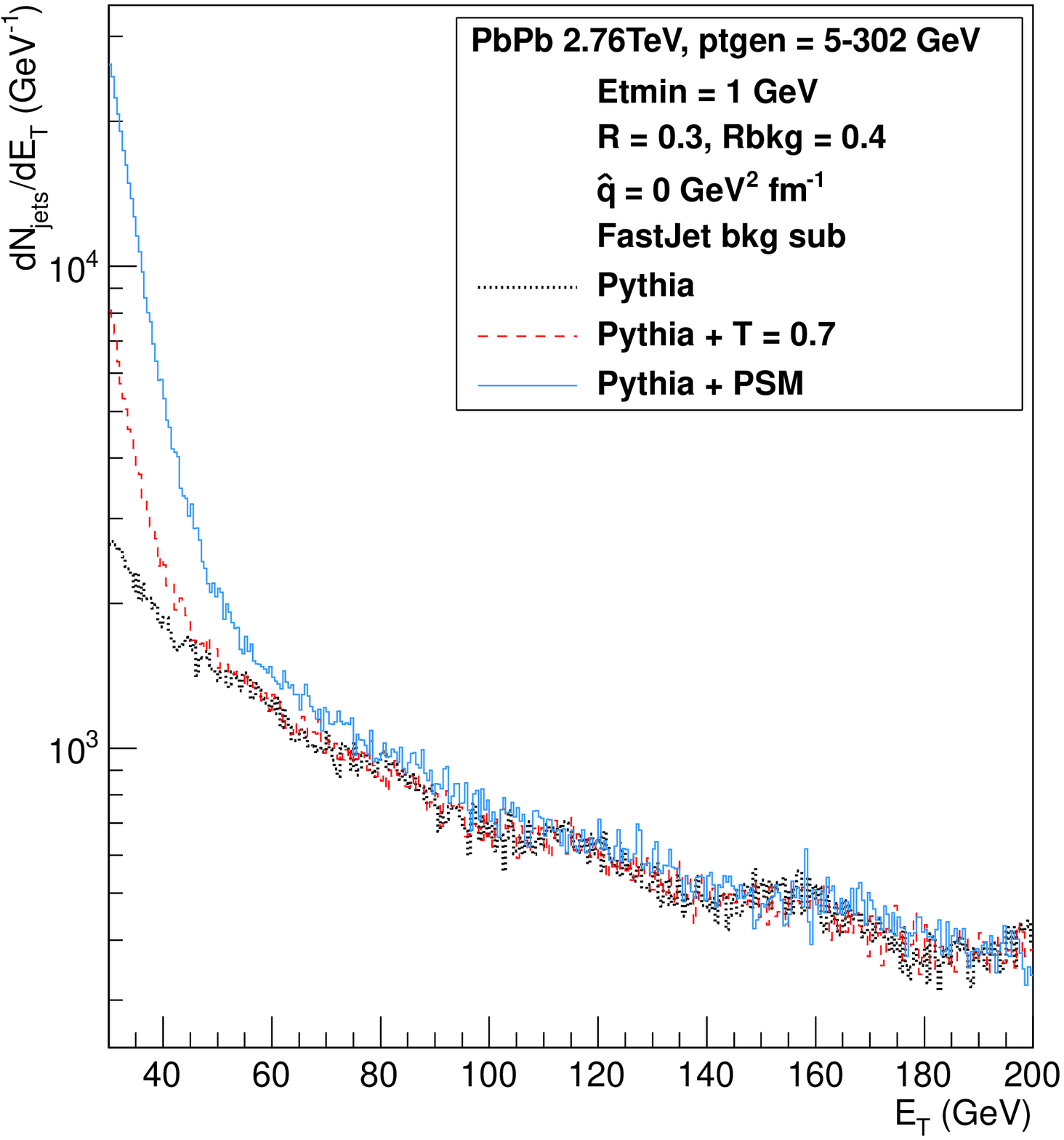}
		}
		\subfigure[Jet subtracted spectrum using a pedestal technique with $\kappa=2.2$.]{
			\label{fig:2Spec_PSMvsT07_CMS2}
			\includegraphics[width=0.30\textwidth]{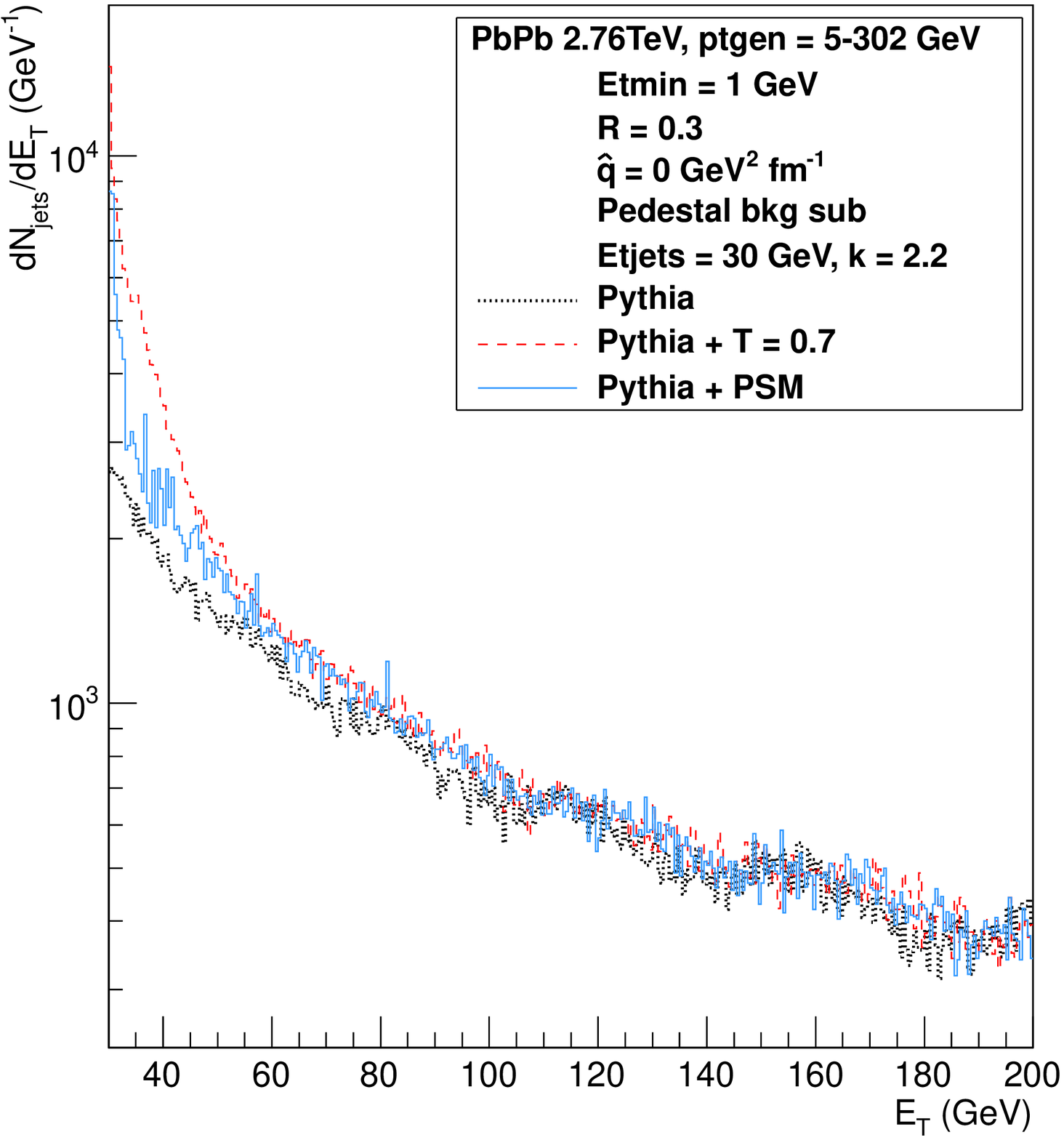}
			}
					\subfigure[Jet subtracted spectrum using a pedestal technique with $\kappa=1$.]{
			\label{fig:2Spec_PSMvsT07_CMS}
			\includegraphics[width=0.30\textwidth]{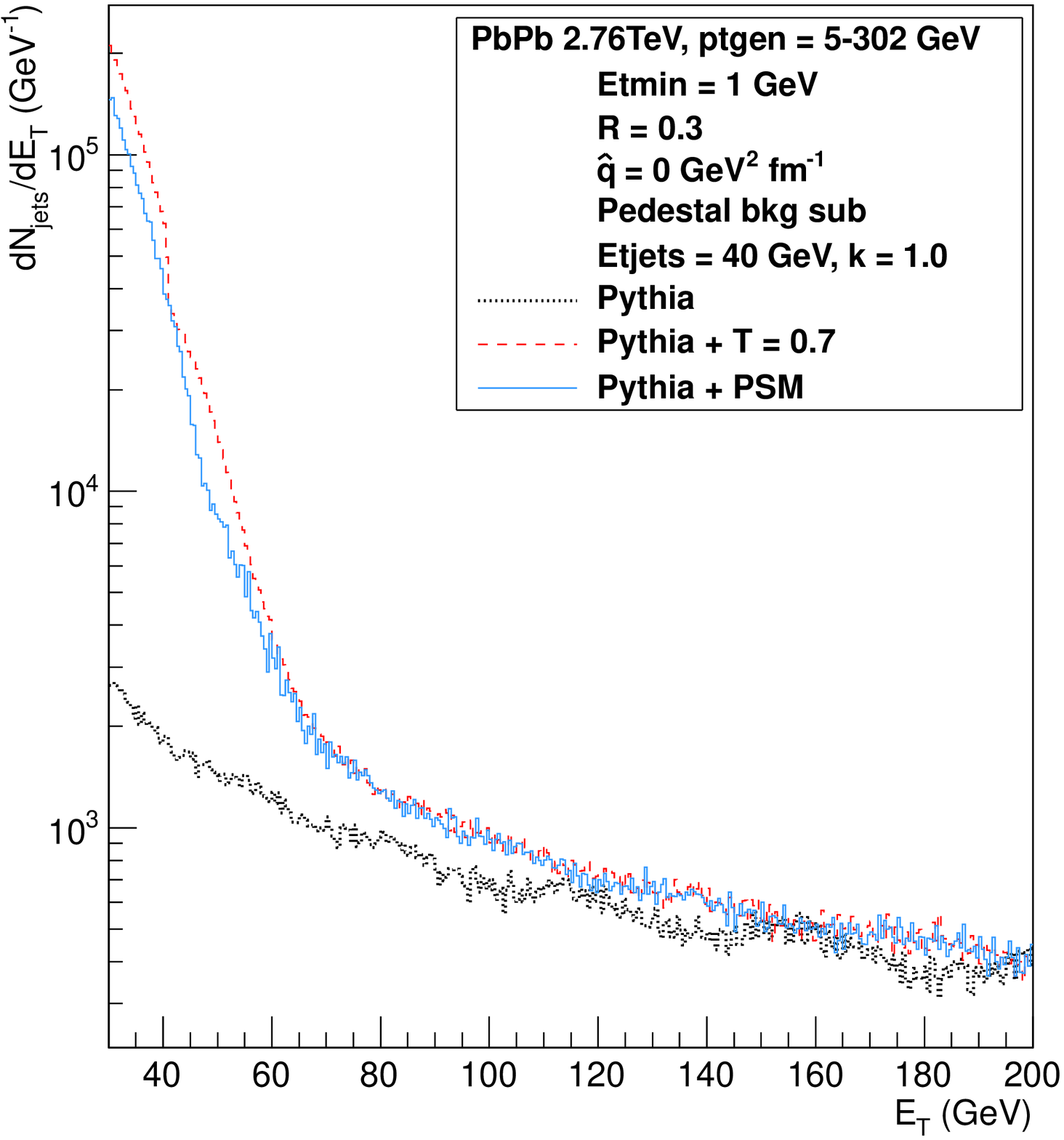}

		}
		\caption{Spectrum of pure PYTHIA jets (black dotted lines) and of subtracted jets using PYTHIA with  PSM as background model (blue solid lines) and with the toy model with  $T = 0.7$ GeV (red dashed lines).}
		\label{fig:2Spec_PSMvsT07}
	\end{center}

\end{figure}

\par For the pedestal subtraction method, a discontinuity in the subtracted spectrum appears in Figure  \ref{fig:SpecCMSappb}. As discussed there, this is related to the cut used in the subtraction procedure, $E_{T,jets}$. But the cut only appears for a uniform background (without the clustering structures coming from a more realistic underlying event). We test the same background subtraction technique using  PSM for the background simulation (see Figures \ref{fig:2Spec_PSMvsT07_CMS2} and \ref{fig:2Spec_PSMvsT07_CMS}), where the best value for $E_{T,jets}$ is the same, and this discontinuity does not appear. Still, the results are quite identical.

\par Furthermore, no significant deviations are found when we compare the jet subtracted spectrum, momentum imbalance and azimuthal correlation obtained using both background models, for both background subtraction methods. From this, we  conclude that our toy model is reliable for our jet studies. 

\section{Second procedure for fixing the parameters in the pedestal subtraction method: fixing $\kappa=1$, varying $E_{T,jets}$}
\label{appb}

\par In this Appendix we elaborate on the pedestal subtraction method when we fix $\kappa=1$ (as in \cite{Kodolova:2007hd,Chatrchyan:2011sx,Chatrchyan:2012nia}) and determine an optimal value of $E_{T,jets}$ by comparing the values of the background estimates in all  $\eta$ stripes ($\left\langle E_{T}^{tower} (\eta) \right\rangle$  and $\left\langle \sigma_{T}^{tower} (\eta) \right\rangle$) given by the subtraction method, to the corresponding values when the toy model is purely underlying event, without generation of  a hard component. An example is shown in Figure \ref{fig:CMS} for $T = 0.9$ GeV. One can see that, although the average values of $\left\langle E_{T}^{tower} (\eta) \right\rangle$ and $\left\langle \sigma_{T}^{tower} (\eta) \right\rangle$ are in more or less agreement with those in \cite{Kodolova:2007hd}, we can only have a satisfactory match\footnote{The match is done by minimizing $\left| \left[\frac{1}{N_{\eta\  bins}}\sum_{\eta \ bins} \frac{\rho_T^{tower}(\eta,{\rm with \  jets})}{\rho_T^{tower}(\eta,{\rm without \  jets)}}\right]-1\right|$.} between the input background parameters (black dashed) and the final estimation parameters (green solid) when $E_{T,jets} = 60$ GeV. For temperatures of $T = 0.7$ GeV, and $T=1.2$ GeV, the optimal cuts are found to be $E_{T jets} = 40$ GeV and $E_{T,jets} = 70$ GeV respectively. These values are sizably higher than the $E_{T jets}$ of order  10 GeV quoted in \cite{Kodolova:2007hd}

\begin{figure}[htbp]
	\begin{center}
		\subfigure[Transverse energy $\left\langle E_{T}^{tower} (\eta) \right\rangle$.]{
			\label{fig:SpecFJ2}
			\includegraphics[width=0.48\textwidth]{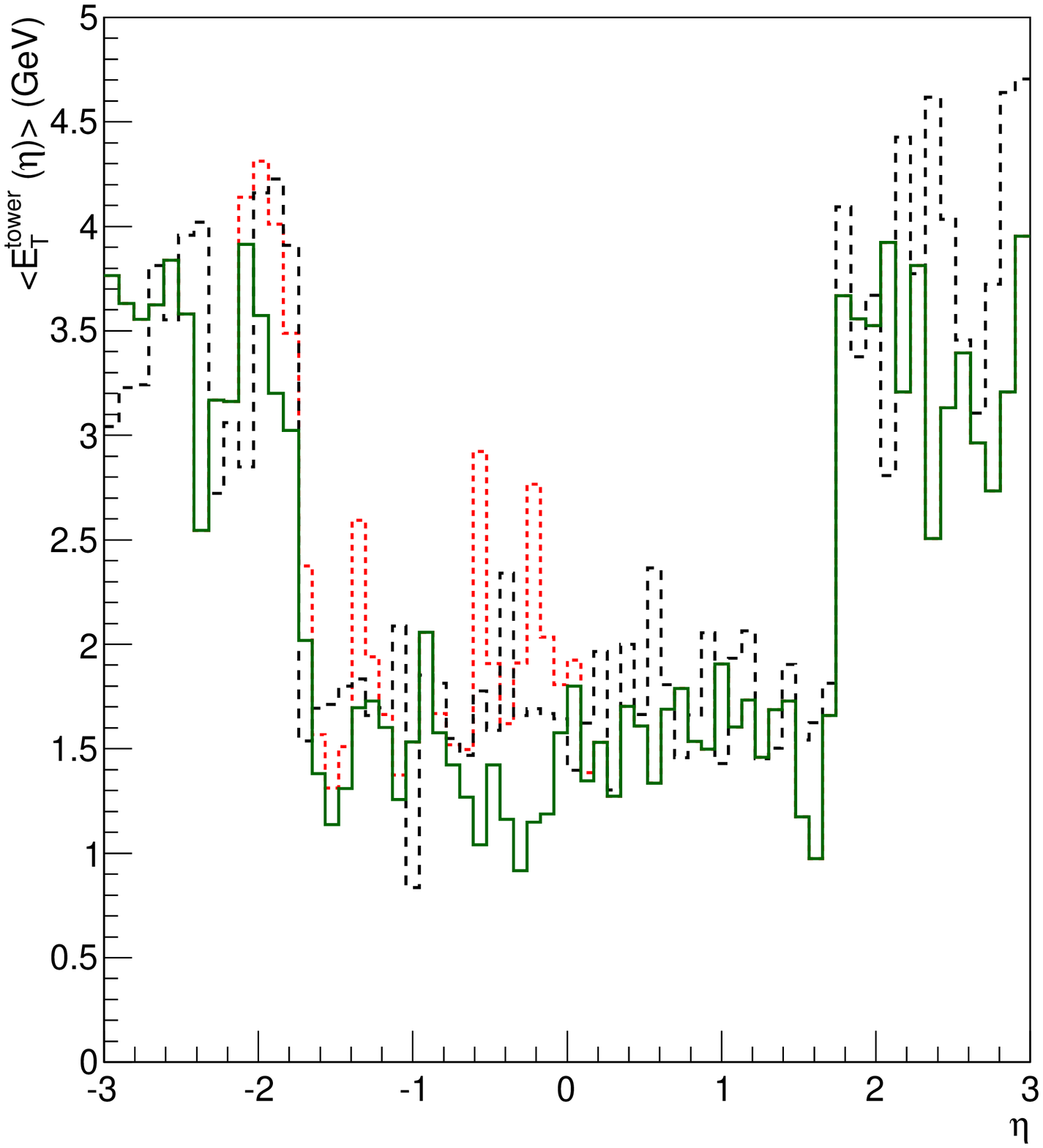}
		}
		\subfigure[Dispersion $\left\langle \sigma_{T}^{tower} (\eta) \right\rangle$.]{
			\label{fig:SpecCMS2}
			\includegraphics[width=0.48\textwidth]{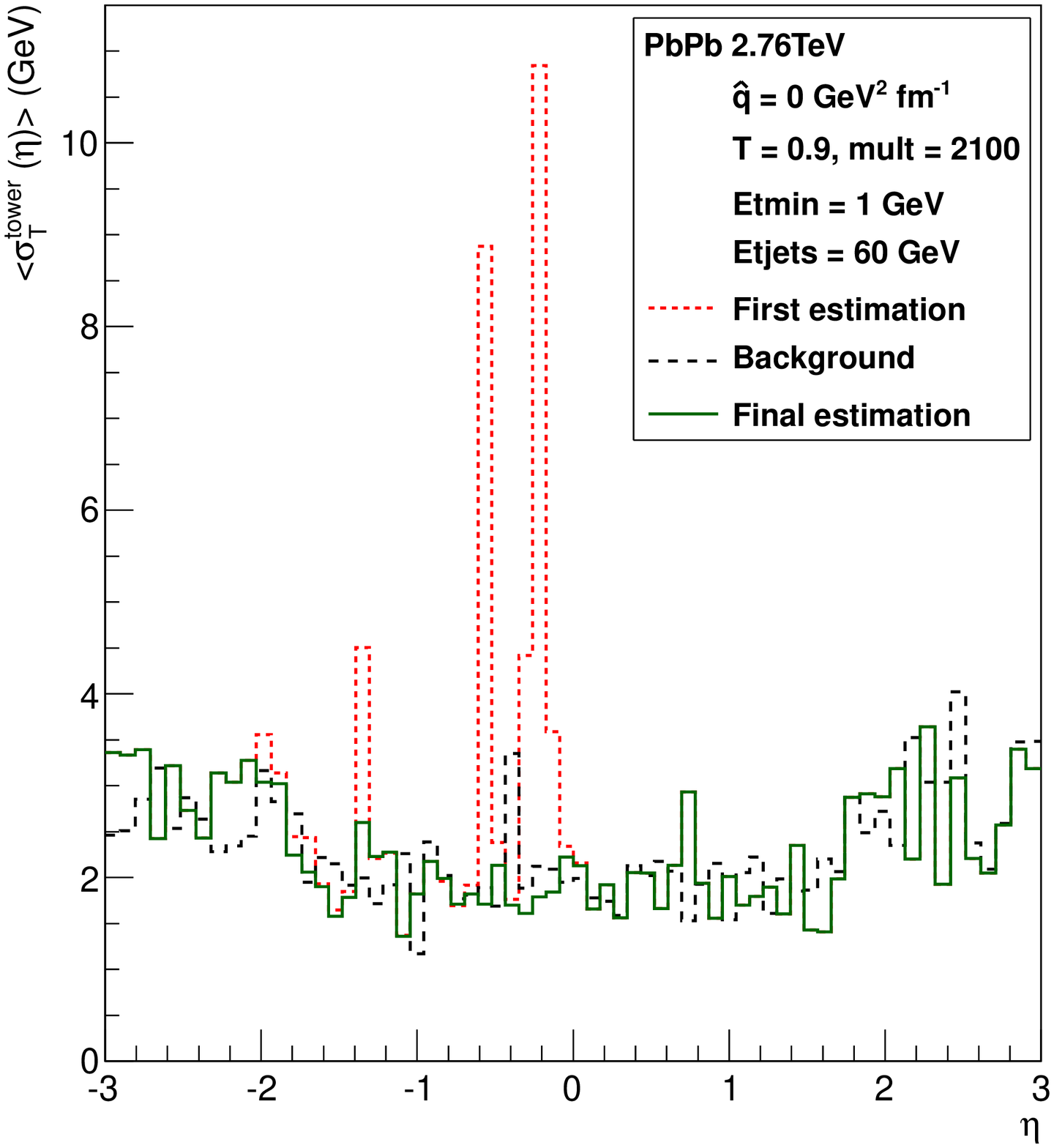}
		}
		\caption{Dependence of the $\phi$-averaged background parameters on pseudo-rapidity $\eta$, for a simulation using Q-PYTHIA with $\hat{q} = 0$ (i.e. unquenched jets) embedded in a background with $T = 0.9$ GeV. The red dotted line corresponds to the first estimation of background parameters and the green solid to the final one. The black dashed line corresponds to the background parameters when using simulated events containing only background.}
		\label{fig:CMS}
	\end{center}
\end{figure}

\begin{figure}[htbp]
	\begin{center}
			\includegraphics[width=0.48\textwidth]{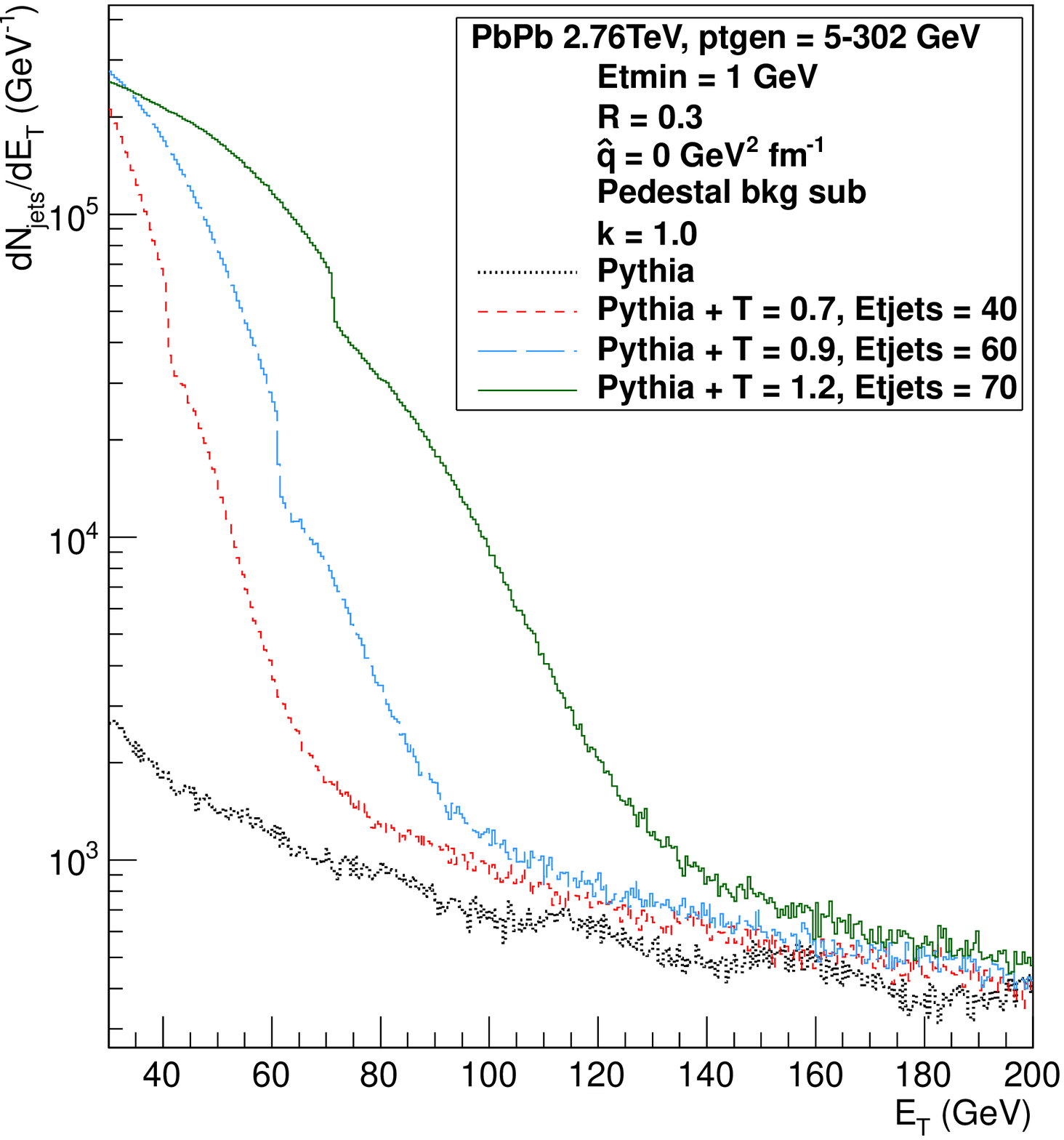}
		\caption{Inclusive jet spectrum of pure (unquenched) PYTHIA events (black dotted line) and spectrum of subtracted jets for PYTHIA embedded in  the toy model background  with a $T = 0.7$ (red dashed line), $T = 0.9$ (blue long dashed line) and $T = 1.2$ GeV (green solid line), reconstructed using a pedestal technique with $\kappa=1$.}
		\label{fig:SpecCMSappb}
	\end{center}
\end{figure}

\begin{figure}[htbp]
	\begin{center}
		\subfigure[Dijet asymmetry $A_J$.]{
			\label{fig:AjFluctCMSappb}
			\includegraphics[width=0.48\textwidth]{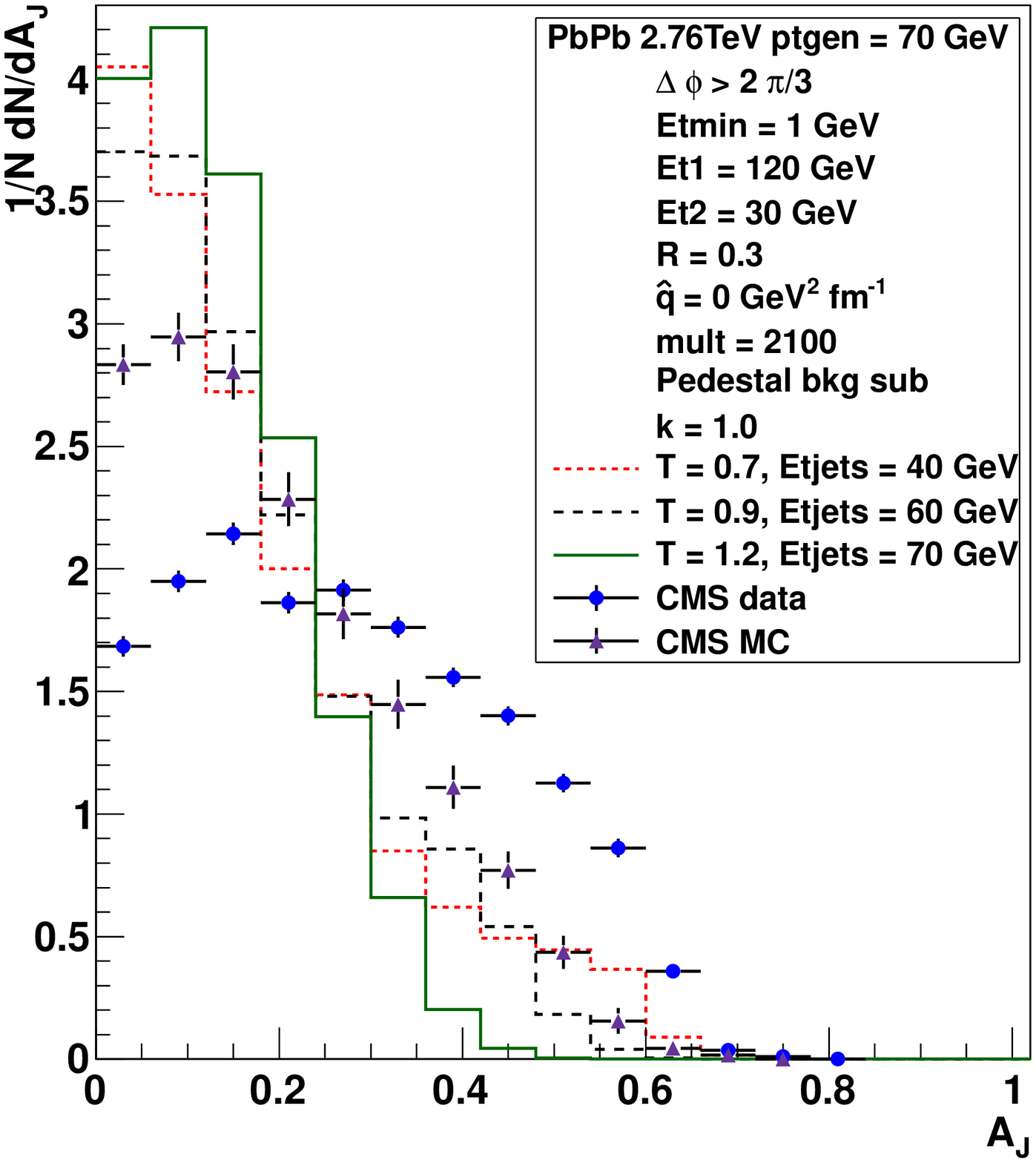}
		}
		\subfigure[Dijet azimuthal correlation.]{
			\label{fig:DphiFluctCMSappb}
			\includegraphics[width=0.48\textwidth]{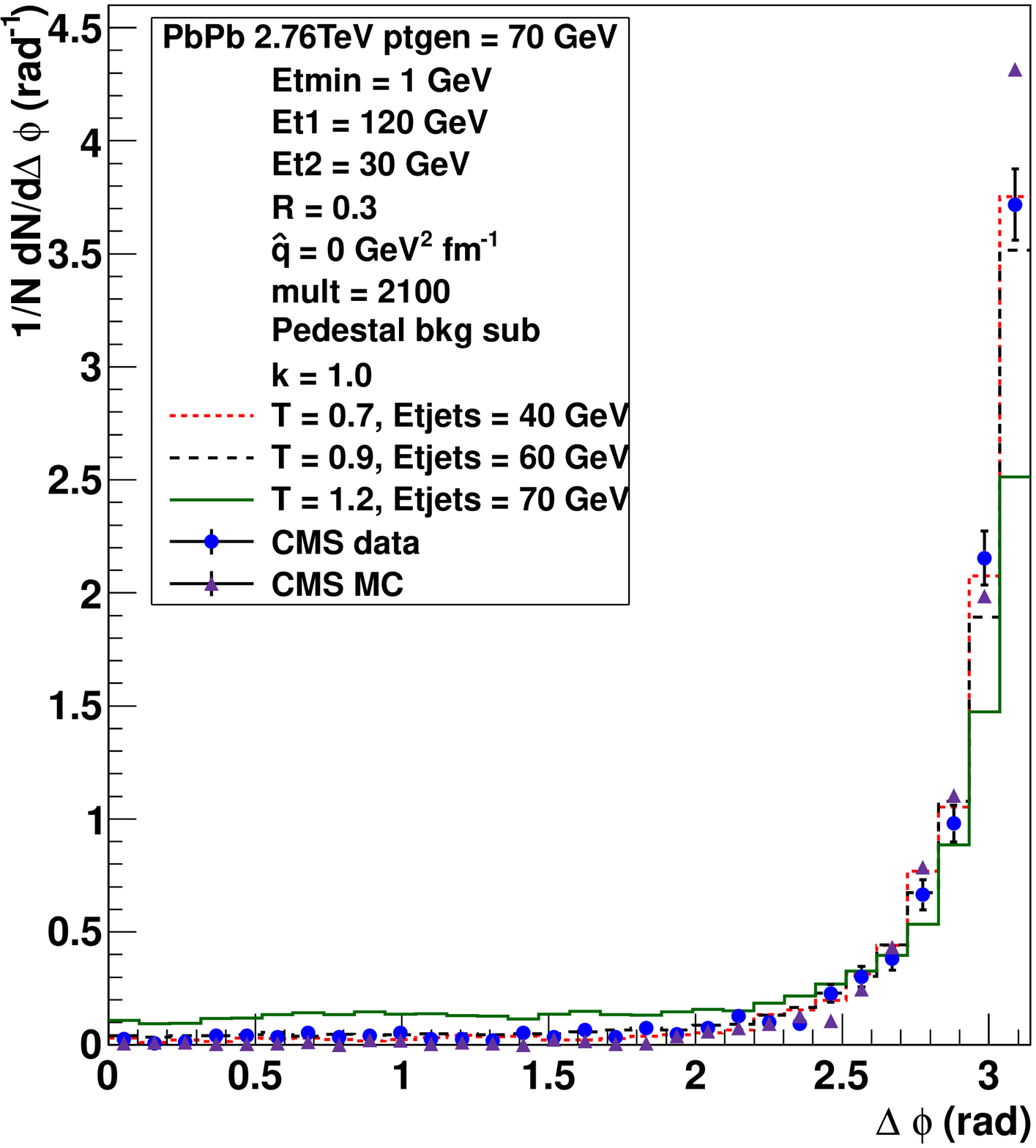}
		}
		\caption{Dijet observables for a simulation using Q-PYTHIA with $\hat{q} = 0$ embedded in a background with different $T$'s. The red dotted lines correspond to a background with $T = 0.7$, the black dashed ones to a $T = 0.9$ and the green solid ones to a $T = 1.2$ GeV. The blue points are the CMS data with the corresponding error bars and the purple triangles the CMS Monte Carlo \cite{Chatrchyan:2012nia}. A pedestal technique with $\kappa=1$ has been used for background subtraction.}
	\end{center}
\end{figure}

\par The inclusive spectra for the case of the pedestal subtraction method with $\kappa=1$ is shown in Figure \ref{fig:SpecCMSappb}. The impact of background fluctuations is stronger than for the other methods or procedures discussed previously. The jet yield is enhanced with respect to PYTHIA up to $E_{T} \sim 200$ GeV for all considered temperatures. {But this effect becomes smaller and the approach to the un-embedded spectrum happens at smaller $E_T$, for lower $T$'s}. The effect of the value of $E_{T,jets}$ used in each case is visible as a discontinuity in the spectrum for $E_T\simeq E_{T,jets}$ (but  only  in the case of a uniform background, see in Appendix \ref{appA} the results using a non-uniform one coming from PSM).

\par While we have not found a satisfactory explanation of this fact, let us note that the two methods make their respective background estimations in a rather different way. For the pedestal one it is made in an early stage (at the particle level) and it is linked with the granularity of the calorimeter in $\eta$, while in FastJet $\rho$ and $\sigma_{jet}$ are determined using the information from a list of jets defined using the $k_t$ algorithm that does not result in a rigid $\eta\times \phi$ shape. As a consequence, the background parameters coming from the pedestal method may become more sensitive of the background structure regarding region-to-region fluctuations than the ones coming from FastJet. All in all, it looks as if the subtraction of the background using FastJet works for large $E_T$ jets while gets a clear contribution from back-reaction at lower $E_T$, while for the pedestal subtraction we see some shift at large $E_T$ that may come from an underestimation of the background due to the zeroing of negative cells in steps 2, 5 in Subsection \ref{cmsproc}.

\par In Figure \ref{fig:AjFluctCMSappb} the dijet asymmetry is shown for different values of $T$. In contrast to the other methods or procedures discussed previously, we observe that the asymmetry is reduced with increasing $T$. This may be linked to the increasing shift of the single inclusive distributions discussed above.

\par The effect of the background subtraction on the azimuthal correlation of the dijet pair is  shown in Figure \ref{fig:DphiFluctCMSappb}. The azimuthal correlation shows a pedestal in the whole $\Delta \phi$ range for the highest fluctuations, more pronounced in this case of $\kappa=1$ than for the previously discussed procedure with $\kappa=2.2$. A tentative explanation is that this pedestal comes from the fake, combinatorial jets which are uncorrelated with the hard PYTHIA dijet. This effect may be more pronounced for the pedestal method as the zeroing procedure leaves some regions of phase space for jet reconstruction empty. Thus, the azimuthal correlation between some reconstructed jets may become washed out.

\begin{figure}[htbp]
	\begin{center}
		\subfigure[Dijet asymmetry $A_J$.]{
			\label{fig:AjFlowappb}
			\includegraphics[width=0.48\textwidth]{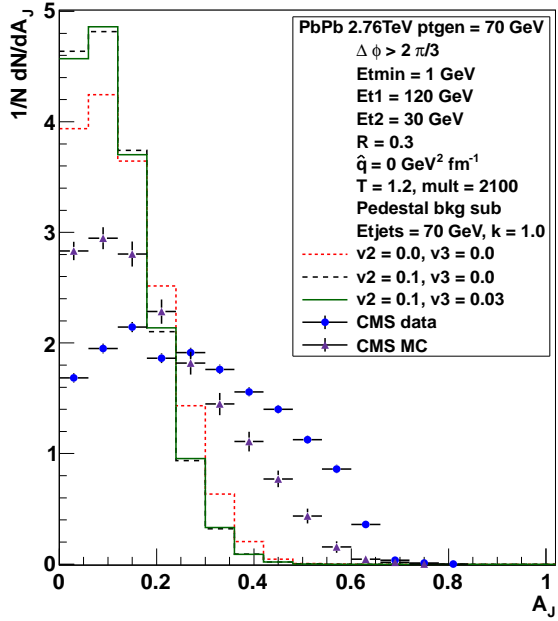}
		}
		\subfigure[Dijet azimuthal correlation.]{
			\label{fig:DphiFlowappb}
			\includegraphics[width=0.48\textwidth]{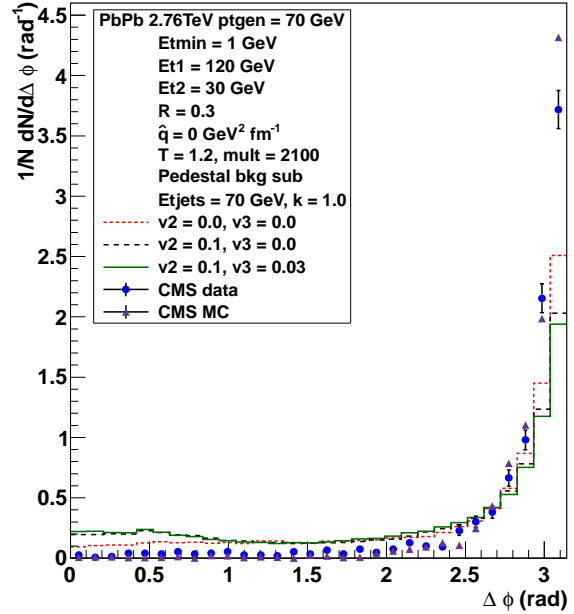}
		}
		\caption{Dijet observables for a simulation using Q-PYTHIA with $\hat{q} = 0$ embedded in a background with $T = 1.2$ GeV. The red dotted lines correspond to a simulation without flow, the black dashed ones with an elliptic flow component ($v_2 = 0.1$) and the green solid ones with an additional triangular flow component ($v_2 = 0.1, v_3  = 0.03$). The blue dots are the CMS data with the corresponding error bars and the purple triangles the CMS Monte Carlo \cite{Chatrchyan:2012nia}. A pedestal technique with $\kappa=1$ has been used for background subtraction.}
	\end{center}
\end{figure}

\par Figure \ref{fig:AjFlowappb} shows the dijet asymmetry for three extreme cases ($v_{2}=0$,$v_{3}=0$;$v_{2}=0.1$,$v_{3}=0$;$v_{2}=0.1$,$v_{3}=0.03$). 
The dijet asymmetry shows negligible dependence on flow, as it was the case for previous procedures of background subtraction.

\begin{figure}[htbp]
	\begin{center}
		\subfigure[Results for $v_2 = 0, v_3 = 0$.]{
			\label{fig:Corr1appb}
			\includegraphics[width=0.31\textwidth]{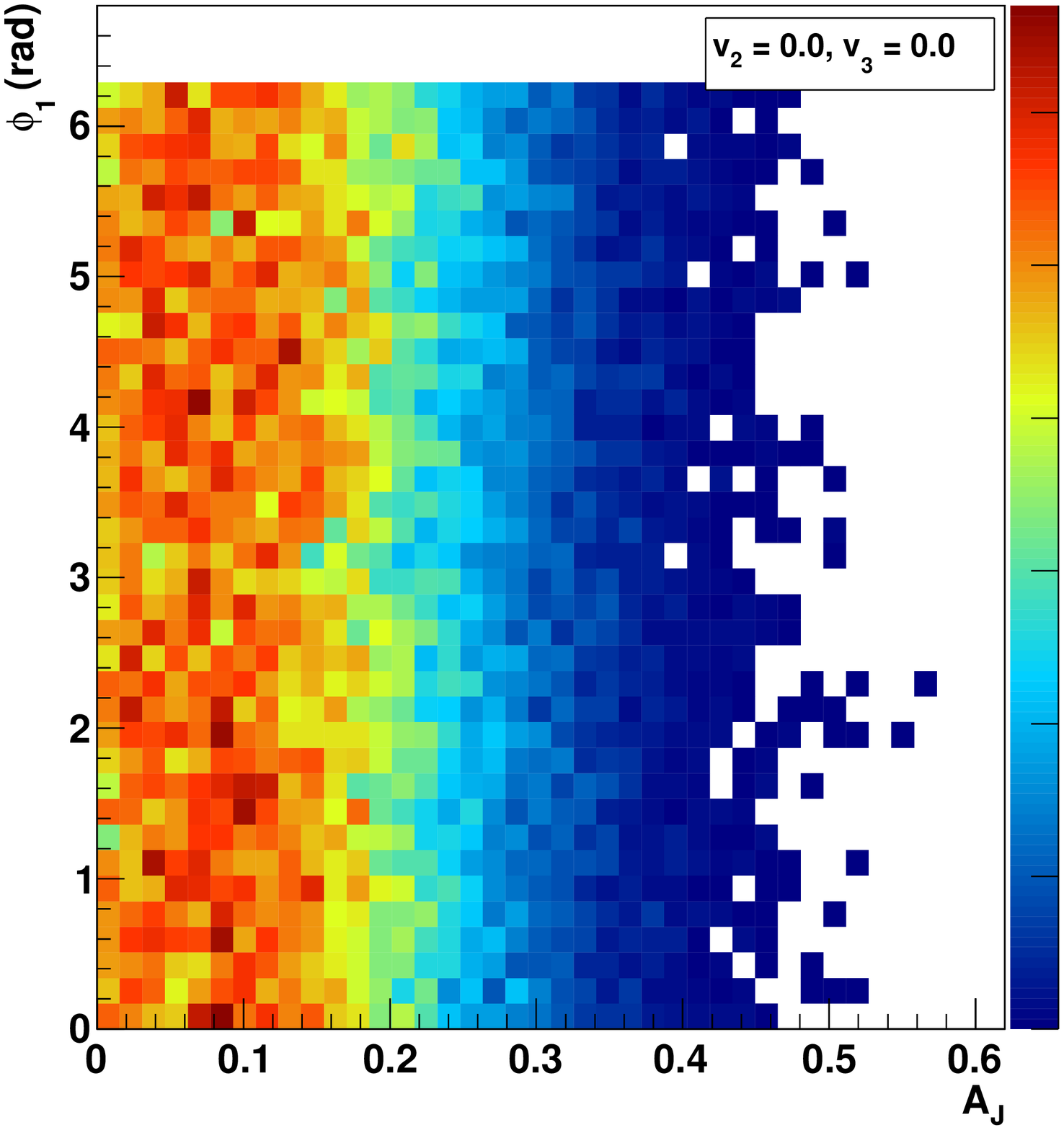}
		}
		\subfigure[Results for $v_2 = 0.1, v_3 = 0$.]{
			\label{fig:Corr2appb}
			\includegraphics[width=0.31\textwidth]{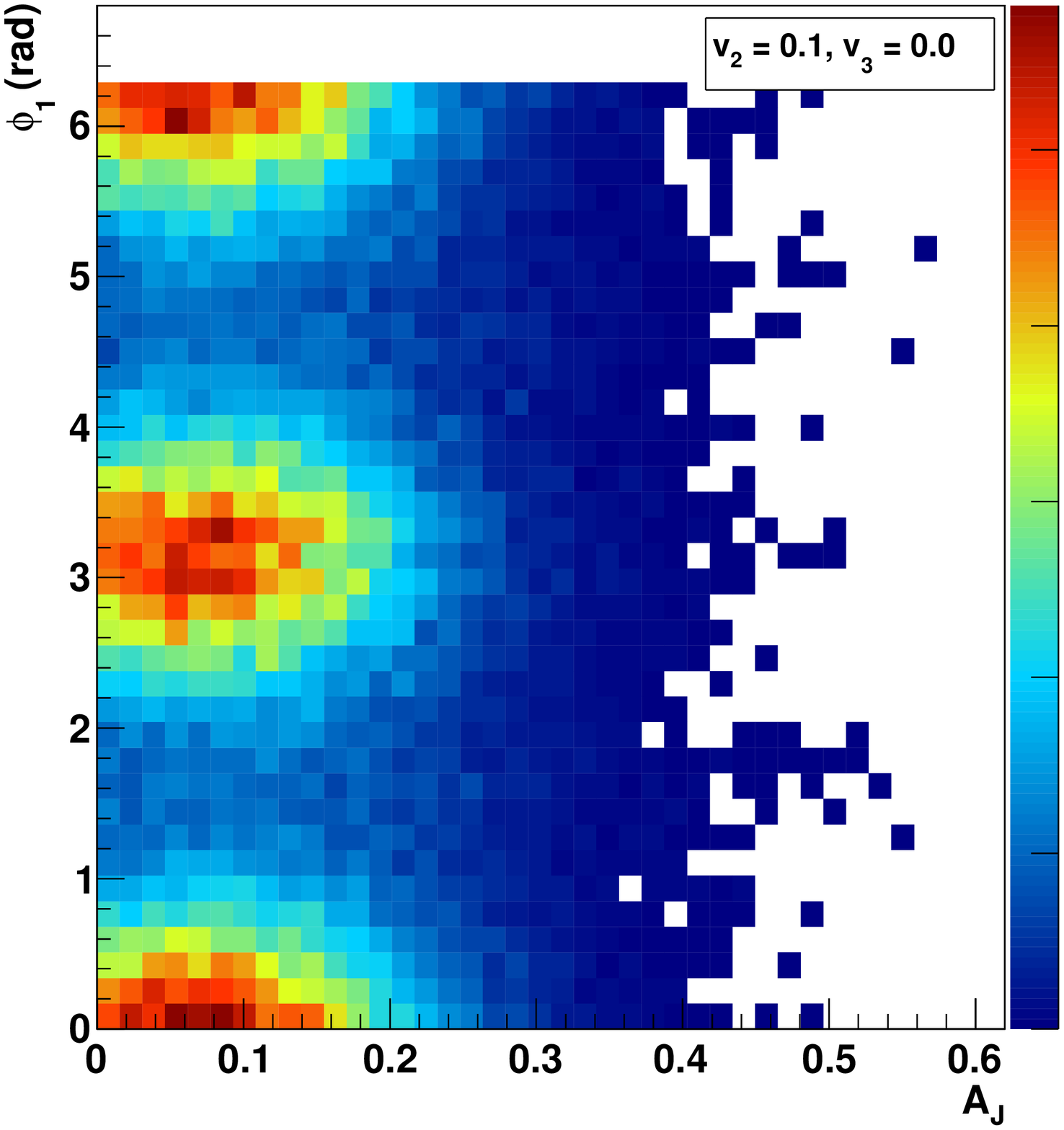}
		}
		\subfigure[Results for $v_2 = 0.1, v_3 = 0.03$.]{
			\label{fig:Corr3appb}
			\includegraphics[width=0.31\textwidth]{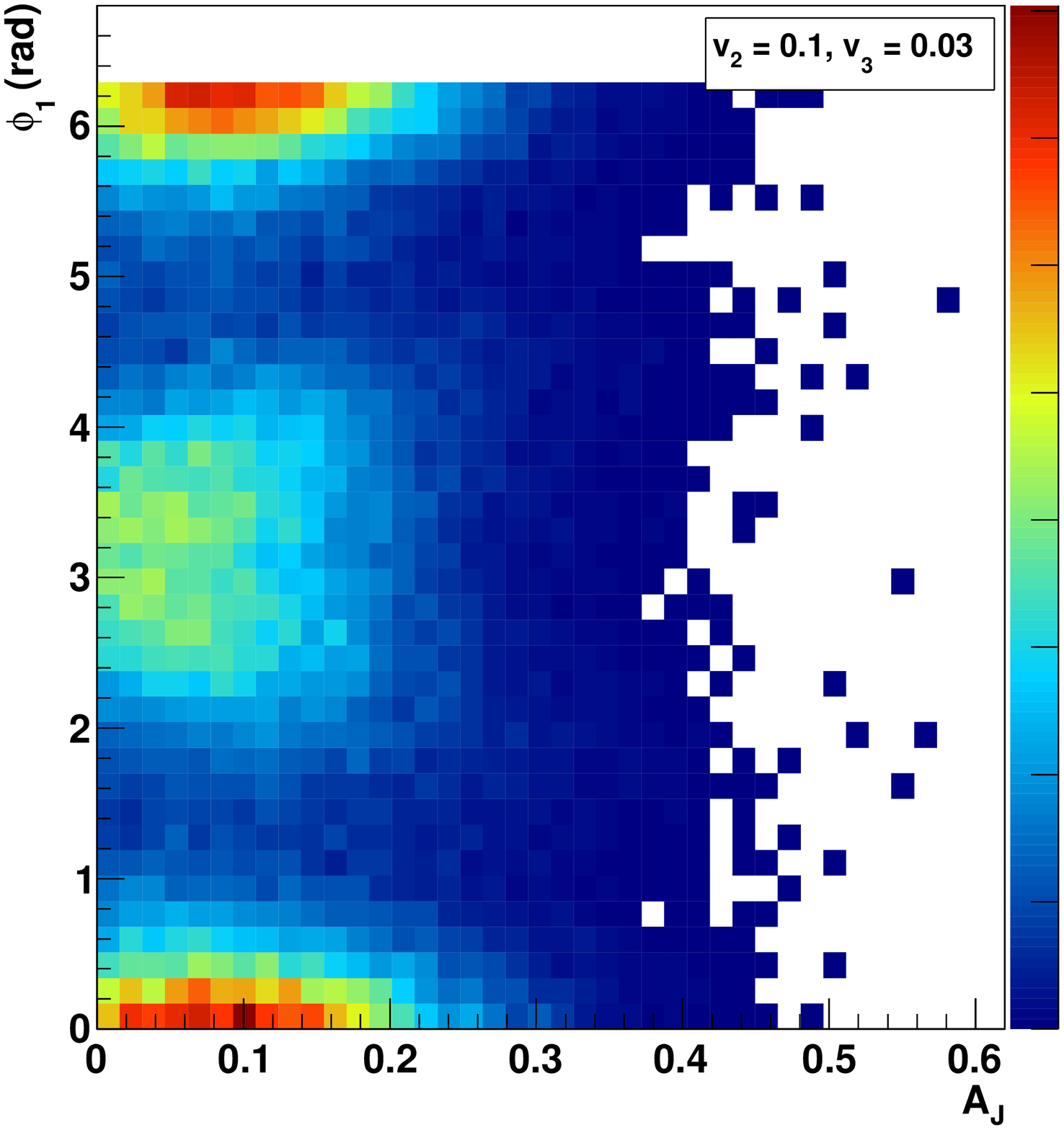}
		}
		\caption{Correlation between $A_J$ and $\phi_1$ for a simulation using Q-PYTHIA with $\hat{q} = 0$ embedded in a background with $T = 1.2$ GeV with a reaction plane fixed to $\phi_{RP} = 0$. The background subtraction was made using a pedestal method with $\kappa=1$. The colour gradation from deep blue to deep red denotes increasing correlations.}
		\label{fig:Corrappb}
	\end{center}
\end{figure}

\begin{figure}[htbp]
	\begin{center}
		\subfigure[{Dijet asymmetry $A_J$.}]{
			\label{fig:Aj_Quench_CMSappb}
			\includegraphics[width=0.48\textwidth]{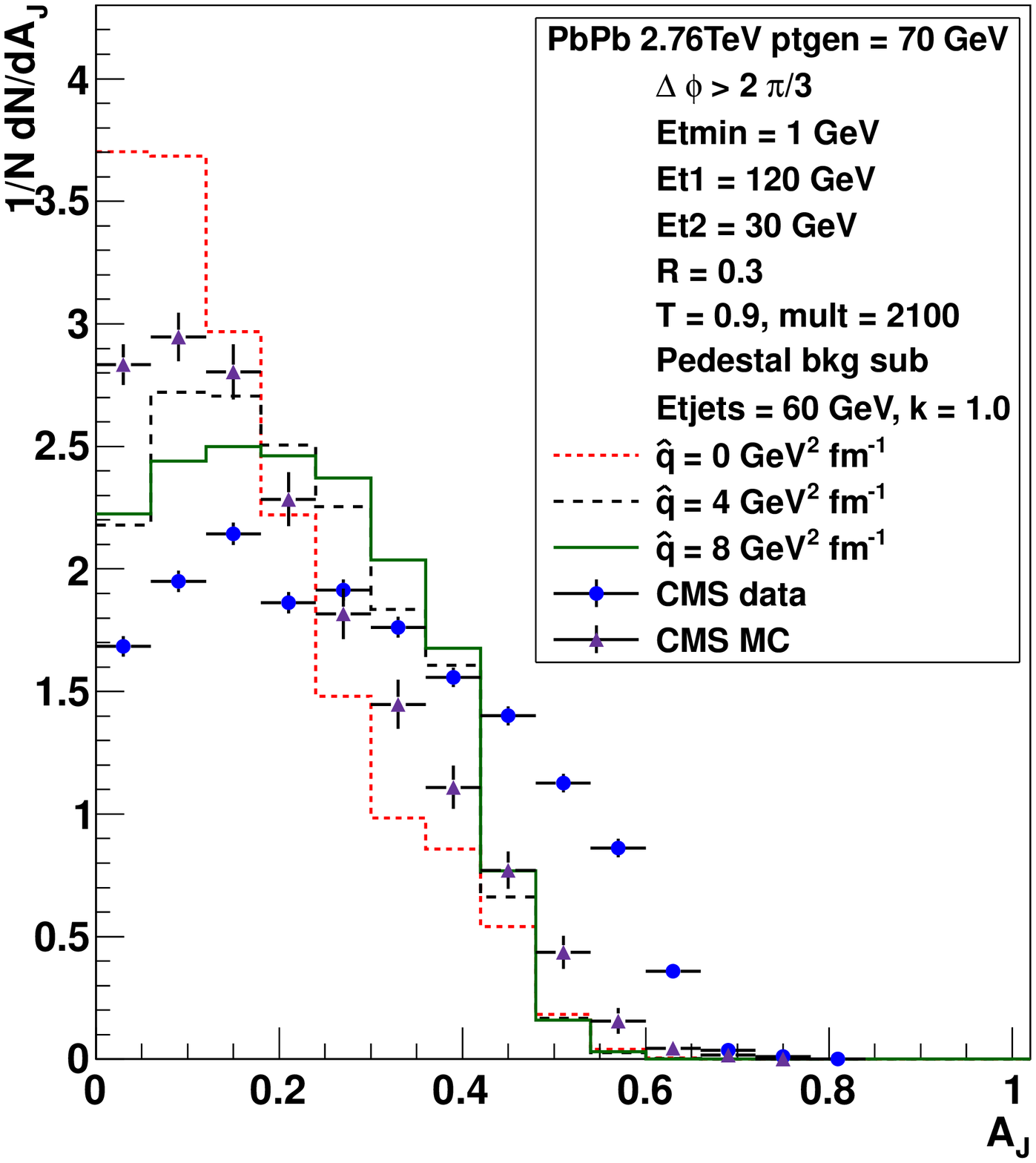}
		}
		\subfigure[{Dijet azimuthal correlation.}]{
			\label{fig:Dphi_Quench_CMSappb}
			\includegraphics[width=0.48\textwidth]{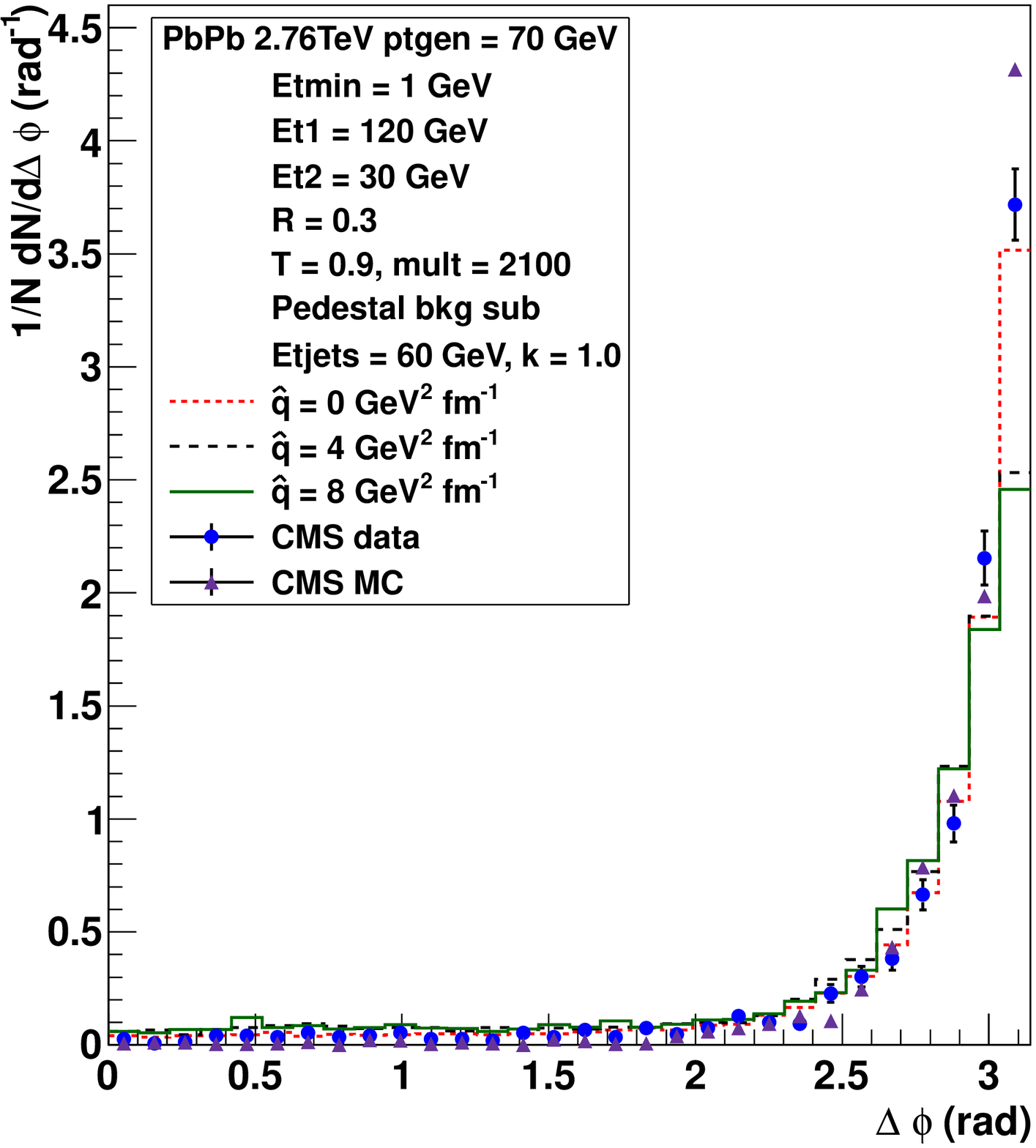}
		}
		\caption{{Dijet observables for a simulation using Q-PYTHIA with different $\hat{q}$   embedded in a background with $T = 0.9$ GeV ($\sigma_{jet} \simeq 11$ GeV). The red dotted lines corresponds to  $\hat{q} = 0$, the black dashed ones to $\hat{q} = 4$ GeV$^{2}$ fm$^{-1}$ and the green solid ones to $\hat{q} = 8$ GeV$^{2}$ fm$^{-1}$. The blue dots are the CMS data with the corresponding error bars, and the purple triangles the CMS Monte Carlo. The background subtraction was made using a pedestal method with $\kappa=1$.}}
		\label{fig:Quench_CMSappb}
	\end{center}
\end{figure}

\par On the other hand and more pronouncedly than for previously discussed methods, the azimuthal correlation, see Figure \ref{fig:DphiFlowappb}, shows a bump near $\Delta \phi \sim 0$. The origin is the same as discussed in Subsection \ref{flow}, see Figure \ref{fig:Corrappb}.

\par Finally, the effects of quenching are shown in Figure \ref{fig:Quench_CMSappb}. The momentum imbalance induced by quenching exists, but it is less significant than for the previously discussed  cases and the agreement with data is consequently poorer even for $\hat{q}=8$ GeV$^2$ fm$^{-1}$. The azimuthal correlations are slightly broader. An explanation of the latter fact would go along the lines developed in Subsection \ref{fluct}, with the broader quenched jets producing additional soft particles that will be distributed over a wide range of the phase space. Hence, the azimuthal position of the reconstructed jets becomes less correlated.

\clearpage






\bibliographystyle{JHEP}
\bibliography{Bibliography}

\end{document}